# AGU Advances

# Oxygen False Positives on Habitable Zone Planets Around Sun-Like Stars

Joshua Krissansen-Totton[1,2,3] , Jonathan J. Fortney[1] , Francis Nimmo[4] , and Nicholas Wogan[2,5]

[1]Department of Astronomy and Astrophysics, University of California, Santa Cruz, Santa Cruz, CA, USA, [2]NASA Nexus for Exoplanet System Science, Virtual Planetary Laboratory Team, University of Washington, Seattle, WA, USA, [3]NASA Sagan Fellow, [4]Department of Earth and Planetary Sciences, University of California, Santa Cruz, Santa Cruz, CA, USA, [5]Department of Earth and Space Sciences, University of Washington, Seattle, WA, USA

**Author Contributions:**
**Conceptualization:** Joshua Krissansen-Totton, Jonathan J. Fortney
**Formal analysis:** Joshua Krissansen-Totton
**Methodology:** Joshua Krissansen-Totton, Francis Nimmo, Nicholas Wogan
**Software:** Joshua Krissansen-Totton, Nicholas Wogan
**Supervision:** Jonathan J. Fortney
**Writing – original draft:** Joshua Krissansen-Totton

**Abstract** Oxygen is a promising exoplanet biosignature due to the evolutionary advantage conferred by harnessing starlight for photosynthesis, and the apparent low likelihood of maintaining oxygen-rich atmospheres without life. Hypothetical scenarios have been proposed for non-biological oxygen accumulation on planets around late M-dwarfs, where the extended pre-main sequence may favor abiotic $O_2$ accumulation. In contrast, abiotic oxygen accumulation on planets around F, G, and K-type stars is seemingly less likely, provided they possess substantial non-condensable gas inventories. The comparative robustness of oxygen biosignatures around larger stars has motivated plans for next-generation telescopes capable of oxygen detection on planets around sun-like stars. However, the general tendency of terrestrial planets to develop oxygen-rich atmospheres across a broad range of initial conditions and evolutionary scenarios has not been explored. Here, we use a coupled thermal-geochemical-climate model of terrestrial planet evolution to illustrate three scenarios whereby significant abiotic oxygen can accumulate around sun-like stars, even when significant noncondensable gas inventories are present. For Earth-mass planets, we find abiotic oxygen can accumulate to modern levels if (1) the $CO_2$:$H_2O$ ratio of the initial volatile inventory is high, (2) the initial water inventory exceeds ~50 Earth oceans, or (3) the initial water inventory is very low (<0.3 Earth oceans). Fortunately, these three abiotic oxygen scenarios could be distinguished from biological oxygen with observations of other atmospheric constituents or characterizing the planetary surface. This highlights the need for broadly capable next-generation telescopes that are equipped to constrain surface water inventories via time-resolved photometry and search for temporal biosignatures or disequilibrium combination biosignatures to assess whether oxygen is biogenic.

**Plain Language Summary** Next-generation telescopes will search for life on exoplanets by looking for the spectral signatures of biogenic gases. Oxygen has been considered a reliable biosignature gas, especially for planets around sun-like stars where non-biological, photochemical production is unlikely. This motivates plans for future telescopes specifically designed for oxygen detection. Here, we develop a coupled model of the atmosphere-interior evolution of terrestrial planets to show that lifeless planets in the habitable zone can develop oxygen-rich atmospheres relatively easily. These false positives for biological oxygen could be distinguished from inhabited planets using other contextual clues, but their existence implies next-generation telescopes need to be capable of characterizing planetary environments and searching for multiple lines of evidence for life, not merely oxygen.

## 1. Introduction

The search for life beyond Earth is a key motivator for exoplanet astronomy. Among the various life detection approaches that have been proposed, atmospheric oxygen is arguably the most promising biosignature. This is because any organism that adapts to exploit free energy from starlight will have a competitive advantage over organisms that are limited by geochemical sources of free energy. The earliest incontrovertible evidence for life on Earth around 3.5 Ga coincides with fossilized photosynthetic stromatolites (Buick, 2008). While it is unknown whether the evolution of oxygenic photosynthesis is contingent—the metabolism emerged only once in Earth's history (Mulkidjanian et al., 2006)—oxygenic photosynthesis confers a unique evolutionary advantage over other forms of photosynthesis since the required substrates, carbon dioxide

Writing – review & editing: Joshua
Krissansen-Totton, Jonathan J. Fortney,
Francis Nimmo, Nicholas Wogan

($CO_2$) and water ($H_2O$) are likely ubiquitous on habitable worlds. Moreover, the accumulation of biogenic oxygen in planetary atmospheres, as has occurred on Earth (Holland, 2006; Lyons et al., 2014), is readily detectable over interstellar distances thanks to absorption features in the visible, near IR, and (for ozone) thermal infrared (reviewed by Meadows et al., 2018). Indeed, three of the four proposed mission concepts currently under consideration by the 2020 Astrophysics Decadal Survey are specifically designed to be capable of life detection via oxygen or ozone (Fischer et al., 2019; Gaudi et al., 2020; Meixner et al., 2019). Ground-based Extremely Large Telescopes may also be capable of oxygen detection via high resolution spectroscopy (Leung et al., 2020; Rodler & López-Morales, 2014; Snellen et al., 2013). The James Webb Space Telescope (JWST) could conceivably detect oxygen/ozone biosignatures for nearby transiting planets, although this would probably require a prohibitively large number of transits for known targets (Fauchez et al., 2020; Krissansen-Totton, Garland, et al., 2018; Lustig-Yaeger et al., 2019; Wunderlich et al., 2019). Finally, oxygen is often considered to be a good biosignature because—apart from the exceptions discussed below—it is seemingly difficult for habitable zone terrestrial planets to maintain oxygen-rich atmospheres without life (reviewed by Meadows et al., 2018).

The terrestrial planets that will be accessible to spectroscopic characterization with JWST and ground-based ELTs will orbit M-dwarfs due to the favorable signal-to-noise of M-dwarf transits. This is unfortunate from the standpoint of recognizing oxygen biosignatures because several features of M-dwarfs make them susceptible to non-biological oxygen accumulation. In particular, the extended pre-main sequence of late M-dwarfs could yield habitable zone terrestrial planets with hundreds or thousands of bar $O_2$ from XUV-driven hydrogen loss (Luger & Barnes, 2015). At least some of this oxygen will likely dissolve in a surface magma ocean and be sequestered in the mantle, but retaining oxygen-rich atmospheres is still possible, especially for highly irradiated terrestrial planets (Barth et al., 2020; Schaefer et al., 2016; Wordsworth et al., 2018). The early development of abiotic, oxygen-rich atmospheres could prevent the subsequent emergence of life by precluding prebiotic chemistry (Wordsworth et al., 2018). It has also been argued that the spectral energy distributions of M-dwarfs are favorable for the photochemical accumulation of $O_2$ and $O_3$ (Domagal-Goldman et al., 2014; Gao et al., 2015; Harman et al., 2015; Tian et al., 2014). While recent assessments of the role of lightning in such photochemical models (Harman et al., 2018) and improved near-UV water cross sections (Ranjan et al., 2020) may preclude some of these scenarios, photochemical runaways yielding $O_2$-CO rich atmospheres remain a strong possibility for late M dwarfs (Hu et al., 2020). More fundamentally, however, whether M-dwarf terrestrial planets can sustain habitable surface conditions, or indeed any kind of substantial atmosphere, for billions of years given the harsh stellar radiation environment is unknown (Airapetian et al., 2017; Dong et al., 2018; Shields et al., 2016; Zahnle & Catling, 2017). Premature claims of biogenic oxygen on M-dwarf planets will rightfully be met with skepticism.

These challenges to M-dwarf habitability and potential ambiguities with oxygen biosignatures motivates the need to characterize habitable zone planets around F, G, and K-type stars (Arney et al., 2019; National Academies of Sciences & Medicine, 2018). The relatively short pre-main sequence of larger stars mean that the accumulation of non-biological oxygen is less likely. Indeed, the most plausible mechanism for non-biological $O_2$ accumulation on habitable zone planets around sun-like stars is due to small atmospheric inventories of non-condensable species like $N_2$: this weakens the water cold trap and may produce high stratospheric water abundances and therefore elevated H escape rates (Kleinböhl et al., 2018; Wordsworth & Pierrehumbert, 2014). Whether terrestrial planets are likely to form with low nitrogen atmospheric inventories is unknown, but in principle, it might be possible to rule out low non-condensable scenarios by looking for the spectral signatures of abundant nitrogen (Schwieterman et al., 2015) or via inferring background gas abundances from retrieved surface pressure (Feng et al., 2018).

The broader question as to whether habitable zone terrestrial planet evolution tends toward anoxic atmospheres, or whether non-biological oxygen production via water photodissociation and hydrogen escape can overwhelm geochemical sinks, has not been explored. The robustness of oxygen biosignatures rests on the assumption that for temperate planets with effective cold traps, small abiotic oxygen source fluxes from H escape will be overwhelmed by geological sinks. To test this assumption, it is necessary to model the redox evolution of terrestrial planet from formation onwards. This is because planetary redox evolution depends on both the initial state of the atmosphere and mantle after the magma ocean has solidified, and on the subsequent internal evolution and atmospheric state. Interior evolution dictates crustal production rates and outgoing fluxes, which determine the efficiency of geologic sinks of oxygen. Moreover, self-consistent modeling of surface climate and weathering processes is necessary because these modulate surface volatile inventories and oxygen production via atmospheric H escape.

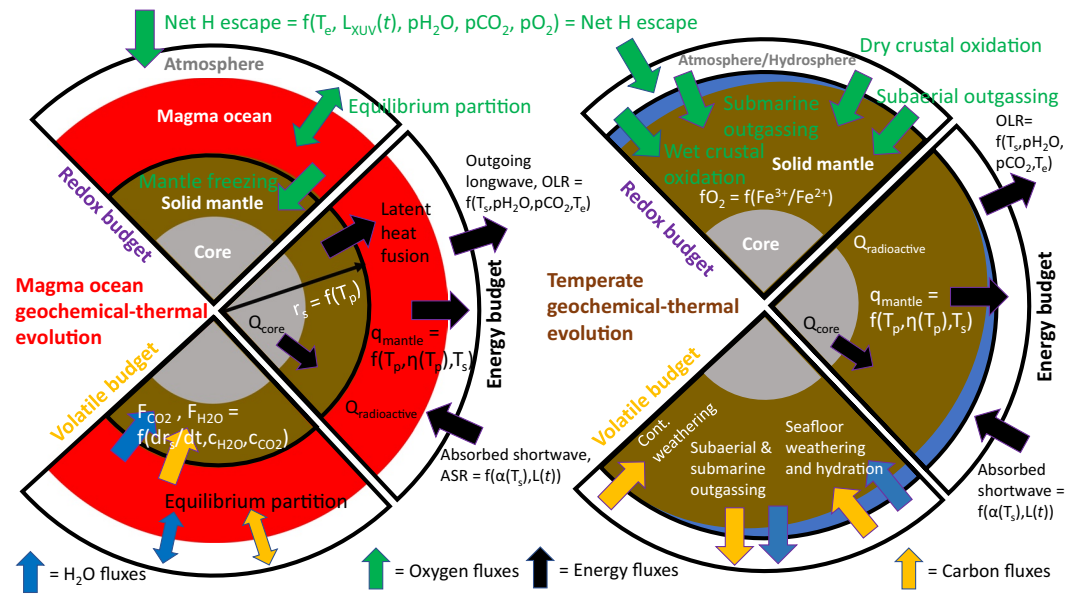

**Figure 1.** Schematic of terrestrial planet geochemical evolution model. The planetary redox budget, thermal-climate evolution, and volatile budget are modeled from initial magma ocean (left) through to temperate geochemical cycling (right). Oxygen fluxes are shown by green arrows, energy fluxes by black arrows, carbon fluxes by orange arrows, and water fluxes by blue arrows. Note that the net loss of H to space adds oxygen to the atmosphere. During the magma ocean phase, the radius of solidification, $r_s$, begins at the core-mantle boundary and moves toward the surface as internal heat is dissipated. The rate at which this occurs is controlled by radiogenic heat production, $Q_{radioactive}$, convective heatflow from the mantle to the surface, $q_{mantle}$, and core heatflow, $Q_{core}$. This internal heatflow balances the difference between outgoing longwave radiation, OLR, and incoming shortwave radiation, ASR. The oxygen fugacity of the mantle, $fO_2$, and the water and carbon content mantle and surface reservoirs are tracked throughout.

Here, we explore the tendency of terrestrial planets to produce abiotic, oxygen-rich atmospheres with a fully coupled model of planetary redox, climate, and thermal evolution. Given the vagaries of planetary accretion, migration, and core formation, it is likely that habitable zone terrestrial planets exist with a broad range of initial volatile inventories (Raymond et al., 2004, 2013; Righter, 2015). Indeed, both planet formation models (Raymond et al., 2004) and exoplanet demographics suggest that water-rich terrestrial planets are abundant (Zeng et al., 2019), and several candidate waterworlds are amenable to JWST characterization (Benneke et al., 2019; Grimm et al., 2018). In contrast, transit and thermal phase curve observations show that the terrestrial exoplanet LHS 3844b is either a bare rock or possesses a thin, high mean-molecular weight atmosphere susceptible to stellar wind erosion (Diamond-Lowe et al., 2020; Kreidberg et al., 2019), and modeling implies it must have formed with less volatiles than the Earth (Kane et al., 2020). Relative abundances of volatiles may also be variable; C/H mass ratios in carbonaceous chondrites vary from ~1.6 to 6.4 (Nittler et al., 2004). We thus apply our model to explore the possibility of non-biological oxygen accumulation for a range of initial volatile inventories. Crucially, we avoid varying background $N_2$ abundances because our focus is on investigating oxygen false positives on planets with comparatively high background non-condensables inventories, that is, sufficient to maintain a cold trap under temperate conditions.

## 2. Materials and Methods

Our model is summarized schematically in Figure 1. A complete description is provided in the supporting information and the Python code is open source (https://doi.org/10.5281/zenodo.4539040). Here, we summarize the salient features of the model necessary for interpreting key findings. There are no biological sources and sinks of volatiles because we are testing the capacity of lifeless planets to accumulate oxygen. Unless stated otherwise, all calculations assume an Earth-mass planet at 1 AU.

Planetary evolution is divided into an initial magma ocean phase, and a subsequent solid-mantle phase, as shown in Figure 1, although a planet may transition between magma ocean and solid mantle multiple times. The model is initialized with a fully molten mantle and some complement of volatiles, radiogenic inventory, and an initial mantle oxygen fugacity (i.e., after core formation, Figure 1, left). Magma ocean solidification follows previous models such as Lebrun et al. (2013) and Schaefer et al. (2016): the magma ocean freezes from the core, upwards, as governed by the following equation:

$$V_{\text{mantle}}\rho_m Q_{\text{radioactive}} - 4\pi q_m r_p{}^2 + Q_{\text{core}} + 4\pi \rho_m H_{\text{fusion}} r_s{}^2 \frac{dr_s}{dt} = V_{\text{mantle}}\rho_m c_p \frac{dT_p}{dt} \tag{1}$$

Here, $V_{\text{mantle}}$ is the volume of the molten mantle, $\rho_m$ is the average density of the mantle, $Q_{\text{radioactive}}$ is radiogenic heat production per unit mass, $r_p$ is planetary radius, $H_{\text{fusion}}$ is the latent heat of fusion of silicates, $r_s$ is the solidification radius, $c_p$ is the specific heat of silicates, $Q_{\text{core}}$ is the heatflow from the metallic core, and $T_p$ is mantle potential temperature. The heatflow from the interior, $q_m$, is calculated using 1-D convective parameterization, with temperature-dependent magma ocean viscosity, $\nu(T_p)$, (see Figure S3 and supporting information Section A.3):

$$q_m = C_{q_m}\left(\frac{(T_p - T_{\text{surf}})^4}{\nu(T_p)}\right)^{1/3} \tag{2}$$

Here, $T_{\text{surf}}$ is mean surface temperature, and $C_{q_m}$ is a constant that depends on thermal conductivity, thermal diffusivity, critical Rayleigh number, gravity, and thermal expansivity. Supporting information Section A.1 describes this convection parameterization in more detail. Equation 1 continues to govern the thermal evolution of the mantle after the magma ocean has solidified with $dr_s/dt = 0.0$.

During magma ocean solidification H, C, and O are partitioned between dissolved melt phases, crystalline phases, and the atmosphere by assuming chemical equilibrium (supporting information Section A.7). The rate at which the mantle freezes is controlled by outgoing longwave radiative (OLR), which is balanced by heat from interior and absorbed shortwave radiation (ASR) from the host star at every timestep (see climate model description below):

$$q_m + \text{ASR} = \text{OLR} \tag{3}$$

During the magma ocean phase, planetary oxidation may occur from the loss of hydrogen to space (less oxygen drag):

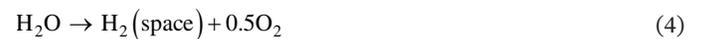

$$\text{H}_2\text{O} \rightarrow \text{H}_2(\text{space}) + 0.5\text{O}_2 \tag{4}$$

Free oxygen produced via H escape is dissolved in the melt and may be transferred to the solid mantle as the magma ocean solidifies (Schaefer et al., 2016). We parameterized atmospheric escape as either diffusion limited or XUV-limited, depending on the composition of the stratosphere and the stellar XUV flux (supporting information Section A.9). During XUV-driven escape of a steam-dominated atmosphere, the hydrodynamic escape of H may drag along O (and even CO₂) following Odert et al. (2018). In contrast, if the stratosphere is mostly dry, then the escape of H will be limited by the rate at which H₂O can diffuse through the cold trap (Wordsworth & Pierrehumbert, 2013), and nothing heavier than H escapes. Standard parameterizations of solar bolometric luminosity (Baraffe et al., 1998, 2002) and XUV luminosity evolution are adopted (Tu et al., 2015), as described in supporting information Section A.4.

A radiative-convective climate model is used to self-consistently calculate surface temperature, OLR, ASR, the water vapor profile, and surface liquid water inventory (if any) during both the magma ocean phase and subsequent temperate evolution. OLR is a function of the surface H₂O and CO₂ inventories, and is calculated using the publicly available correlated-k radiative transfer code of Marcq et al. (2017). To obtain

OLR in the presence of condensable water vapor, a dry adiabat to moist adiabat to isothermal atmospheric structure is assumed (Kasting, 1988). To calculate ASR across a wide range of temperatures, we adapted the albedo parameterization described in Pluriel et al. (2019). Refer to supporting information Section A.5 for full details of radiative transfer calculations along with example outputs.

When heat from accretion and short-lived radiogenics is sufficiently dissipated—the timescale for which is controlled by insolation and greenhouse warming from outgassed volatiles—a planet's surface temperature may drop below the solidus and the magma ocean phase is over. At this point, the model transitions to solid-state mantle convection and temperate geochemical cycling (Figure 1, right). The redox budget during solid-state evolution is modeled as follows: the only source of oxygen is still atmospheric escape using the same parameterization as described above. However, there are now numerous crustal sinks for oxygen including (i) subaerial and submarine outgassing of reduced species (e.g., $H_2$, CO, and $CH_4$), (ii) water-rock serpentinizing reactions that generate $H_2$ (the "wet crustal" sink), and (iii) direct oxidation of surface crust by atmospheric oxygen (the "dry crustal" sink):

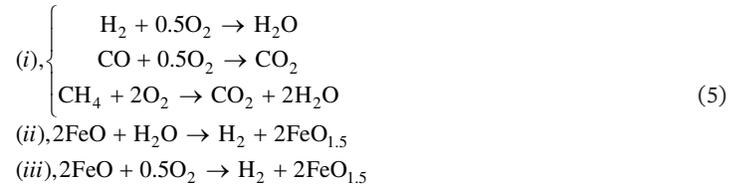

$$(i), \begin{cases} H_2 + 0.5O_2 \rightarrow H_2O \\ CO + 0.5O_2 \rightarrow CO_2 \\ CH_4 + 2O_2 \rightarrow CO_2 + 2H_2O \end{cases}$$
$$(ii), 2FeO + H_2O \rightarrow H_2 + 2FeO_{1.5}$$
$$(iii), 2FeO + 0.5O_2 \rightarrow H_2 + 2FeO_{1.5}$$
$$(5)$$

The sizes of these three oxygen sinks are self-consistently calculated from the planetary interior evolution and mantle volatile content. Outgassing fluxes are calculated using the melt-gas equilibrium outgassing model of Wogan et al. (2020); outgassing fluxes depend on mantle oxygen fugacity, degassing overburden pressure, the volatile content of the mantle, specifically $H_2O$ and $CO_2$ content, and the rate at which melt (new crust) is produced (see supporting information Section A.10). The possible influence of pressure overburden on redox evolution has been discussed previously (Wordsworth et al., 2018), and is quantifiable within our outgassing model framework. Dry and wet crustal sinks for oxygen similarly depend on crustal production rates, and are described in full in supporting information Sections A.12 and A.13. Crustal production is calculated from interior heatflow, which is modulated by temperature-dependent mantle viscosity and radiogenic heat production (supporting information Section A.10). We assume plate tectonics when calculating crustal production rates to maximize crustal sinks of oxygen.

Weathering processes (Krissansen-Totton, Arney, et al., 2018) and the deep hydrological cycle (Schaefer & Sasselov, 2015) are explicitly modeled because climate and surface volatile inventories control crustal oxygen sinks and atmospheric escape processes. Our model incorporates a rudimentary carbon cycle, which is a simplified version of that described in Krissansen-Totton, Arney, et al. (2018). Carbon is added to the atmosphere via magmatic outgassing (described above), and returned via continental weathering and seafloor weathering, whose relative contributions depend on climate and the total surface water inventory. Our carbon cycle parameterization is described in supporting information Section A.11 and the deep hydrological cycle in supporting information Section A.12. Self-consistent climate modeling enables an assessment of whether abiotic oxygen can coexist with a habitable surface climate, which would be especially problematic for unambiguous biosignature gas interpretations.

The time evolution of volatile reservoirs during both the magma ocean phase and the solid state evolution is governed by the following equations (see supporting information Section A.6 for details):

$$\frac{dM_{Solid-\Lambda_i}}{dt} = 4\pi\rho_m k_{\Lambda_i} fr_{\Lambda_i} r_s^2 \frac{dr_s}{dt} + F_{ingas-\Lambda_i} - F_{outgas-\Lambda_i}$$
$$\frac{dM_{Fluid-\Lambda_i}}{dt} = -4\pi\rho_m k_{\Lambda_i} fr_{\Lambda_i} r_s^2 \frac{dr_s}{dt} - F_{ingas-\Lambda_i} + F_{outgas-\Lambda_i} - Esc_{\Lambda_i}$$
$$(6)$$

Here, $\Lambda_i$ represents a generic volatile species (e.g., $H_2O$, $CO_2$, free O). The first term on the right hand side represents the transfer of volatiles from the fluid phases (magma ocean + atmosphere) to the solid mantle as the magma ocean solidifies: $k_{\Lambda_i}$ is the melt-solid partition coefficient for species $\Lambda_i$, and $fr_{\Lambda_i}$ represents the mass fraction of the volatile species in the magma. The remaining fluxes are subaerial plus submarine outgassing from the mantle to the atmosphere, $F_{\text{outgas}-\Lambda_i}$, ingassing from the atmosphere to the mantle (e.g., crustal oxidation or hydration), $F_{\text{ingas}-\Lambda_i}$, and escape to space $Esc_{\Lambda_i}$. Because we are assuming a plate tectonics regime, the model does not separately track volatile reservoirs in the crust and mantle. Instead, we assume carbon, oxygen, and water added to crust is immediately subducted into the mantle; a single, well-mixed "interior" reservoir is used to represent storage of volatiles in solid silicates. The extremely broad range of crustal hydration and crustal oxidation efficiency factors sampled (see below) can accommodate differing subduction and arc volcanism efficiencies.

Note that the model only tracks C, H, and O-bearing species, as well as $Fe^{2+}/Fe^{3+}$ speciation in the interior. Nitrogen fluxes are not modeled, and we instead assumed a 1 bar $N_2$ background partial pressure in all model runs. This conservative assumption ensures that, for temperature surface conditions, there are always sufficient non-condensables to maintain a cold trap and prevent excessive water loss; any oxygen accumulation that results is due to other processes.

The model does not include any explicit photochemistry; it tracks fluxes of oxygen into/out of the combined atmosphere-ocean reservoir, and all outgassed reductants are assumed to instantaneously deplete atmospheric oxygen. This simplification is adequate for estimating oxygen accumulation because if oxygen sources exceed oxygen sinks, then oxidant build-up will occur; neither photolysis reactions nor spontaneous reactions can add net reducing power the atmosphere-ocean system. Our model cannot predict low steady state oxygen abundances in predominantly anoxic atmospheres, however; atmospheric $O_2$ is truncated at a lower limit of 0.1 Pa for numerical efficiency. Importantly, our modeling approach is agnostic on the plausibility of the various photochemical scenarios that have been proposed for abiotic $O_2$ atmosphere, such as $O_2$- and CO- rich atmospheres maintained by continuous $CO_2$ photodissociation (Gao et al., 2015; Hu et al., 2020). Studies of these scenarios enforce global redox balance at the boundaries of the atmosphere-ocean system and determine whether appreciable atmospheric oxygen exists in the resultant photochemical steady state (e.g., Harman et al., 2015). Here, we are instead using a time-dependent model to investigate whether slight imbalances in atmosphere-ocean boundary fluxes can result in atmospheric oxygen accumulation on long timescales (cf., Luger & Barnes, 2015; Schaefer et al., 2016; Wordsworth et al., 2018).

There are many uncertain parameterizations and parameter values in our model, and so all results are presented as Monte Carlo ensembles that randomly sample a wide range of uncertain parameter values. Parameter ranges and their justifications are described in full in the supporting information (Table S1). We sampled a range of temperature-dependent mantle viscosities, efficiencies of XUV-driven escape, uncertain early sun XUV fluxes, carbon cycle feedbacks, deep hydrological cycle dependencies, and albedo parameterizations. Unknown parameters that are particularly important for oxygen false positives include the dry crustal oxidation efficiency, $f_{\text{dry}-\text{oxid}}$, which is the fraction of $Fe^{2+}$ in newly produced crust that is oxidized to $Fe^{3+}$ in the presence of an oxidizing atmosphere via non-aqueous reactions. This parameter is sampled uniformly in log space from $10^{-4}$ to 10% (Gillmann et al., 2009). Another important parameter is the XUV-driven escape efficiency, $\varepsilon_{\text{lowXUV}}$, which is the fraction of stellar XUV energy that drives H-escape. This is sampled uniformly from 0.01 to 0.3, and the portion of energy that goes into escape once the XUV flux exceeds what is required for O-drag is an additional free parameter.

## 2.1. Model Validation

To validate the model, we first show that it can successfully reproduce the atmospheric evolution of Earth and Venus. Venus results are described in detail in supporting information Section C, and here we summarize key results for Earth. Figure 2 shows Monte Carlo model outputs over a range of Earth-like volatile inventories, specifically an initial water content of 1–10 Earth oceans, an initial $CO_2$ content of 20–2,000 bar. Moreover, only initial $CO_2$ inventories less than the initial water inventory by mass are permitted, and an initial (post core-formation) mantle redox state around the Quartz-Fayalite-Magnetite (QFM) buffer is assumed. There is evidence for more reducing Hadean continental crust (Yang et al., 2014), and other terrestrial planets such as Mars likely have reducing mantles (Wadhwa, 2001). While a rapidly oxidized mantle (e.g., Zahnle et al., 2010) is assumed in all nominal calculations, the sensitivity of our results to initial mantle redox is explored in supporting information Section G.

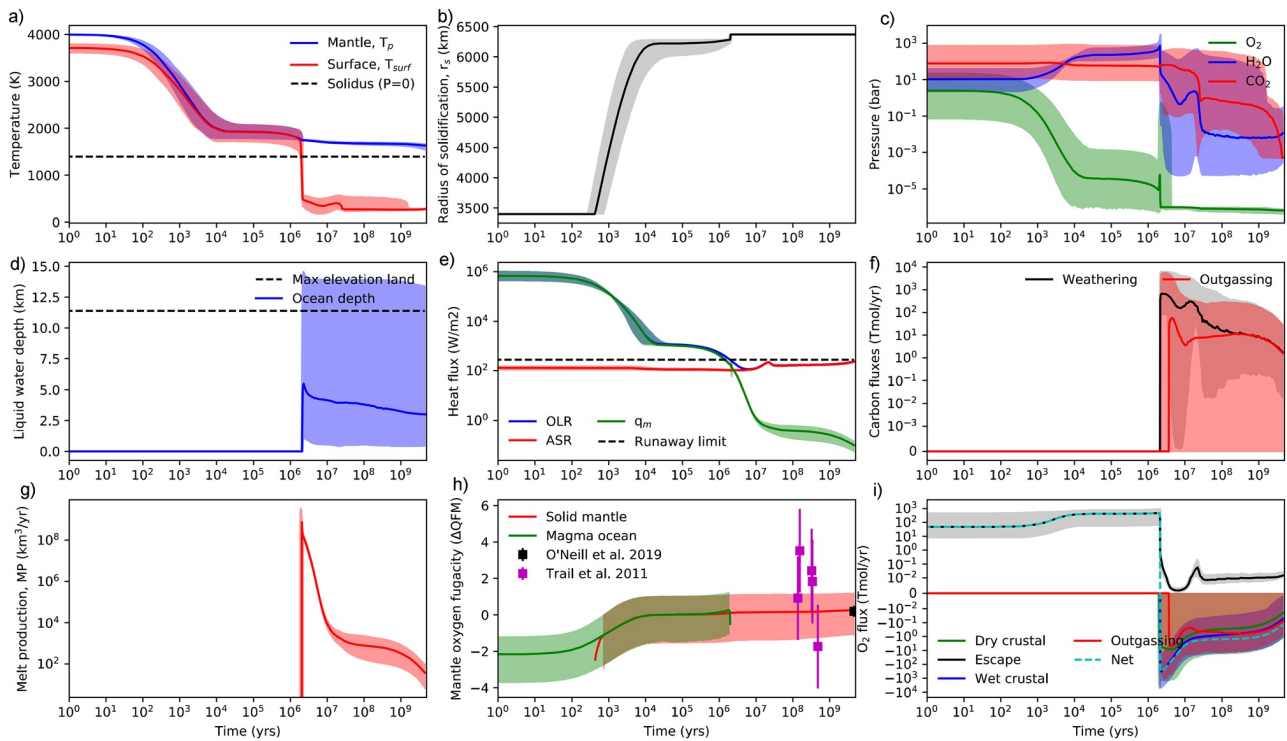

**Figure 2.** Earth's coupled redox-thermal-climate evolution (without life). The model is applied to the Earth from magma ocean to present with initial water inventories ranging from 1 to 10 Earth oceans, and initial $CO_2$ inventories ranging from 20 to 2,000 bar. Additionally, we only plot model runs where the initial water inventory exceeds the initial $CO_2$ inventory. The lines are median values and shaded regions denote 95% confidence intervals across 3,000 model runs. In the absence of life, Earth's atmosphere after 4.5 Ga is always anoxic (c) because outgassing and crustal hydration sinks overwhelm oxygen production via photolysis and diffusion-limited hydrogen escape (i). (a, b) The magma ocean persists for a few million years, consistent with previous studies. (e) The magma ocean ends when the planet's interior cools such that heatflow from the interior drops below the runaway greenhouse limit. (d) When this occurs, liquid water oceans condense onto the surface, (f) a temperate carbon cycle commences. (c) There is sometimes a brief spike in atmospheric oxygen following magma ocean solidification due to the persistence of a steam atmosphere and hydrogen escape, (i) but this oxygen is rapidly drawn down by geological sinks. (g) Volatile cycling is controlled by the rate at which fresh crust is produced. Mantle redox evolution is plotted (h) alongside proxy estimates (O'Neill et al., 2018; Trail et al., 2011).

Figure 2a shows the time-evolution of mantle potential temperature and surface temperature, Figure 2b shows the solidification of the magma ocean from core to surface, which takes several Myr, Figure 2c shows the evolution of atmospheric volatile inventories, Figure 2d shows the globally averaged depth of liquid water oceans. The inflection in temperature evolution and solidification radius around $10^4$ years reflects the transition from a low viscosity, rapidly convecting magma ocean, to more solid-like magma mush convection (Lebrun et al., 2013). Note the transfer of water from steam atmosphere (Figure 2c) to liquid water ocean (Figure 2d) following magma ocean solidification at around $10^6$ years. Figure 2e shows the planetary energy budget, Figure 2f shows carbon outgassing and weathering fluxes, Figure 2g shows total crustal production, Figure 2h shows the evolution of (solid) mantle oxygen fugacity relative to the QFM buffer, and Figure 2i shows oxygen fluxes into/out of the atmosphere, excluding loss of oxygen to the magma ocean, which is modeled as instantaneous melt-atmosphere equilibration rather than a continuous sink flux; this temperature-dependent equilibrium partitioning controls atmospheric $pO_2$ for the first few million years (Figure 2c).

Our modeled early Earth atmosphere-thermal-climate evolution is broadly consistent with semiquantitative reconstructions of Hadean atmospheric evolution (Zahnle et al., 2007, 2010). We also find that in virtually every model run, after 4.5 Gyr of atmospheric evolution, the atmosphere is anoxic (Figure 2c). This is unsurprising. Hydrogen escape during the initial ∼ Myr magma ocean does not add free oxygen to the atmosphere but instead oxidizes the interior, as has been described previously (Hamano et al., 2013). In some cases, small amounts of abiotic oxygen are produced in the post magma-ocean steam atmosphere, but this atmospheric oxygen is rapidly overwhelmed by outgassing and other crustal sinks. Subsequent oxygen production via diffusion-limited escape is small, and so there are no further opportunities for abiotic accumulation so long as the planet remains geologically active.

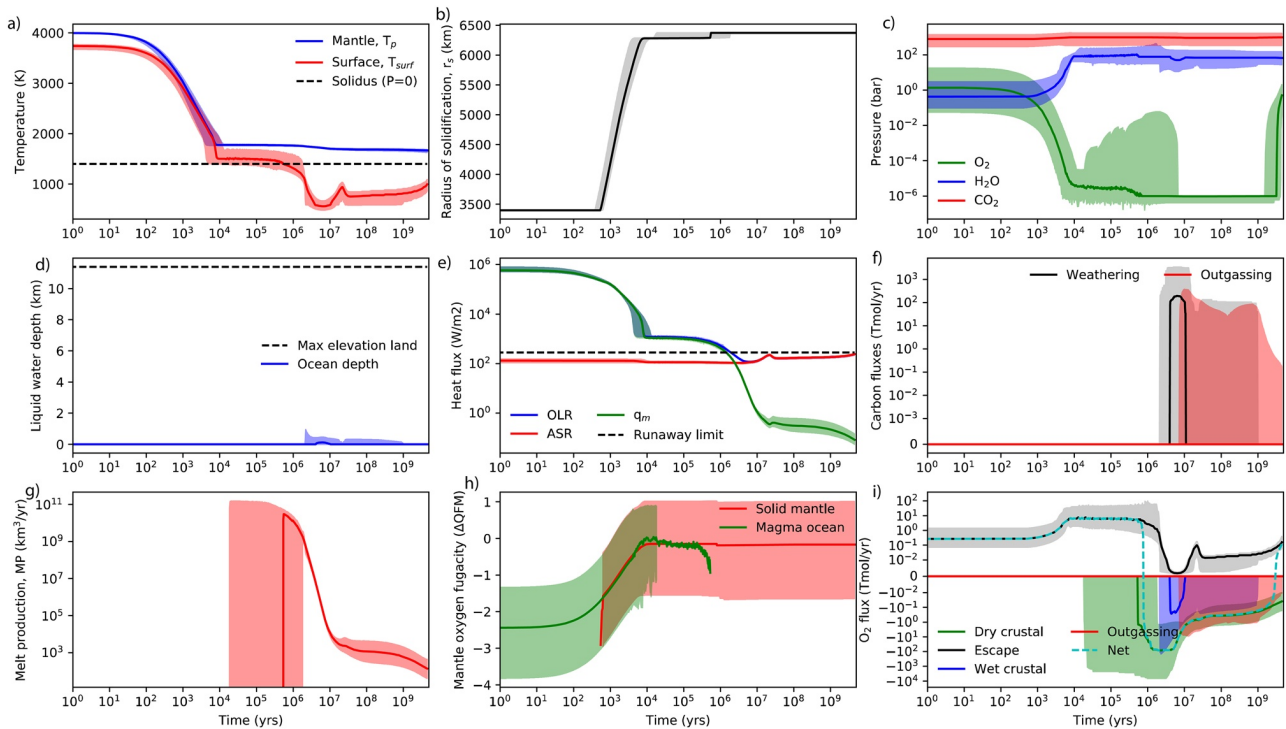

**Figure 3.** Oxygen false positives from high initial $CO_2$:$H_2O$ inventories (Scenario 1). The model is applied to the Earth from magma ocean to present with randomly sampled initial water inventories ranging from ~0.1 to 10 Earth oceans, and initial $CO_2$ inventories ranging from ~20 to 2,000 bar (implying $CO_2$:$H_2O$ ranging from 0.01 to 100 by mass). Only model outputs with modern day atmospheric oxygen exceeding $10^{17}$ kg (>~0.02 bar) are plotted. Subplots are the same as in Figure 2, and shaded regions denote 95% confidence intervals. (a) High atmospheric $CO_2$ ensures the surface temperature always exceeds the critical point of water after the pre-main sequence, and so permanent liquid water oceans do not condense. The lack of surface water, low volatile content of the mantle, and high surface pressure increasing volatile solubility in partial melts all limits oxygen sinks. (a, i) The largest atmospheric sink is dry crustal oxidation, which diminishes with time as the interior cools. (c) Atmospheric oxygen produced via H escape may start to accumulate after several Gyr of evolution.

For Venus, the model can recover current atmospheric conditions assuming the initial water inventory is small, and that crustal sinks of oxygen are efficient (see supporting information Section C). Venusian histories in which the surface was never habitable and in which the surface was habitable for several billion years can both be reconciled with the current atmosphere, which is broadly consistent with previous modeling of Venus' atmospheric evolution (Chassefière et al., 2012; Kasting & Pollack, 1983; Way et al., 2016).

## 3. Results

If Earth's initial volatile inventories are varied, then oxygen-rich atmospheres may be possible. Here, we outline three scenarios whereby an abiotic Earth could have accumulated an oxygen-rich atmosphere after 4.5 Gyr. None of these scenarios guarantee an oxygen-rich atmosphere; instead, oxygenated atmospheres are a possible outcome that is dependent on the efficiency of oxygen crustal sinks and atmospheric escape.

### 3.1. Scenario 1: High $CO_2$:$H_2O$ Initial Inventory Leading to Perpetual Runaway Greenhouse

Figure 3 shows selected model outputs for planets with initial $CO_2$:$H_2O$ volatile inventories greater than one by mass and atmospheric $O_2 > 10^{17}$ kg at present ($P_{O_2}$ > ~0.02 bar). For these planets, the greenhouse warming from a dense $CO_2$ atmosphere ensures that the surface temperature is above the critical point of water; liquid water never condenses on the surface at 1 AU, except briefly, and in small amounts, during early solar evolution (Figure 3d). These high $CO_2$:$H_2O$ perpetual runaway atmospheres have been described previously (Marcq et al., 2017; Salvador et al., 2017). The lack of liquid surface water precludes $CO_2$-drawdown via silicate weathering (Figure 3f). Reactions between supercritical water and silicates will be severely kinetically limited by sluggish solid state diffusion, and are therefore assumed to be negligible (Zolotov et al., 1997). Consequently, a dense $CO_2$ atmosphere and supercritical surface temperature persist indefinitely (Figure 3a), despite the planet residing in the habitable zone. Moreover, there is sufficient steam in the atmosphere to ensure diffusion-limited hydrogen escape provides an appreciable source flux of oxygen (Figure 3i).

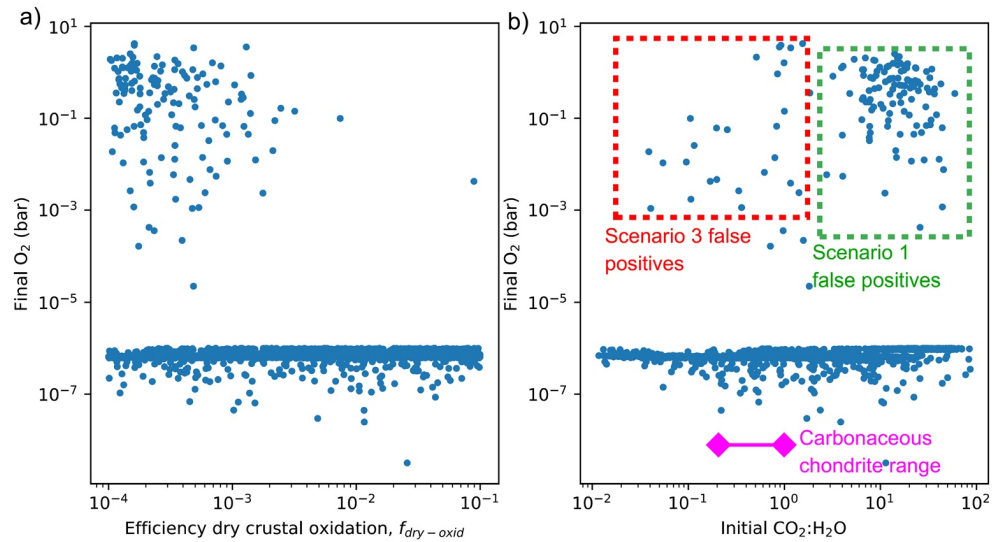

**Figure 4.** Conditions required for Scenario 1 oxygen false positive. Each dot denotes a single model run, and model runs are shown for uniformly sampled initial volatile abundances: $10^{20}$–$10^{22}$ kg $CO_2$ and $10^{20}$–$10^{22}$ kg $H_2O$. Atmospheric oxygen at 4.5 Gyr is plotted as a function of (a) dry crustal oxidation efficiency, and (b) the initial $CO_2$:$H_2O$ inventory by mass. Note that Scenario 1 oxygen accumulation (high $CO_2$, perpetual runaway greenhouse atmospheres) requires both an initial $CO_2$:$H_2O$ ratio >1 (green box) and for dry crustal oxidation to be relatively inefficient, with <0.1% of $Fe^{2+}$ in newly produced crust oxidized. The observed range in carbonaceous chondrite $CO_2$:$H_2O$ ratios (purple interval) is shown in (b) as a rough proxy for Earth's initial volatile inventory. The outliers with high oxygen (red box) are Scenario 3 (desertworld) false positives, which are examined in Section 3.3. Anoxic atmospheres truncate at ~$10^{-6}$ bar for numerical efficiency (see supporting information); these model runs represent outcomes with essentially no atmospheric oxygen.

Oxygen accumulation also requires limited oxygen sinks, and this may occur on high $CO_2$:$H_2O$ worlds for two reasons. First, since permanent oceans do not condense, it is difficult to sequester outgassed volatiles left over from the magma ocean in the interior; hydration reactions and carbonatization reactions do not occur without liquid surface water. Limited mantle regassing after magma ocean outgassing implies low mantle volatile content, which inhibits the capacity of outgassed reductants to draw down oxygen. Second, the high pressure from the dense $CO_2$ atmosphere, while not enough to significantly increase the silicate solidus, does increases the solubility of volatiles in partial melts. In combination with low mantle volatile content, this ensures limited outgassing of reducing species (Figures 3f and 3i, cf., Gaillard & Scaillet [2014]).

However, even with limited outgassing, new crust is still being produced that may be directly oxidized by gaseous $O_2$ (Figure 3g). Figure 4a shows atmospheric oxygen abundances after 4.5 Gyr as a function of dry crustal oxidation efficiency, $f_{dry-oxid}$, for a large number of model runs sampling $10^{20}$–$10^{22}$ kg initial $CO_2$ and $H_2O$ (or ~20–2,000 bar and 0.1–10 Earth oceans). The efficiency parameter is necessarily quite low (<0.1%) in all the model runs that produce significant oxygen. The plausibility of such inefficient crustal oxidation is explored in the discussion. Figure 4b shows 4.5 Gyr atmospheric oxygen as a function of the initial $CO_2$:$H_2O$ ratio, and confirms that high oxygen accumulation only occurs when $CO_2$:$H_2O$ exceeds unity.

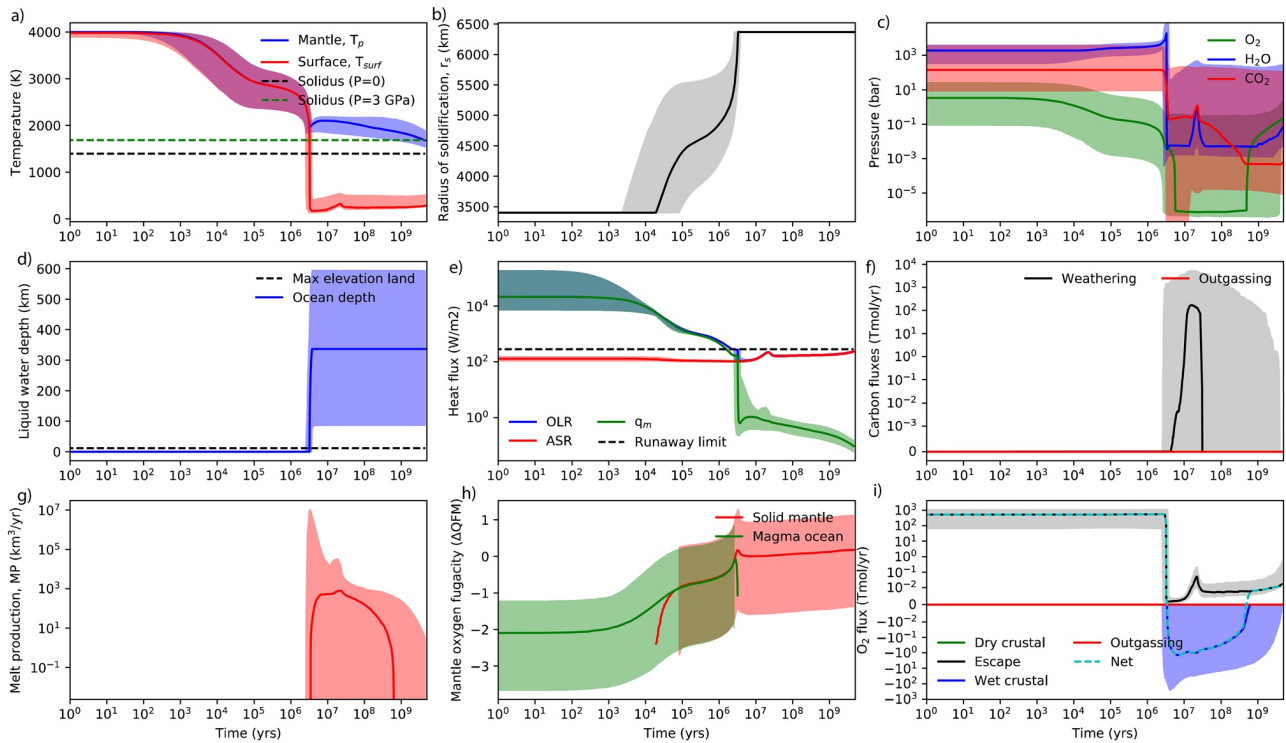

**Figure 5.** Oxygen false positives on waterworlds (Scenario 2). The model is applied to the Earth from magma ocean to present with randomly sampled initial water inventories ranging from 10 to 230 Earth oceans, and initial $CO_2$ inventories ranging from ~20 to 6,000 bar. Only model outputs with modern day atmospheric oxygen exceeding $10^{17}$ kg (>~0.02 bar) are plotted. Subplots are the same as in Figure 2, and shaded regions denote 95% confidence intervals. (d) The large surface volatile inventory increases the (g) mantle solidus such that melt production and tectonics shut off shortly after formation. (i) This shuts down all oxygen sinks and (c) allows for the gradual accumulation of oxygen via diffusion-limited H escape over several Gyr.

### 3.2. Scenario 2: Waterworlds

Figure 5 shows selected model outputs for planets with $H_2O$ volatile inventories between 10 and 230 Earth oceans. For these planets, a liquid water ocean condenses out of the atmosphere after a few million years and any oxygen left over from the post-formation steam-dominated atmosphere is typically removed by oxygen sinks (Figure 5i), just as in the nominal Earth model (Figure 2). However, the pressure overburden from the large surface water inventory dramatically increases the solidus of the silicate mantle. This effect, which has been described previously (Kite & Ford, 2018; Noack et al., 2016), causes fresh crustal production to cease completely after a few billion years when the mantle potential temperature drops below the solidus (Figure 5g). The cessation of crustal production suppresses all geological oxygen sinks; hydration reactions stop as there is no melt production or new crust to oxidize (Figure 5i). The source flux of oxygen is low in this scenario. An effective cold trap ensures oxygen production rates of ~0.01 Tmol/yr via diffusion-limited escape. However, since oxygen sinks are negligible, this small source flux is sufficient to accumulate modern Earth-like oxygen abundances over several Gyr (Figure 5c). There are also a small number of model runs where oxygen persists from the magma ocean, since equilibrium oxygen fugacities are high at the elevated solidus temperature under high pressures (Figure 5c).

Figure 6 shows the oxygen abundance after 4.5 Gyr for individual model runs as a function of the initial water inventory. Waterworld oxygen false positives only begin to become likely when the initial water inventory exceeds around 50 Earth oceans or $10^{23}$ kg (for Earth-sized planets). There is significant scatter in results due to uncertainty in the temperature-dependent mantle viscosity, which controls the duration of tectonics.

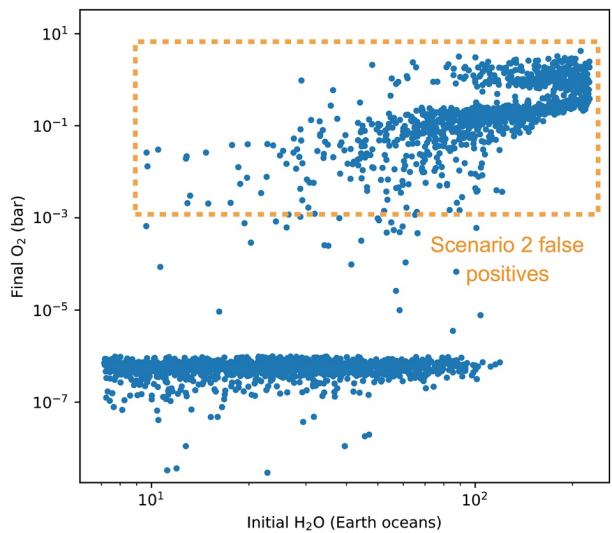

**Figure 6.** Prevalence of Scenario 2 waterworld false positives. Each dot denotes a single model run, and model runs are shown for uniformly sampled initial volatile abundances: $10^{20}$–$10^{22.5}$ kg $CO_2$ and $10^{22}$–$10^{23.5}$ kg $H_2O$. Atmospheric oxygen at 4.5 Gyr is plotted as a function of the initial planetary water inventory. Waterworld oxygen false positives are unlikely unless the initial water inventory exceeds $10^{23}$ kg (~50 Earth oceans). There is a high probability of abiotic oxygen accumulation for water inventories exceeding 100 Earth oceans. Clustering around 0.1 bar occurs because this is the amount of oxygen accumulated after 4.5 Gyr for temperate surface conditions and Earth-like stratospheric water abundances. Warmer surface oceans (~400–600 K) result in more stratospheric water vapor and thus enable higher oxygen levels.

### 3.3. Scenario 3: Desertworlds

The final scenario whereby abiotic oxygen could accumulate on habitable zone planets around Sun-like stars occurs for planets with extremely small initial volatile inventories (initial water inventory < ~0.3 Earth oceans). Figure 7 shows selected model outputs representing this desertworld false positive. The required sequence of events are as follows: the low volatile inventory ensures that the magma ocean freezes quickly (typically ~$10^5$ years, Figure 7b), even though the planet is still in a runaway greenhouse state due to high heatflow from the interior (Figure 7e). A steam-dominated atmosphere can therefore persist for a few million years, and oxygen may accumulate during this time because there is no surface magma ocean to dissolve the oxygen. Dry crustal oxidation will remove some oxygen during this steam atmosphere phase, but oxidation will be limited by the rate at which oxygen can diffuse into extrusive lava flows (Figures 7i and S1d). When a shallow ocean does eventually condense out as heatflow from the interior drops below the runaway greenhouse limit (Figures 7d and 7e), oxygen may persist for billions of years if oxygen sinks are small (Figure 7c). Outgassing sinks are limited by the low volatile inventory of the planet, but inefficient dry crustal oxidation is also required for the oxygen to persist for 4.5 Gyr; the efficiency parameter, $f_{dry-oxid}$, must be <0.1% (Figure S1). This habitable scenario is qualitatively different to the uninhabitable Scenario 1 false positives: in the former, atmospheric oxygen accumulates early and gradually declines due to crustal sinks, whereas in the latter, oxygen accumulation takes several Gyrs. Desertworld false positives also require efficient XUV-driven hydrodynamic escape (>10%) during the steam atmosphere phase to produce large atmospheric oxygen abundances after the magma ocean has solidified (Figure S1).

## 4. Discussion

The modeling approach adopted in this paper has several important caveats and limitations. First, we consider the reasons why abiotic oxygen accumulation could be underestimated in our model.

### 4.1. Assumptions That May Underestimate Abiotic Oxygen Accumulation

As noted above, the model does not track nitrogen fluxes and instead assumes 1 bar $N_2$ partial pressure throughout. This limits oxygen accumulation by providing a non-condensable background gas to throttle hydrogen escape at the cold trap (Kleinböhl et al., 2018; Wordsworth & Pierrehumbert, 2014). Nitrogen atmospheric evolution for terrestrial planets is highly uncertain, and the evolution of Earth's atmospheric $N_2$ inventory is poorly constrained (Johnson & Goldblatt, 2018; Stüeken et al., 2016). However, for terrestrial planets that form with low nitrogen inventories, or with most of their nitrogen sequestered in the interior (Wordsworth, 2016), then oxygen accumulation on temperate planets could be a more common outcome than our modeling suggests.

Our nominal outgassing model may overestimate fluxes of reduced gases per unit mass partial melt. This is because we do not account for graphite saturation and redox-dependent partitioning of carbon-bearing species between crystalline and melt phases; we instead assume a constant partition coefficient for relating solid mantle $CO_2$ content and total melt plus gas phase concentrations (e.g., Lebrun et al., 2013). Models of Martian outgassing (Grott et al., 2011), and more generalized terrestrial outgassing models (Ortenzi et al., 2020) both show that reducing mantles tend to outgas fewer volatiles by mass than more oxidized mantles for the same amount of crustal production. Our model does, however, account for the greater reducing power of volcanic outgassing on planets with lower mantle oxygen fugacities (Gaillard & Scaillet, 2014; Wogan et al., 2020). Consequently, our conservative approach maximizes fluxes of outgassed reductants, and may underestimate oxygen accumulation. Sensitivity tests which consider more reducing mantles and graphite saturated melts are described below.

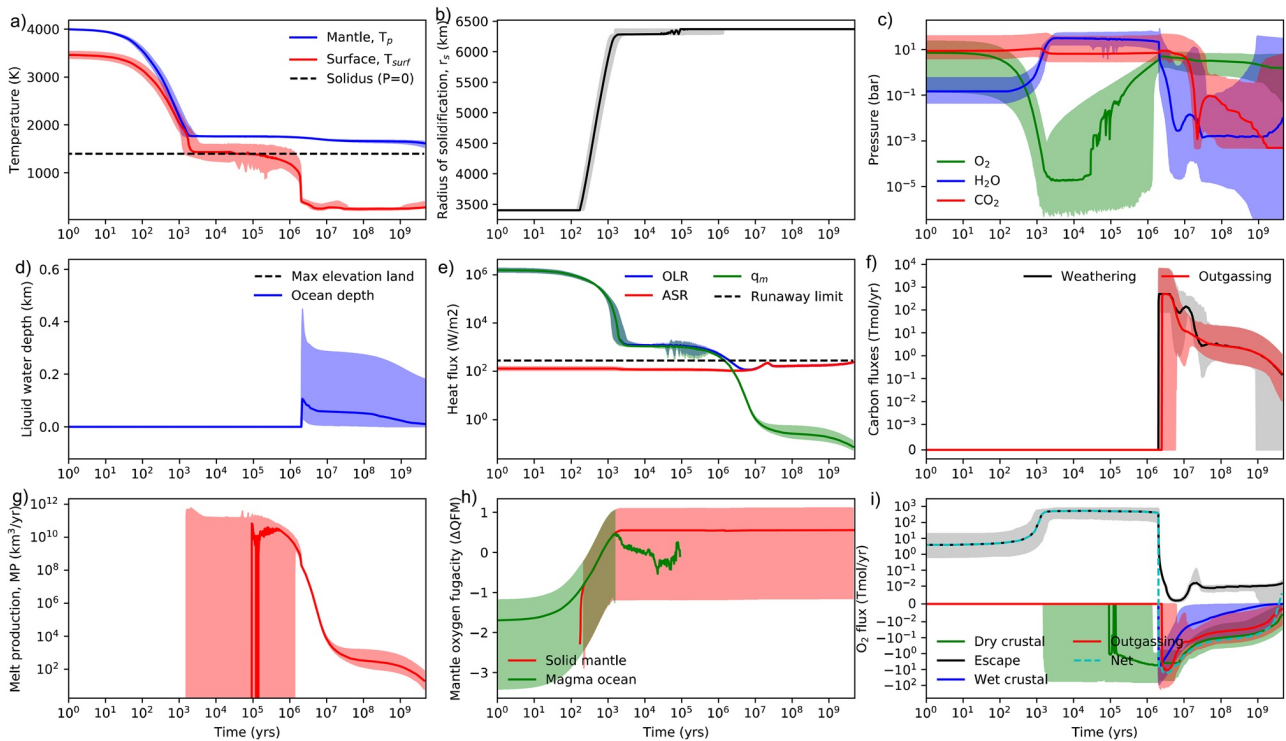

**Figure 7.** Oxygen false positives on desertworlds (Scenario 3). The model is applied to the Earth from magma ocean to present with randomly sampled initial water inventories ranging from ~0.05 to 0.35 Earth oceans, and initial $CO_2$ inventories ranging from ~6 to 60 bar. Only model outputs with modern day atmospheric oxygen exceeding $10^{17}$ kg (>~0.02 bar) are plotted. Subplots are the same as in Figure 2, and shaded regions denote 95% confidence intervals. (b) The low initial volatile inventory ensures the magma ocean solidifies before the runaway greenhouse is over, (c) allowing for significant oxygen accumulation in the ~Myr steam atmosphere that (e) persists until the heatflow from the interior drops below the runaway greenhouse limit. If subsequent oxygen sinks are low, then the few bar oxygen that accumulate early on may persist for billions of years.

Finally, our model may underestimate the duration of steam atmospheres following magma ocean solidification, and therefore underestimate oxygen accumulation prior to ocean condensation. Based on the time required to precipitate an Earth ocean and the apparent absence of stable climate states at the runaway greenhouse limit (Figure S4), it is typically assumed that the time required to transition from steam atmosphere to surface water ocean is ~$10^3$ years (Abe, 1993; Zahnle et al., 2007). In our model, the atmosphere-ocean system is assumed to be in radiative equilibrium with a negligible heat capacity (Lebrun et al., 2013). However, for waterworlds with hundreds of Earth oceans, the time required for a steam atmosphere to condense is long enough for abiotic oxygen accumulation to be significant. Moreover, stable climate states may exist in between a molten surface and a temperate surface ocean, especially if clouds—which we ignore—are included in radiative transfer calculations (Marcq et al., 2017 their Figure 6). Finally, in our model, oxygen is partitioned between the magma ocean and the atmosphere assuming chemical equilibrium. This is a reasonable assumption for high temperature/low-viscosity magma oceans with very short mixing times, but as the surface temperature approaches the solidus, the "magma ocean mush" will behave more like a solid (Lebrun et al., 2013; Salvador et al., 2017), and oxygen produced during the final stages of the magma ocean may not be efficiently sequestered in the melt.

## 4.2. Assumptions That May Overestimate Abiotic Oxygen Accumulation

Next, we consider model assumptions that might cause abiotic oxygen accumulation to be overestimated. The omission of a nitrogen cycle means the model ignores the removal of atmospheric oxygen in the ocean as dissolved nitrate. Nitrate formation is thermodynamically favorable at temperate surface conditions (Krissansen-Totton et al., 2016), and the reaction may occur via lightning. Indeed NO-formation via lightning has previously been established as an important mechanism for removing photochemically produced atmospheric oxygen and inhibiting oxygen false positives (Harman et al., 2018). However, nitrogen fixation via lightning is unlikely to prevent abiotic oxygen accumulation on waterworlds. In the absence of any other sources or sinks nitrate-formation via lightning could draw down the modern Earth's atmospheric oxygen reservoir in 20–200 Myr (Krissansen-Totton et al., 2016) and could therefore remove 1 bar of nitrogen every 0.17–1.7 Gyr. Once most atmospheric nitrogen is converted to nitrate in the ocean, oxygen may accumulate rapidly due to the lack of a non-condensible cold trap (Wordsworth & Pierrehumbert, 2014), and nitrogen will not be significantly replenished by outgassing on waterworlds due to the pressure overburden effect precluding new crustal production. Waterworlds with large initial nitrogen atmospheric inventories (10 s of bar) could avoid abiotic oxygen accumulation if oxygen drawdown via lightning exceeds oxygen production via diffusion-limited hydrogen escape. However, it is debated whether nitrate is the kinetically stable form of nitrogen on habitable worlds (Hu & Diaz, 2019; Ranjan et al., 2019; Wong et al., 2017). Efficient conversion of nitrate back to molecular nitrogen via abiotic chemodentrification might return oxygen to the atmosphere and prevent its accumulation in the ocean, regardless of the initial $N_2$ volatile inventory. In summary, it is unlikely that nitrate sinks will always preclude abiotic oxygen accumulation on waterworlds, but better constraints on aqueous nitrogen chemistry would improve this assessment.

Another important limitation of our model is that it ignores infrared cooling of the upper atmosphere and the throttling of hydrogen escape that results from a cool stratosphere. In our nominal model, an isothermal 210 K stratosphere is assumed. However, $CO_2$-rich atmospheres may efficiently radiate in the IR, cooling the stratosphere and enhancing the water cold trap (Wordsworth & Pierrehumbert, 2013). To test the sensitivity of our results to stratospheric temperature, we repeated our calculations and introduced an additional stratospheric temperature variable, which was randomly sampled from 150 to 250 K. The full results of these calculations are shown in Figure S13. To summarize, neglecting the stratospheric radiation budget does not affect the viability of Scenarios 2 (waterworlds) or 3 (desertworlds) because, in the former, the oxygen source is diffusion-limited escape through a $N_2$-dominated atmosphere at modern Earth-like rates, whereas for the later, oxygen accumulation occurs predominantly during the early, steam-dominated atmosphere, and so stratospheric temperature does not have a strong influence on escape rates (Figures S13b and S13c). However, oxygen accumulation via Scenario 1 (high $CO_2$:$H_2O$ perpetual runaway greenhouse) is unlikely if the stratosphere is cooler than 200 K (Figure S13a). Photochemically produced ozone or hazes may offset the cooling effect of $CO_2$, and so a full assessment of Scenario 1 requires more detailed radiative-photochemical modeling.

One additional caveat is that our model assumes an anhydrous solidus. This simplification is probably reasonable for post-magma ocean mantle conditions (Kite & Ford, 2018), but it is possible to imagine a scenario whereby waterworld mantles become increasingly hydrated via subduction after the magma ocean phase, and that this hydration offsets the pressure overburden effect to maintain geologic activity, and therefore oxygen sinks, for much longer than our nominal model suggests. To test this possibility, we conducted a sensitivity test where we accounted for mantle hydration decreasing the solidus (Katz et al., 2003). The results of this sensitivity test are discussed in detail in supporting information Section D, but in summary, mantle hydration does not have a large effect on oxygen accumulation on waterworlds; it merely shifts the ocean mass threshold for oxygen accumulation.

Sensitivity tests were also conducted to assess whether the delivery of reducing material such as metallic iron and FeO via impacts (Zahnle et al., 2020), a more reducing initial mantle, or larger planet-star separations could inhibit abiotic oxygen accumulation. These sensitivity test results are described in full in supporting information Sections E, G, and H, respectively. In summary, we find that high impactor fluxes could preclude desertworld oxygen accumulation assuming all impactor material is completely oxidized (supporting information Section E). Impactors may thus prevent Scenario 3 (desertworld) false positives in some cases, although this is not guaranteed because it is possible to imagine planetary formation pathways with smaller impactor fluxes and/or where the majority of impactor material is buried or lost to space as suggested by some impact simulations (Marchi et al., 2018). Scenarios 1 (high $CO_2$:$H_2O$ perpetual runaway greenhouse) and 2 (waterworlds) are viable under a more reducing (iron-wüstite buffer) initial mantle (supporting information Section G). This counterintuitive result occurs because, even though degassed volatiles are likely to be more reducing, total volatile concentrations in the melt phase are typically lower due to graphite saturation (Grott et al., 2011; Ortenzi et al., 2020). Moreover, crustal sinks are precluded by high overburden pressure, regardless of the redox state of the crustal material. Although Scenario 3 (desertworlds) is seemingly not excluded by a more reducing mantle, evaluating this would require more complete radiative transfer and photochemical modeling of $CO-H_2$ dominated atmospheres. Finally, when nominal calculations are repeated at 1.3 AU, both Scenario 2 and Scenario 3 oxygen false positives still occur frequently (supporting information Section H). Scenario 1 false positives do not occur at large planet-star separations because a high $CO_2:H_2O$ atmosphere cannot maintain a perpetual runaway greenhouse state after magma ocean solidification.

For Scenario 1 and 3 to be viable, the dry crustal oxidation parameter, $f_{dry-oxid}$, must be relatively small (<0.1%). This contrasts with Venus where $f_{dry-oxid}$ probably needs to exceed >0.1% to remove virtually all $O_2$ from the atmosphere (see Venus validation in supporting information). This parameter is challenging to definitively constrain because it represents a range of physical processes including the diffusion of oxygen into extrusive lava flows (Gillmann et al., 2009), direct oxidation of small grain erosion products (Arvidson et al., 1992), and various other gas-solid redox reactions (Zolotov, 2019). Even if the oxidation of fresh crust is typically efficient, low dry crustal oxidation efficiencies cannot be ruled out because tectonic regimes where most magmatic activity is intrusive and isolated from the atmosphere are possible. In any case, the uncertainties in crustal oxidation processes highlights need for future missions to Venus to better constrain its redox evolution. Mars' crust is more oxidized than its upper mantle, but this oxidation cannot necessarily be used to constrain dry crustal oxidation efficiency since it might be attributable to early hydrous alteration (Herd et al., 2002; Wadhwa, 2001). The presence of gray, reduced sediments mere centimeters below more oxidized Martian regolith, as revealed by Curiosity, argues against efficient post depositional gas-solid oxidation under oxic conditions (Ming et al., 2014).

Finally, we note that in some cases abiotic oxygen accumulation is contingent on highly uncertain atmospheric escape physics. This uncertainty does not affect the viability of the waterworld (Scenario 2) false positives because the required H escape flux is comparable to the modern Earth's diffusion-limited escape flux (Catling & Kasting, 2017, p. 148). Oxygen accumulation only occurs in this case because of the pressure overburden suppression of oxygen sinks. Scenario 1 (high $CO_2:H_2O$ perpetual runaway greenhouse) is similarly unaffected. However, for Scenario 3 (desertworlds), oxygen accumulation only occurs because of efficient XUV-driven escape of hydrogen, $\varepsilon_{lowXUV} > 0.1$. If H-escape is photochemically limited (e.g., Wordsworth et al., 2018) then oxygen accumulation may be limited by the photochemical dissociation of water by UV photons, and the rate and which $H_2O$ recombination reactions occur. However, the oxygen source fluxes required in our desertworld scenario (~500 Tmol $O_2$/yr, Figure 7i) are comparable to the water loss rates inferred for a steam-only early Earth atmosphere calculation using a photochemical model (Wordsworth et al., 2018, their Figure 9). Future work ought to couple the geochemical evolution model here to a photochemical model that includes C-bearing species to better assess the potential for oxygen accumulation on desertworlds. Finally, it is possible that non-thermal O loss (Airapetian et al., 2017) or photochemically modulated stoichiometric escape of H and O (McElroy, 1972) could lessen oxygen accumulation. Upcoming JWST observations of highly irradiated terrestrial planets may constrain escape processes and improve predictions of oxygen accumulation on more temperate planets.

Sulfur outgassing and burial may have played an important role in the oxygenation of Earth's atmosphere (Gaillard et al., 2011; Olson et al., 2019). Sulfur species are ignored in our nominal model because their bulk abundances are probably too small to qualitatively change our oxygenation scenarios (Wordsworth et al., 2018). Supporting information Section I explores the consequences of adding reduced sulfur species to our outgassing model, and confirms that, for Earth-like sulfur mantle abundances, total oxygen sinks are comparable to when sulfur is neglected. With that said, mantle sulfur abundances are contingent on formation processes (e.g., Grewal et al., 2019) and could be highly variable. Incorporating a complete sulfur cycle into a redox evolution model to investigate the sensitivity of oxygenation to initial sulfur content is an opportunity for future research. Note, however, that mantle sulfur abundances are irrelevant for waterworlds (Scenario 2) where all crustal production is suppressed.

In summary, there are several unknowns that preclude definitive predictions of how frequently the three scenarios outlined in this study might occur, but none can be ruled-out with current knowledge.

### 4.3. Implications for Future Observations

How might future observations discriminate between the three abiotic oxygen scenarios described above and oxygen produced by a biosphere? In principle, high $CO_2$:$H_2O$ atmospheres should be possible to diagnose via direct imaging spectral observations because they are not habitable. A clear atmosphere is likely since the coexistence of atmospheric $H_2O$, $O_2$, $O_3$, and abundant $OH$ radicals may preclude the accumulation of photochemical hazes. Strong $CO_2$ absorption features ought to be visible, as should pressure-sensitive CIA features from the high pressures; more detailed photochemical and spectroscopic simulations will be required to determine the best false positive discriminants for these worlds.

Because waterworlds and desertworlds are habitable, they may be more challenging to discriminate from inhabited terrestrial planets. Crucially, waterworld false positives would be ruled out by a detection of sub-aerial land because, for Earth-like gravity, the presence of emerged continents limits the maximum ocean depth to around 10 km (Cowan & Abbot, 2014), or equivalently, a few Earth oceans by mass. This limit arises because silicates cannot support their own weight with greater topography. Consequently, the detection of an ocean-continent dichotomy using time-resolved photometric mapping (Cowan et al., 2009; Farr et al., 2018; Fujii et al., 2010; Kawahara & Fujii, 2010; Lustig-Yaeger et al., 2018) could rule out a waterworld false positive, assuming alternative explanations for dichotomies in surface maps could be excluded. This highlights the need for large aperture direct imaging mission to ensure sufficient time-resolution to map the surface over a planet's rotation. Alternatively, independent mass and radius constraints from radial velocity observations and thermal infrared direct imaging (Quanz et al., 2019), respectively, could also help rule out large (few wt.%) water inventories based on bulk density.

Desertworlds are likely the most challenging scenario to disambiguate from biological oxygen. Time-resolved photometric surface maps and/or the lack of ocean glint could help evaluate the surface water inventory and might be suggestive of a small water inventory (Lustig-Yaeger et al., 2018; Robinson et al., 2010). There are potentially other diagnostic spectral signatures of desertworlds such as spatial variation in atmospheric water vapor and photochemistry that could be tested using general circulation models and photochemical models. The presence of long-lived sulfuric acid hazes (Loftus et al., 2019) has been proposed as putting an upper bound on surface water abundances, but the desertworlds considered here likely have larger surface water inventories than this threshold.

Broadly speaking, the scenarios outlined in this study emphasize that no single observation, including oxygen detection on habitable zone planets around sun-like stars, will be uniquely diagnostic of life. It will be necessary to design future telescopes that are capable of both constraining the full planetary/stellar context and identifying multiple lines of evidence for life (Catling et al., 2018; Walker et al., 2018). For example, oxygen detection on an ostensibly habitable terrestrial planet would be persuasive if accompanied by surface biosignature detections (Schwieterman et al., 2018), temporal biosignatures (Olson et al., 2018), or co-existing reducing gases in atmospheric disequilibrium (Krissansen-Totton et al., 2016). The coexistence of oxygen and methane remains an excellent biosignature and would not be expected for any of the oxygen false-positive scenarios described above. Indeed, it is difficult to produce large methane abundances in habitable planet atmospheres without life, even in anoxic atmospheres (Krissansen-Totton, Olson, et al., 2018; Wogan et al., 2020). It should also be noted that the scenarios in this study were illustrated for habitable zone planets around sun-like stars, but they may also be applicable to habitable zone planets around M-dwarfs.

## 5. Conclusions

The redox evolution of habitable zone terrestrial planets is strongly dependent on initial volatile inventories and the efficiency of crustal sinks. Uninhabited, Earth-sized planets within the habitable zone of G-type stars are very unlikely to accumulate abiotic oxygen if their initial volatile inventories are Earth-like. However, if initial volatile inventories differ dramatically from that of the Earth, then non-biological oxygen accumulation is possible, even when atmospheric noncondensable inventories are large. This may occur when either (i) the initial $CO_2$:$H_2O$ ratio exceeds one, which suppresses oxygen sinks due to the low mantle volatile content and because surface conditions are too hot for aqueous reactions, or (ii) the initial $H_2O$ inventory is very large, thereby halting crustal production after a few billion years and shutting off all oxygen sinks, or (iii) the planet is very volatile-poor, in which case oxygen may accumulate during the steam atmosphere that persists after magma ocean solidification. Inefficient dry crustal oxidation is required for scenarios (i) and (iii) to yield large oxygen abundances, and scenario (i) is sensitive to stratospheric temperature. Fortunately, observational discriminants exist for all three of these scenarios; scenario (i) planets are uninhabitable, whereas the ability to constrain surface water inventories using time-resolved photometry would be useful for ruling out scenarios (ii) and (iii). More generally, the possible existence of these oxygen false positive scenarios highlights the need for a systems approach to biosignature assessment where biogenicity is judged not by the presence or absence of a single biosignature gas, but by multiple lines of evidence from both spectrally resolved and temporally resolved observations.

## Conflict of Interest

The authors declare no conflicts of interest relevant to this study.

## Data Availability Statement

The Python code for our model is open source, https://doi.org/10.5281/zenodo.4539040.

**Acknowledgments**
The interdisciplinary modeling approach used in this study was made possible by many conversations with colleagues. We thank David Catling, Sonny Harman, Callie Hood, Owen Lehmer, Emmanuel Marcq, Laura Schaefer, Edward Schwieterman, and Kevin Zahnle for their valuable insights and suggestions. The manuscript was also greatly improved by constructive comments from Fabrice Gaillard, Robin Wordsworth, and the anonymous reviewer. Joshua Krissansen-Totton was supported by the NASA Sagan Fellowship and through the NASA Hubble Fellowship grant HF2-51437 awarded by the Space Telescope Science Institute, which is operated by the Association of Universities for Research in Astronomy, Inc., for NASA, under contract NAS5-26555. Nicholas Wogan was supported by the NASA Astrobiology Program Grant Number 80NSSC18K0829. We also acknowledge support from NASA's NExSS Virtual Planetary Laboratory, funded under NASA Astrobiology Institute Cooperative Agreement Number NNA13AA93A, and the NASA Astrobiology Program grant 80NSSC18K0829.

## References From the Supporting Information

**Oxygen false positives on habitable zone planets around sun-like stars**

Joshua Krissansen-Totton[1,2,3,*], Jonathan J. Fortney[1], Francis Nimmo[4], Nicholas Wogan[2,5]

[1]Department of Astronomy and Astrophysics, University of California, Santa Cruz, CA

[2]Virtual Planetary Laboratory

[3]NASA Sagan Fellow

[4]Department of Earth and Planetary Sciences, University of California, Santa Cruz, CA

[5]Department of Earth and Space Sciences, University of Washington, Seattle, WA

**Contents of this file**

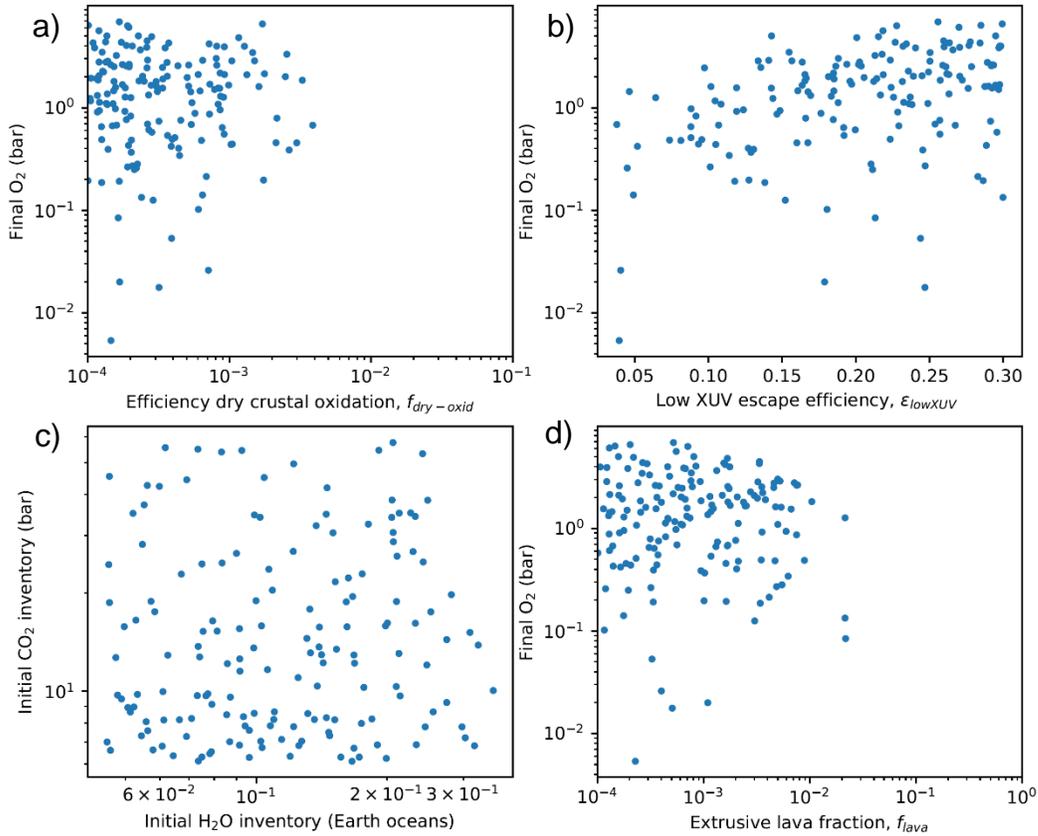

**Fig. S1. Conditions required for Scenario 3 oxygen false positive.** Each dot denotes a single model run, and model runs are shown for all model runs in Fig. 7. Atmospheric oxygen at 4.5 Gyrs is plotted as a function of (a) dry oxidation efficiency, and (b) XUV-driven escape efficiency. Subplot (c) shows the initial water and carbon dioxide inventories that result in desertworld false positives, and (d) shows the maximum fractional planetary area that is continuously molten (available for oxidation) after the magma ocean has solidified.

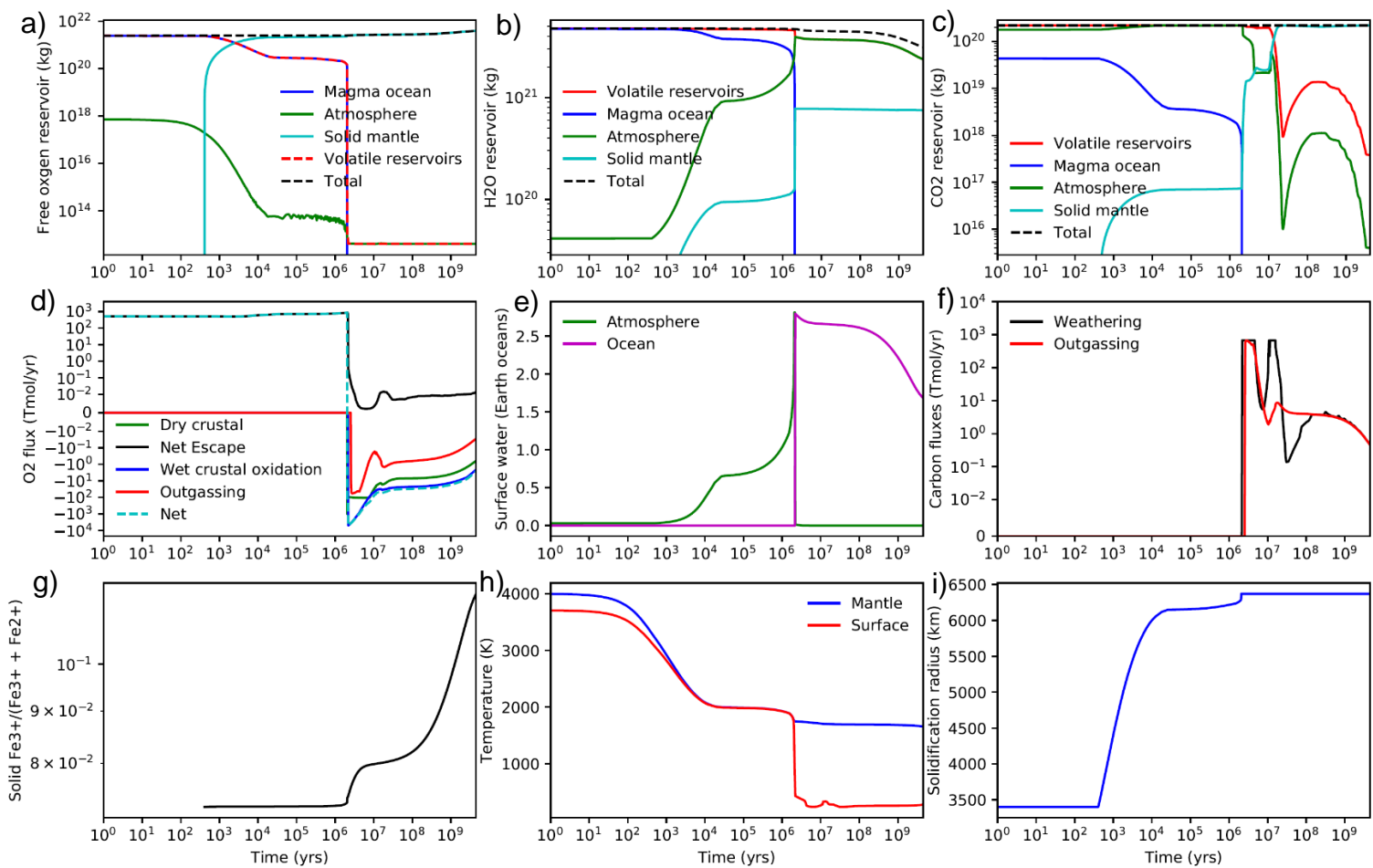

**Fig. S2: Illustrative model outputs.** Selected model outputs from a single, representative model run taken from Fig. 2 in the main text, the nominal Earth model. Subplots illustrate (a) free oxygen reservoirs, (b) water reservoirs, (c) carbon dioxide reservoirs, (d) free oxygen fluxes, (e) atmosphere-ocean partitioning of water, (f) carbon outgassing and weathering fluxes, (g) iron oxidation in the solid mantle, (h) surface and mantle potential temperature, and (i) the solidification radius. Early fluctuations in surface temperature and volatile fluxes are due to non-monotonic evolution of solar luminosity during the pre-main sequence. Total free oxygen (a) increases and total water decreases (b) as H is preferentially lost to space, whereas total $CO_2$ remains constant (c).

**Text A. Model description**

<u>A.1) Thermal evolution:</u>

Planetary thermal evolution is specified by energy budget and temperature-dependent viscosity. The time-evolution of mantle potential temperature, $T_p$ (K), is determined by the following equations, representing the magma ocean and solid-state convection phases, respectively:

$$\frac{4\pi\rho_m Q_{radioactive}\left(r_p^{\,3}-r_c^{\,3}\right)}{3}-4\pi q_m r_p^{\,2}+Q_{core}+4\pi\rho_m H_{fusion}r_s^{\,2}\frac{dr_s}{dt}=\frac{4\pi\rho_m c_p\left(r_p^{\,3}-r_s^{\,3}\right)}{3}\frac{dT_p}{dt}$$

$$(1)$$

$$\frac{4\pi\rho_m Q_{radioactive}\left(r_p^{\,3}-r_c^{\,3}\right)}{3}-4\pi q_m r_p^{\,2}+Q_{core}=\frac{4\pi\rho_m c_p\left(r_p^{\,3}-r_c^{\,3}\right)}{3}\frac{dT_p}{dt} \qquad (2)$$

Here, $Q_{radioactive}$ (J/kg/s) is the heat production via radiogenic isotopes. Earth's radiogenic inventories are taken from Lebrun et al. (2013), and a range of 0.33-3.0 times this nominal, Earth-like inventory is sampled in Monte Carlo calculations. Additionally, $\rho_m$ = 4000 kg/m³ is the assumed average density of mantle material, $r_p$ is the total radius of the planet, $r_c$ is the radius of the core, $r_s$ is the radius of solidification, $H_{fusion} = 4\times10^5$ J/kg is the latent heat of fusion of silicates, $c_p$ = 1200 J/kg/K is the heat capacity of silicates. We adopt a simple exponential decay for core heatflow (J/s):

$$Q_{core} = 20\times10^{12}\times\exp(-0.2\times(t-4.5\,Gyrs)) \qquad (3)$$

Note that we do not account for uncertainty in core heatflow since we are already sampling an order of magnitude range in radiogenic inventories (see above). Moreover, by choosing a core heatflow history at the higher end of literature estimates (Nimmo 2007; O'Rourke & Stevenson 2016), we are effectively maximizing crustal recycling and subsequent oxygen sinks. Tidal heating is ignored; if Earth-moon system tidal heating were included then the duration of the magma ocean could be extended by a few million years, potentially providing more time for oxidation via H escape (Zahnle et al. 2015). Equation (1) governs thermal evolution except in rare cases where transition to runaway greenhouse causes surface temperature to increase above mantle potential temperature, in which case a conduction regime is adopted (see below).

The heatflow from the convecting interior, $q_m$ (W/m²), is parameterized as follows:

$$q_m = \left(\frac{k}{d_{convect}}\right)\left(T_p-T_{surf}\right)\left(Ra/Ra_{crit}\right)^{\beta}$$

$$Ra = \alpha g\frac{\left(T_p-T_{surf}\right)d_{convect}^{\,3}}{\kappa\nu(T_p)} \qquad (4)$$

$$d_{convect} = \begin{cases} r_p-r_s,\ \ r_s<r_p \\ r_p-r_c,\ \ r_s=r_p \end{cases}$$

Here, $\alpha = 2 \times 10^{-5}$ K$^{-1}$ is the thermal expansion coefficient for silicates, $g$ (m/s$^2$) is surface gravity, $Ra_{crit} = 1100$ is the critical Rayleigh number, $k = 4.2$ W/m/K is the thermal conductivity of silicates, $\beta = 1/3$ and $\kappa = 10^{-6}$ m$^2$/s is thermal diffusivity (Lebrun et al. 2013; Schaefer et al. 2016). Heatflow is dependent on the kinematic viscosity, $\nu$, which is a function of mantle potential temperature (see below). This parameterization is adopted for both solid state and magma ocean phases.

A.2) Solidus parameterization and magma ocean freezing:

The solidus controls both the freezing of the magma ocean and the production of partial melt (and outgassing) during temperate geochemical evolution. The adiabatic mantle temperature profile, solidus, and liquidus are parameterized as follows.

$$T(r) = T_p\left(1 + \alpha \frac{g}{c_p}\left(r_p - r\right)\right)$$

$$T_{solidus}(r, P_{Overburden}) = \frac{T_1(r)\exp\left(10^{-5}(-r_p + r + 10^5)\right) + T_2(r)\exp\left(10^{-5}(r_p - r - 10^5)\right)}{\exp\left(10^{-5}(-r_p + r + 10^5)\right) + \exp\left(10^{-5}(r_p - r - 10^5)\right)}$$

$$T_1 = 1420 + 104.42 \times 10^{-9}\, g\,\rho_m\left(r_p - r\right) + 104.42 \times 10^{-9} P_{overburden}$$

$$T_2 = 1825 + 26.53 \times 10^{-9}\, g\,\rho_m\left(r_p - r\right) + 26.53 \times 10^{-9} P_{overburden}$$

$$T_{liquidus}(r, P_{Overburden}) = T_{solidus}(r, P_{Overburden}) + 600$$

(5)

Here, $T_1$ is a linear fit to the solidus for low pressure dry peridotite, and $T_2$ is a linear fit to the solidus for the high pressure lower mantle (Hirschmann 2000; Schaefer et al. 2016). A smooth function between them is assumed for $T_{solidus}$ so that an analytic derivative exists at all radii (see below). Following Schaefer et al. (2016), the liquidus is assumed to be 600 K warmer than the solidus at all pressures. We also allow for modulation of the solidus and liquidus by the pressure overburden of surface volatiles. Here, $P_{overburden}$ is the pressure from all H$_2$O (liquid and gaseous), CO$_2$, and O$_2$ at the surface. The pressure overburden is only accounted for after the magma ocean has solidified and after the mantle has degassed.

The time evolution of the solidification radius is determined by a similar method to Schaefer et al. (2016). The rate of change in the solidification radius can be obtained by noting that the time derivative of solidus evolution and the time derivative of the adiabatic temperature profile must be equal at the solidification radius:

$$\frac{dT_{solidus}(r_s)}{dt} = \frac{dT(r_s)}{dt}$$

$$\frac{d}{dt}\left(\frac{T_1(r_s)\exp\left(10^{-5}(-r_p + r_s + 10^5)\right) + T_2(r_s)\exp\left(10^{-5}(r_p - r_s - 10^5)\right)}{\exp\left(10^{-5}(-r_p + r_s + 10^5)\right) + \exp\left(10^{-5}(r_p - r_s - 10^5)\right)}\right) = \frac{d}{dt}\left(T_p\left(1 + \alpha \frac{g}{c_p}\left(r_p - r_s\right)\right)\right)$$

(6)

Equation (6) can be expanded using the chain rule and rearranged to obtain an expression for the solidus evolution in terms of the potential temperature time-derivative:

$$\frac{dr_s}{dt} = \begin{cases} 0, T(r_c) > T_{solidus}(r_c) \\ 0, T_{surf} < T_{solidus}(r_p) \\ h(r_s)\dfrac{dT_p}{dt}, otherwise \end{cases}$$

$$where, h(r_s) = \frac{1 + \alpha(g/c_p)(r_p - r_s)}{\alpha g T_p / c_p + Z(r_s)},$$

$$Z(r_s) = \frac{S_1(r_s) - S_2(r_s)}{\left(e^{10^{-5}(-r_p + r_s + 100000)} + e^{10^{-5}(r_p - r_s - 100000)}\right)^2},$$

$$S_1(r_s) = \left(e^{10^{-5}(-r_p + r_s + 100000)} + e^{10^{-5}(r_p - r_s - 100000)}\right)$$

$$\times\left(-a_1 g \rho_m(r_p - r_s) + T_1(r_s) \times 10^{-5} e^{10^{-5}(-r_p + r_s + 100000)} - a_2 g \rho_m e^{10^{-5}(r_p - r_s - 100000)} - T_2(r_s) \times 10^{-5} e^{10^{-5}(r_p - r_s - 100000)}\right),$$

$$S_2(r) = \left(T_1(r) e^{10^{-5}(-r_p + r_s + 100000)} + T_2(r) e^{10^{-5}(r_p - r_s - 100000)}\right)\left(10^{-5} e^{10^{-5}(-r_p + r_s + 100000)} - 10^{-5} e^{10^{-5}(r_p - r_s - 100000)}\right)$$

$$(7)$$

Note that the solidification radius must remain constant when the core-mantle boundary temperature exceeds the solidus temperature and when the surface temperature drops below the solidus to ensure the solidification radius and mantle potential temperature begin evolving together when $T(r_c) = T_{solidus}(r_c)$ and stop evolving when the magma ocean freezes. This procedure was checked against simply numerically solving for the solidification radius at every timestep:

$$T_{solidus}(r_s) = T(r_s)$$

$$\frac{T_1(r_s)\exp\left(10^{-5}(-r_p + r_s + 10^5)\right) + T_2(r_s)\exp\left(10^{-5}(r_p - r_s - 10^5)\right)}{\exp\left(10^{-5}(-r_p + r_s + 10^5)\right) + \exp\left(10^{-5}(r_p - r_s - 10^5)\right)} = T_p\left(1 + \alpha\frac{g}{c_p}(r_p - r_s)\right)$$

$$(8)$$

This approach is more computationally expensive but yields an identical solidification radius evolution to the analytic expression eq. (7).

### A.3) Mantle Viscosity Parameterization:

Our viscosity parameterization needs to have several properties. First, it must successfully reproduce the modern Earth's heatflow, melt production, and plate velocity. Second, it needs to transition smoothly from low viscosity magma ocean, to magma mush, to solid state convection. Our parameterization is a variation of those assumed in other magma ocean-to-solid interior evolution models (Lebrun et al. 2013; Salvador et al. 2017; Schaefer et al. 2016). However, the parameterizations in these studies needed modification because they predict a low viscosity magma-ocean or mush at the modern

Earth's potential temperature (~1620 K). Noting that there is a large uncertainty in the critical melt fraction that controls the transition from solid-like to fluid-like convection (Costa et al. 2009), we adopted a parameterization that ensures this transition occurs at a temperature that exceeds the modern Earth's potential temperature. This is illustrated in Fig. S3, which compares our viscosity parameterization to others in the literature.

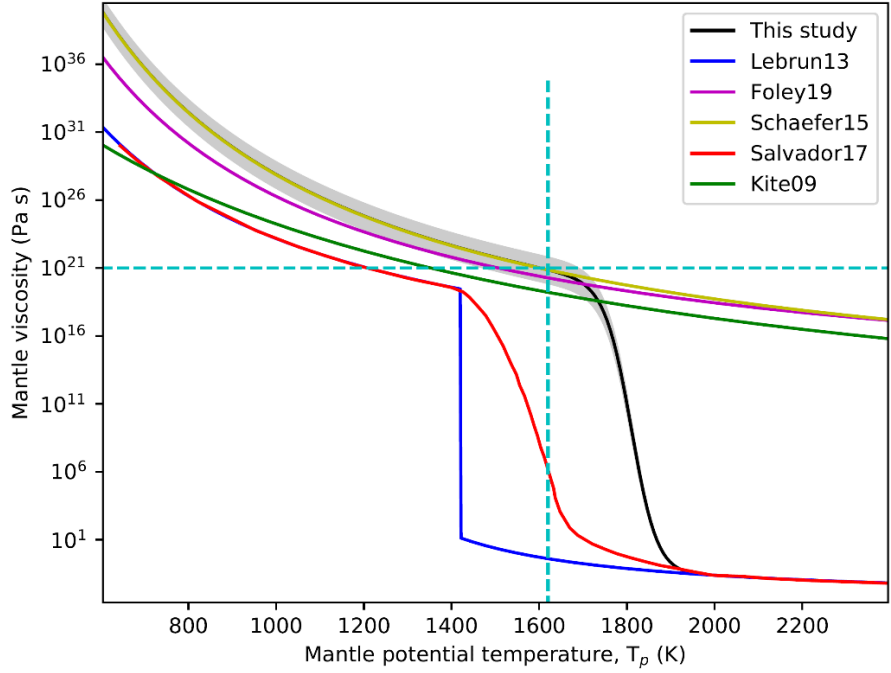

Fig. S3: Assumed viscosity parameterizations compared to other parameterizations from the literature (solid lines). The dashed-cyan lines represent the mantle potential temperature and viscosity required to reproduce the modern Earth's melt production and plate velocity.

$$\log_{10}(\nu) = \begin{cases} \dfrac{\log_{10}(\nu_{rock}) \times 0.2 \times \left(T_p - T_{solidus}\right)^5 + \log_{10}(\nu_{magma}) \times 0.8 \times \left(T_{liquidus} - T_p\right)^5}{0.2 \times \left(T_p - T_{solidus}\right)^5 + 0.8 \times \left(T_{liquidus} - T_p\right)^5}, T_{solidus} \le T_p \le T_{liquidus} \\ \log_{10}(\nu_{rock}), T_p < T_{solidus} \\ \log_{10}(\nu_{magma}), T_p > T_{liquidus} \end{cases}$$

$$\nu_{rock} = V_{coef}\, 3.8 \times 10^7 \exp\left(\frac{350000}{8.314 T_p}\right) / \rho_m$$

$$\nu_{magma} = 0.00024 \exp\left(\frac{4600}{\left(T_p - 1000\right)}\right) / \rho_m$$

(9)

Here, $V_{coef}$ = $10^1$ to $10^3$ Pa s is a randomly sampled parameter that accounts for uncertainty in solid-state viscosity.

### A.4) Stellar evolution

We assume solar bolometric luminosity evolution, $L(t)$ (W), for all model runs (Baraffe et al. 1998; Baraffe et al. 2002). For the evolution of stellar XUV luminosity, we follow the empirical fit developed in Tu et al. (2015). The early sun's rotation rate, $\Omega_0$, is an unknown parameter sampled uniformly from 1.8 to 45 (relative to modern) in log space. From the early sun's rotation rate, the time for the early sun to fall out of saturation, $t_{sat}$ (Myr), is given by:

$$t_{sat} = 2.9\Omega_0^{1.14} \tag{10}$$

For the chosen range of rotation rates, sampled saturation times range from 6 to 226 Myrs. We assume that at saturation the sun's XUV luminosity is $10^{-3.13}L(t)$. To retrieve the modern XUV flux, we define the exponent, $\beta = 0.86/(0.35\log_{10}(\Omega_0) - 0.98)$, and specify the XUV evolution as follows:

$$L_{XUV}(t) = \begin{cases} 10^{-3.13}L(t), \ t < t_{sat} \\ 10^{-3.13}(t/t_{sat})^{\beta}L(t), \ t \geq t_{sat} \end{cases} \tag{11}$$

The flux received at each planet's orbital distance, $D_{planet-star}$ (m), is calculated using Earth-sun and Venus-sun separations. We begin our model runs at $t$ = 10 Myrs in the stellar evolution model, but results are insensitive to the choice of zero point.

### A.5) Surface Energy Budget:

At each time-step in the model, the surface temperature is solved numerically by finding the surface temperature that ensures heatflow from the interior plus absorbed shortwave radiation (ASR) is exactly balanced by outgoing longwave radiation (OLR). This assumption ignores the intrinsic heat capacity of the ocean, which is reasonable for Earth-like oceans, but for waterworlds this radiative equilibrium approach may underestimate the transition time from runaway greenhouse to surface ocean (see Discussion). Surface temperature is found by solving the following equation for $T_{surf}$:

$$q_m(T_p, T_{surf}, \nu(T_p)) + ASR(\alpha_{bond}, L(t)) = OLR(T_{surf}, T_e, M_{Fluid-H_2O}, M_{Fluid-CO_2}, [CO_3^{2-}]) \tag{12}$$

Here, $q_m$ is specified by equation (4). Absorbed shortwave radiation is a function of the planetary albedo and stellar luminosity:

$$ASR(\alpha_{bond}, L(t)) = \frac{L(t)(1 - \alpha_{bond})}{16\pi D_{planet-star}^2} \tag{13}$$

Here, $D_{planet-star}$ (m) is the distance between Earth (or Venus) and the sun. Following Pluriel et al. (2019) we assumed the following parameterization for albedo:

$$T_{transition} = \begin{cases} 1000.0 + 200 \times \log_{10}\left(\dfrac{pH_2O}{10^5}\right)^2, \ pH_2O > 10^5 \ Pa \\ 1000.0, \ pH_2O \leq 10^5 \ Pa \end{cases}$$

$$\alpha_{bond} = 0.5\left(A_{cold} - A_{hot}\right)\tanh\left(\frac{T_{transition} - T_{surf}}{400}\right) + 0.5\left(A_{cold} - A_{hot}\right)$$

(14)

Here, $T_{transition}$ (K) is the albedo transition temperature that controls the transition from a hot-state albedo, $A_{hot}$ to a cold-state albedo, $A_{cold}$. The "hot" and "cold" states do not refer to non-glaciated and glaciated states, which we do not consider in our model. Instead, they allow for a transition from low albedo cloud free runaway greenhouse atmospheres at high ($> \sim 1000$ K) temperatures, to a range of cloudy and non-cloudy states under more temperate conditions (Pluriel et al. 2019). This distinction is important for modeling Venus, where the "cold" state albedo is ~0.7, but the albedo during the initial runaway greenhouse phase was potentially much lower. Albedos for hot and cold states are sampled uniformly form 0-0.3 and 0.25-0.35, respectively for Earth and 0-0.3 and 0.2-0.7 for Venus. The albedo of the hot state must always be equal to or less than that of the cold state.

To calculate the outgoing longwave radiation (OLR), we used the publicly available code from Marcq et al. (2017). This code uses DISORT (Stamnes et al. 1988) with four stream longwave radiative transfer. The radiative transfer model only considers opacity due to water vapor and carbon dioxide. Rock vapor opacities are ignored since the time spent at rock-vaporizing temperatures is very short and unlikely to affect long term redox evolution. Correlated k coefficients are calculated from the high resolution molecular absorption spectra computed with *kspectrum* (Eymet et al. 2016), $H_2O$-$H_2O$ continuum absorption is taken from Clough et al. (2005), and $CO_2$-$CO_2$ continuum absorption from fits to Venus observations (Bézard et al. 2011). $H_2O$-$CO_2$ continuum opacity is not considered, and is likely negligible compared to $H_2O$-$H_2O$ and $CO_2$-$CO_2$ continuum absorption (Ma & Tipping 1992). The runaway greenhouse limit calculated using the code of Marcq et al. (2017) closely agrees with line-by-line calculations in Goldblatt et al. (2013).

The atmosphere model calculates atmospheric structure and abundance profiles using the expressions for dry and moist adiabats in Kasting (1988), and the thermodynamic properties of water are taken from steam tables (Haar et al. 1984). Given an assumed surface temperature and total water inventory, the code calculates the atmospheric water vapor profile assuming a dry convective regime (partial pressure of water vapor less than saturation) to moist convective regime (partial pressure of water vapor equals saturation) to isothermal temperature structure, where the isotherm temperature is the planetary skin temperature. The dry convective regime may not be present if water vapor is saturated at the surface. Once the water vapor profile has been calculated, the remainder of the surface water inventory (if any) resides in a surface water ocean.

If a portion of the surface water resides in an ocean, then the partitioning of carbon between the atmosphere and ocean also determines the OLR. Thus, OLR is a function of dissolved carbonate concentrations. This is calculated as follows:

$$\left[CO_3^{2-}\right] = \frac{\Omega \times K_{sp}(T_{surf})}{\left[Ca^{2+}\right]} \quad (15)$$

Here, rather than explicitly track the ocean alkalinity budget, we follow the approach of Schwieterman et al. (2019) and assume that carbonate precipitation will ensure the long-term carbonate saturation state of the ocean, $\Omega$, is constant. Values for $\Omega$ are sampled randomly from 1 to 10 to allow for abiotic supersaturation. Similarly, rather than explicitly track cation weathering budgets, we assumed constant dissolved calcium abundances and sample uniformly in log space from $10^{-4}$ to $3 \times 10^{-1}$ mol/kg. Dissolved calcium concentrations in Earth oceans have varied from $10^{-2}$ to $3 \times 10^{-1}$ mol/kg over Earth history (Halevy & Bachan 2017), but we adopt a broader range to account for different crustal compositions and ocean volumes. For example, Kite and Ford (2018) consider waterworld scenarios with essentially zero $Ca^{2+}$ up to 0.25 mol/kg based on thermodynamic models of basalt-water interaction. Explicitly tracking the ocean alkalinity budget is an opportunity for future research. The temperature dependent solubility product, $K_{sp}$, is the same as in Krissansen-Totton et al. (2018). Once the total carbon reservoir, the ocean size, and the dissolved carbonate concentrations are known, the entire carbonate equilibrium system of equations can be solved to determine atmospheric pCO$_2$ (Krissansen-Totton et al. 2018).

Rather than call the atmospheric radiative transfer code in real time, we precomputed a grid of OLR values as a function of surface temperature (250-4000 K), surface water (10 Pa – 1 GPa), surface carbon dioxide (10 Pa to 0.1 GPa), and planetary effective temperature (150 – 350 K). Within the grid we linearly interpolate between grid points, and on the rare occasion when the model moves beyond the grid, linear extrapolation is adopted. A 1 bar partial pressure of N$_2$ is assumed at every grid point. Since the atmospheric model calculates atmospheric structure, stratospheric mixing ratios are obtained, which are used to determine atmospheric escape rates (see below).

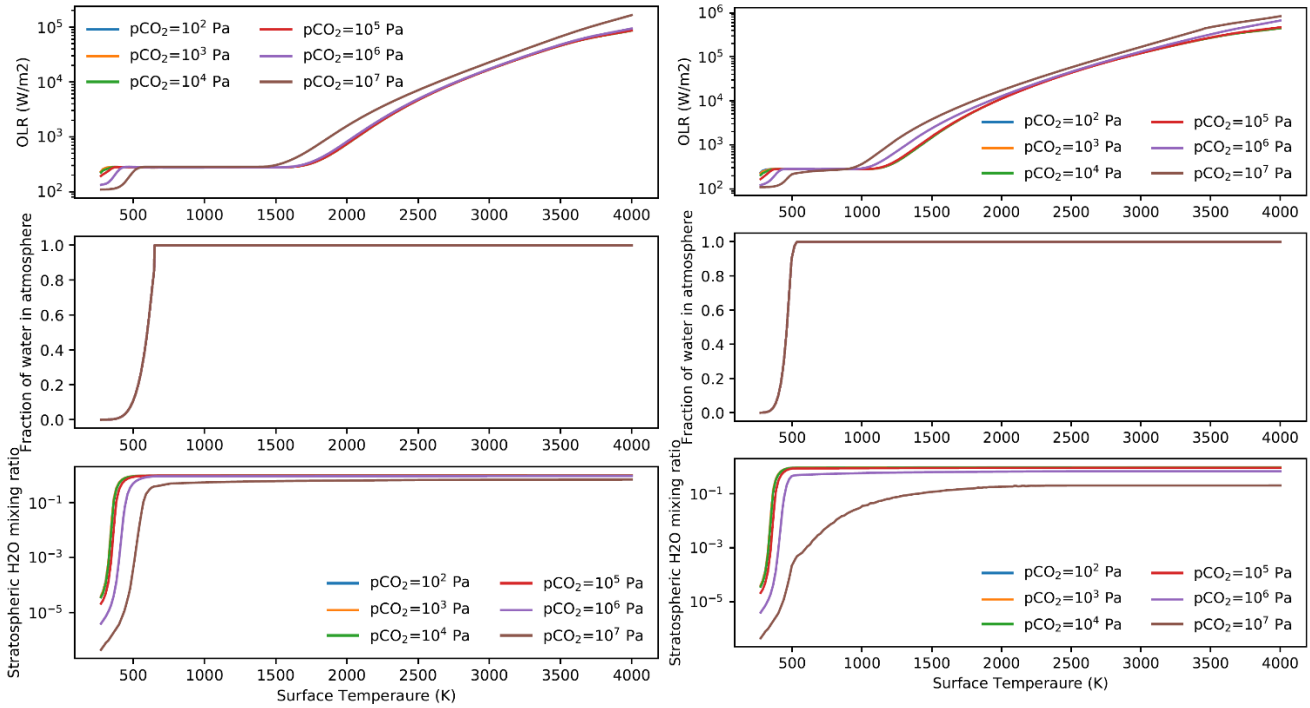

Fig. S4: Illustrative figure showing calculations from OLR grid. Top plots show OLR as a function of surface temperature for varying surface $CO_2$ pressures ($\log_{10}$(Pa)). Here, was have assumed $[CO_3^{2-}]$ = $10^{-3}$ mol/kg. Middle plots show the amount of water in the atmosphere as a fraction of the total surface water inventory, $fr_{atmo-H_2O}$. Bottom plots show the stratospheric water mixing ratio for different $CO_2$ inventories. On the left-hand side 1 Earth ocean is assumed, whereas on the right-hand side 0.1 Earth oceans are assumed.

A.6) Volatile reservoirs and planetary redox budget:

The time evolution of planetary volatile budgets and redox states is determined by the following system of equations:

$$\frac{dM_{Solid-H_2O}}{dt} = 4\pi\rho_m k_{H_2O} fr_{H_2O} r_s^2 \frac{dr_s}{dt} + F_{ingas-H_2O-gain} - F_{outgas-H_2O}$$

$$\frac{dM_{Fluid-H_2O}}{dt} = -4\pi\rho_m k_{H_2O} fr_{H_2O} r_s^2 \frac{dr_s}{dt} - F_{ingas-H_2O-loss} + F_{outgas-H_2O} - 0.5E_H \mu_{H_2O}$$

$$\frac{dM_{Solid-O}}{dt} = 4\pi\rho_m fr_{FeO_{1.5}} r_s^2 \frac{dr_s}{dt} \frac{\mu_O}{2\mu_{FeO_{1.5}}} + F_{oxid-solid}$$

$$\frac{dM_{fluid-O}}{dt} = -4\pi\rho_m fr_{FeO_{1.5}} r_s^2 \frac{dr_s}{dt} \frac{\mu_O}{2\mu_{FeO_{1.5}}} - F_{oxid-fluid} + \left(0.5E_H - E_O\right)\mu_O$$

$$\frac{dM_{Solid-FeO_{1.5}}}{dt} = 4\pi\rho_m fr_{FeO_{1.5}} r_s^2 \frac{dr_s}{dt} + F_{oxid-solid} \frac{2\mu_{FeO_{1.5}}}{\mu_O}$$

$$\frac{dM_{Solid-FeO}}{dt} = 4\pi\rho_m fr_{FeO} r_s^2 \frac{dr_s}{dt} - F_{oxid-solid} \frac{2\mu_{FeO}}{\mu_O}$$

$$\frac{dM_{Solid-CO_2}}{dt} = 4\pi\rho_m k_{CO_2} fr_{CO_2} r_s^2 \frac{dr_s}{dt} + F_{Weather-CO_2} - F_{outgas-CO_2}$$

$$\frac{dM_{Fluid-CO_2}}{dt} = -4\pi\rho_m k_{CO_2} fr_{CO_2} r_s^2 \frac{dr_s}{dt} - E_{CO_2}\mu_{CO_2} - F_{Weather-CO_2} + F_{outgas-CO_2}$$

$$(16)$$

Here, $M_{Solid-i}$ (kg) and $M_{Fluid-i}$ (kg) represent the masses of the i-th species in the solid interior and the fluid magma ocean plus surface reservoirs, respectively. Water, free oxygen, solid FeO and FeO$_{1.5}$ in the mantle, and carbon dioxide are separately tracked in the model. The variables $fr_i$ represents the mass fraction of the i-th species in the magma ocean partial melt. These are calculated at every timestep using the equilibrium relations described below. Key constants include the density of mantle material, $\rho_m$ = 4000 kg/m$^3$, molecular masses of key species, $\mu_i$ (kg/mol), and the assumed partition coefficients for CO$_2$ and H$_2$O, $k_{H_2O}$ = 0.01 and $k_{CO_2}$ = 2×10$^{-3}$, respectively (Lebrun et al. 2013). See the Discussion section in the main text for possible consequences of assuming constant partition coefficients. Escape fluxes for atomic hydrogen, $E_H$ (mol/s$^2$), atomic oxygen $E_O$ (mol/s$^2$), and carbon dioxide, $E_{CO_2}$ (mol/s$^2$) are parameterized below. The remaining fluxes (kg/s) are the ingassing of surface water into the interior $F_{ingas-H_2O-gain}$, outgassing of water from the interior, $F_{outgas-H_2O}$, and the loss of surface water $F_{ingas-H_2O-loss}$ (in general $F_{ingas-H_2O-loss} \neq F_{ingas-H_2O-gain}$ because serpentinizing reactions can remove water from the surface that never reaches the interior because the hydrogen produced is lost to space), oxidation of the interior, $F_{oxid-solid}$, corresponding loss of oxidants from surface reservoirs, $F_{oxid-fluid}$ (typically $F_{oxid-solid} = F_{oxid-fluid}$ except for anoxic atmospheres where reductants are lost to space), outgassing of carbon to the surface, $F_{outgas-CO_2}$, and loss of surface carbon to weathering, $F_{Weather-CO_2}$. Fluxes of ferrous and ferric iron are equal to $F_{oxid-solid}$ scaled by appropriate molecular masses. All fluxes denoted $F_i$ are only non-zero during solid-state convection and are described in detail in their corresponding sections below. Total fluid volatile masses are converted to partial pressures using atmospheric mean molecular weight (no magma ocean) or using the melt solubility relationships described below (magma ocean).

Fig. S2 shows illustrative outputs from a single model run. Subplots Fig. S2a, S2b, and S2c show the evolution free oxygen, water, and carbon dioxide reservoirs, respectively, as governed by equation (16).

A.7) Magma-ocean evolution:

Whilst the magma ocean exists, volatiles in fluid phases are partitioned between the melt, melt crystals, and the atmosphere. For water, this partitioning is described by the following equation (Schaefer et al. 2016):

$$k_{H_2O} fr_{H_2O} M_{crystal} + \left( M_{liquid} - M_{crystal} \right) fr_{H_2O} + \frac{\mu_{H_2O}}{\bar{\mu}} \frac{4\pi r_p^2}{g} \left( \frac{fr_{H_2O}}{3.44 \times 10^{-8}} \right)^{1/0.74} = M_{fluid-H_2O}$$

(17)

Here, $M_{liquid} = 4\rho_m \pi / 3 \left( r_p^3 - r_s^3 \right)$ (kg) is the mass of the magma ocean, $M_{crystal} = (1-\psi) M_{liq}$ is the crystal mass fraction in the magma ocean, which depends on melt fraction, $\psi$ (see below). The molecular mass of the atmosphere is given by $\bar{\mu}$ (mol/kg). The third term on the left-hand side represents the mass of water in the atmosphere, where we adopt the solubility relationship from Papale (1997). An analogous expression can be used to calculate the partitioning of carbon dioxide:

$$k_{CO_2} x_{CO_2} M_{crystal} + x_{CO_2} \left( M_{liquid} - M_{crystal} \right) + \frac{\mu_{CO_2}}{\bar{\mu}} \frac{4\pi r_p^2}{g} \left( \frac{x_{CO_2}}{4.4 \times 10^{-12}} \right) = M_{fluid-CO_2}$$

(18)

Here, we use the solubility relationship from Pan et al. (1991).

For oxygen, equilibrium partitioning is more complicated because both $Fe^{2+}$ and $Fe^{3+}$ melt phases must be included. We adopt the experimental fit in Kress and Carmichael (1991):

$$\ln \left( \frac{x_{Fe_2O_3}}{x_{FeO}} \right) = -1.828 x_{Fe} + 0.196 \ln \left( P_{O_2} \right) + \frac{11492}{T_{surf}} - 6.675 - 2.243 x_{Al_2O_3}$$

$$+ 3.201 x_{CaO} + 5.854 x_{Na_2O} + 6.215 x_{K_2O} - 3.36 \left( 1 - \frac{1673}{T_{surf}} - \ln \left( \frac{T_{surf}}{1673} \right) \right) - 7.01 \times 10^{-7} \frac{P_{surf}}{T_{surf}}$$

$$- 1.54 \times 10^{-10} \frac{P_{surf} \left( T_{surf} - 1673 \right)}{T_{surf}} + 3.85 \times 10^{-17} \frac{P_{surf}^2}{T_{surf}}$$

(19)

Here, $x_{Fe_2O_3}$ and $x_{FeO}$ are the mole fractions of iron-bearing species in the melt, whereas $x_{Fe} = 2 x_{Fe_2O_3} + x_{FeO}$ is the total mole fraction of all iron-bearing species in the mantle (constant). The molar abundances of other species, $x_i$, are assumed to be represent Bulk

Silicate Earth (White 2013). We calculate oxygen solubility using surface temperature and pressure conditions because the interface controls volatile partitioning.

In equation (19) there are three unknowns ($P_{O_2}$, $x_{FeO}$, and $x_{Fe_2O_3}$) and so it is necessary to make appropriate substitutions to solve for $x_{FeO}$. This can be done by expressing the partial pressure of oxygen in terms of the total fluid free oxygen minus the free oxygen dissolved in the magma ocean, and by noting that $x_{FeO} = x_{Fe} - 2x_{Fe_2O_3}$:

$$\ln\left(\frac{x_{Fe_2O_3}}{x_{Fe} - 2x_{Fe_2O_3}}\right) = -1.828 x_{Fe} + 0.196 \ln\left(\frac{\frac{\bar{\mu}}{\mu_{O_2}}\left(M_{fluid-O} - M_{liquid}\left(\frac{0.5\mu_O}{\mu_{FeO_{1.5}}}\right)x_{Fe_2O_3}\frac{\mu_{Fe_2O_3}}{\mu_{sil}}\right)}{4\pi\,r_p^{\,2}/g}\right)$$

$$+\frac{11492}{T_{surf}} - 6.675 - 2.243 x_{Al_2O_3} + 3.201 x_{CaO} + 5.854 x_{Na_2O} + 6.215 x_{K_2O}$$

$$-3.36\left(1 - \frac{1673}{T_{surf}} - \ln\left(\frac{T_{surf}}{1673}\right)\right) - 7.01\times10^{-7}\frac{P_{surf}}{T_{surf}} - 1.54\times10^{-10}\frac{P_{surf}\left(T_{surf} - 1673\right)}{T_{surf}}$$

$$+3.85\times10^{-17}\frac{P_{surf}^{\,2}}{T_{surf}}$$

$$(20)$$

Here, the average molecular weight of silicates is taken to be:

$$\mu_{sil} = x_{MgO}\mu_{MgO} + x_{SiO_2}\mu_{SiO_2} + x_{Al_2O_3}\mu_{Al_2O_3} + x_{CaO}\mu_{CaO} + x_{Fe_2O_3}\mu_{Fe_2O_3} + x_{FeO}\mu_{FeO} \qquad (21)$$

Equation (20) is solved at every timestep in the model, and from the values for melt concentrations of iron-bearing species, we can retrieve the melt fractions (by mass) that dictate the time-evolution of volatile reservoirs:

$$fr_{FeO} = x_{FeO}\frac{\mu_{FeO}}{\mu_{sil}}$$

$$fr_{FeO_{1.5}} = x_{Fe_2O_3}\frac{\mu_{Fe_2O_3}}{\mu_{sil}} \qquad (22)$$

$$P_{O_2} = \frac{\frac{\bar{\mu}}{\mu_{O_2}}\left(M_{fluid-O} - M_{liquid}\left(\frac{0.5\mu_O}{\mu_{FeO_{1.5}}}\right)x_{Fe_2O_3}\frac{\mu_{Fe_2O_3}}{\mu_{sil}}\right)}{4\pi\,r_p^{\,2}/g}$$

Equation (19) is also used to calculate mantle oxygen fugacity for the purposes of outgassing calculations by substituting the solid mantle molar fractions of oxidized and reduced iron.

A.8) Transition from magma ocean to solid mantle convection:
The model switches between magma ocean and solid-state convection freely, as dictated by the radiation and interior heating budget. At each time step, surface temperature is compared to the solidus. For as long as the surface temperature exceeds the solidus, the magma ocean model is adopted (equation (1)), and the solidification radius evolves with time according to equation (7). However, once the surface temperature drops below the solidus, the solidification radius is set to the planetary radius, and solid state interior evolution is dictated by equation (2). Volatiles are instantaneously exchanged between the magma ocean and the solid interior at this transition. The model assumes that when the surface freezes, any volatiles still dissolved in the magma mush remain in the interior (e.g. as basaltic glass, or gas bubbles in melt inclusions), which is reasonable given the short timescale for magma ocean solidification and the high viscosity of the late-stage magma mush. This assumption also maximizes the mantle's capacity for subsequent outgassing of reduced products that remove oxygen from the atmosphere.

During the transition from magma ocean to solid-state convection, volatile inventories undergo the following one-off adjustment:

$$M_{Solid-H_2O} = M_{Solid-H_2O} + \min\left\{M_{solid-H_2O-\max} - M_{Solid-H_2O}, k_{H_2O}F_{H_2O}M_{crystal} + F_{H_2O}\left(M_{liq} - M_{crystal}\right)\right\}$$

$$M_{Fluid-H_2O} = M_{Fluid-H_2O} - \min\left\{M_{solid-H_2O-\max} - M_{Solid-H_2O}, k_{H_2O}F_{H_2O}M_{crystal} + F_{H_2O}\left(M_{liq} - M_{crystal}\right)\right\}$$

$$M_{Solid-O} = M_{Solid-O} + \frac{4}{3}\pi\rho_m F_{FeO_{1.5}}\left(r_p^{\ 3} - r_s^{\ 3}\right)\frac{\mu_O}{2\mu_{FeO_{1.5}}}$$

$$M_{fluid-O} = M_{fluid-O} - \frac{4}{3}\pi\rho_m F_{FeO_{1.5}}\left(r_p^{\ 3} - r_s^{\ 3}\right)\frac{\mu_O}{2\mu_{FeO_{1.5}}}$$

$$M_{Solid-FeO_{1.5}} = M_{Solid-FeO_{1.5}} + \frac{4}{3}\pi\rho_m F_{FeO_{1.5}}\left(r_p^{\ 3} - r_s^{\ 3}\right)$$

$$M_{Solid-FeO} = M_{Solid-FeO} + \frac{4}{3}\pi\rho_m F_{FeO}\left(r_p^{\ 3} - r_s^{\ 3}\right)$$

$$M_{Solid-CO_2} = M_{Solid-CO_2} + k_{CO_2}F_{CO_2}M_{crystal} + F_{CO_2}\left(M_{liq} - M_{crystal}\right)$$

$$M_{Fluid-CO_2} = M_{Fluid-CO_2} - k_{CO_2}F_{CO_2}M_{crystal} - F_{CO_2}\left(M_{liq} - M_{crystal}\right)$$

Note that the water transferred to the mantle cannot exceed the maximum water content of the mantle. Transition from solid state convection to magma ocean, which is typically only relevant for Venus model runs, is similarly calculated.:

$$M_{Solid-i} = M_{Solid-i} - M_{liq}\,M_{Solid-i}/M_{mantle}$$
$$M_{fluid-i} = M_{fluid-i} + M_{liq}\,M_{Solid-i}/M_{mantle}$$

(23)

Fig. S5 shows the fraction of total $CO_2$ and $H_2O$ that reside in the solid mantle immediately after magma ocean solidification for outputs from Fig. 2 in the main text. These mantle fractions are determined by the partitioning of volatiles in the magma ocean as described in Supplementary Section A.7, and the instantaneous retention of leftover melt when surface temperature drops below the solidus, as described in this section. While we do not explicitly model mechanisms of volatile retention in the magma ocean, such as compaction within the moving freezing front, our spread of final mantle volatile fractions is comparable to that of more detailed models (e.g. Hier-Majumder & Hirschmann 2017).

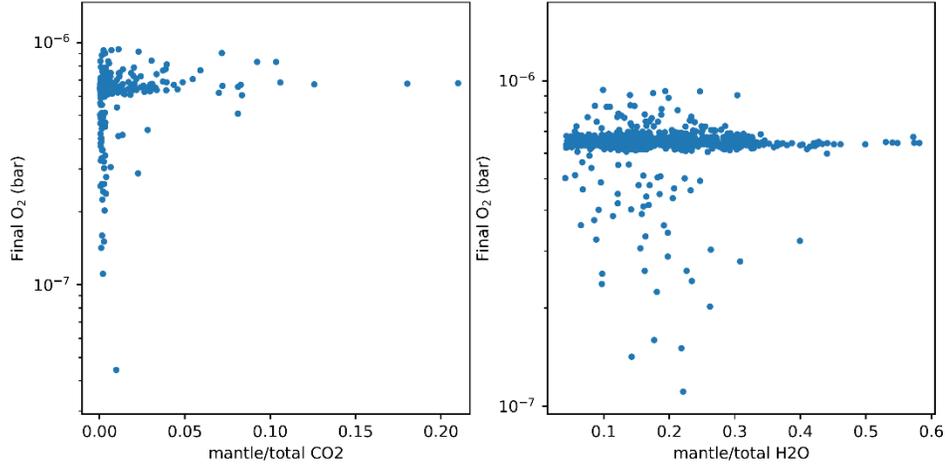

Fig. S5: Volatile fraction in the solid mantle immediately after magma ocean solidification for nominal model calculations (Fig. 2). The left subplot shows the carbon dioxide mantle fraction, whereas the right subplot shows the water mantle fraction.

In a few rare cases, the transition back from solid to magma ocean causes the surface temperature to exceed the mantle potential temperature. When this occurs, we modify the energy budget as follows:

$$ASR(\alpha_{bond}, L(t)) = OLR(T_{surf}, T_e, M_{Fluid-H_2O}, M_{Fluid-CO_2}, [CO_3^{2-}]) + q_c(T_p, T_{surf}) \qquad (24)$$

Here, $q_c$ (W/m$^2$) is the conduction of heat from the surface to the interior and is approximated by the diffusion equation:

$$q_c(T_p, T_{surf}) = k \frac{\left(T_{surf} - T_p\right)}{r_p - r_c} \qquad (25)$$

The time evolution of volatile reservoirs is also modified in this regime to account for the fact that the radius of solidification is moving downward towards the core (see source code for full details). This lasts until surface temperature is less than mantle potential temperature, and the model switches back to a convective mantle.

A.9) Atmospheric Escape Parameterization:

Atmospheric escape controls the source flux of abiotic oxygen. Escape rates are determined by the composition and temperature of the stratosphere, and by the stellar XUV flux. For low stratospheric water abundances, escape is limited by the diffusion of water through background gases, or by the XUV flux from the star (whatever is smaller). As the water content in the upper atmosphere increases, the escape regime transitions to XUV-limited because, for steam-dominated atmosphere, there is no cold trap limiting the supply of water to the upper atmosphere. Our approach does not rigorously capture the complexities of atmospheric escape physics (e.g. Owen 2019), but instead uses plausible parameterizations that incorporate broad parameter ranges to cover a wide range of uncertain physical processes. Moreover, our parameterization collapses to well-established solutions (e.g. the diffusion limit) for end-member cases.

Following Wordsworth and Pierrehumbert (2013), the diffusion of water through non-condensible background gases is given by

$$\varphi_{diff} = b_{H_2O} f_{H_2O} \left( 1/H_n - 1/H_{H_2O} \right)$$

$$b_{H_2O} = \frac{b_{H_2O-CO_2} pCO_2 + b_{H_2O-N_2} pN_2 + b_{H_2O-O_2} pO_2}{pCO_2 + pN_2 + pO_2}$$

$$H_{H_2O} = \frac{8.314 T_{strat}}{\mu_{H_2O} g}$$

$$H_n = \frac{8.314 T_{strat}}{\bar{\mu} g}$$

(26)

Here, $b_{i-j}$ (mol/m/s) is the binary diffusion coefficient of the i-th species through the j-th species (Marrero & Mason 1972; Zahnle & Kasting 1986). These are weighted by the stratospheric mixing ratios of each non-condensible constituent ($CO_2$, $N_2$, and $O_2$), which are obtained from the atmospheric profile calculations (see above). The scale height of water, $H_{H_2O}$ (m), and of the background gases, $H_n$ (m), depend on stratospheric temperature, $T_{strat}$ = 200 K. The stratospheric water vapor mixing ratio is $f_{H_2O}$ and the diffusion limited escape flux is $\varphi_{diff}$ (mol $H_2O$/m$^2$/s). The diffusion-limited flux should arguably be set by the diffusion of atomic hydrogen through background gases since eddy diffusion dominates vertical transport at altitudes where molecular water is more abundant than atomic H and O (Catling & Kasting 2017), but our conservative approach minimizes oxygen accumulation from H escape.

To calculate XUV-driven hydrodynamic escape of H, and associated O and $CO_2$ drag, we follow Odert et al. (2018) and Zahnle and Kasting (1986). The XUV-energy mass loss rate, $\Phi_{XUV}$ (kg/m$^2$/s) is specified by the following equation:

$$\Phi_{XUV} = \frac{\varepsilon(F_{XUV}, X_O, \zeta, \varepsilon_{lowXUV}) F_{XUV} r_p}{4 G M_p}$$

(27)

Here, $F_{XUV}$ is the XUV flux (W/m$^2$) received from the star (see stellar parameterization). The efficiency of hydrodynamic escape, $\varepsilon$, is a function of atmospheric composition and XUV stellar flux, as described below.

In general, the XUV-driven mass flux will be partitioned between H loss, O drag and, very under high XUV fluxes, $CO_2$ drag. The hydrogen escape flux, $\Phi_H$ (molecules H/m$^2$/s), can be obtained by analytically solving equations (4), (5), and (6) in Odert et al. (2018):

$$\Phi_{XUV} + \frac{m_O X_O g (m_O - m_H) b_{H-O}}{k_B T (1 + X_O)} + \frac{m_{CO_2} X_{CO_2} g (m_{CO_2} - m_H) b_{H-CO_2}}{k_B T (1 + b_{H-CO_2} X_O / b_{O-CO_2})}$$

$$- \frac{m_{CO_2} X_{CO_2} b_{H-CO_2} g (m_O - m_H)}{(1 + b_{H-CO_2} X_O / b_{O-CO_2}) k_B T (1 + X_O)} + \frac{m_{CO_2} X_{CO_2} b_{H-CO_2} X_O / b_{O-CO_2}}{1 + b_{H-CO_2} X_O / b_{O-CO_2}}$$

$$= \Phi_H \left( m_H + m_O X_O + \frac{m_{CO_2} X_{CO_2}}{1 + b_{H-CO_2} X_O / b_{O-CO_2}} + \frac{m_{CO_2} X_{CO_2} \left( b_{H-CO_2} X_O / b_{O-CO_2} \right)}{1 + b_{H-CO_2} X_O / b_{O-CO_2}} \right)$$

(28)

Here, $m_i$ (kg) is the mass of the i-th species, $X_i$ is the stratospheric mixing ratio of the i-th species, where we conservatively assume CO₂ is not dissociated to minimize the drag of carbon, $T$ (K) is stratospheric temperature, $k_B$ is Boltzmann's constant, and $b_{i-j}$ is the binary diffusion coefficient of i through j. We refer the reader to the original paper for the details. Crucially, once the loss of hydrogen is known, the oxygen fractionation factor, $\chi_O$, can be obtained:

$$\chi_O = 1 - \frac{g (m_O - m_H) b_{H-O}}{\Phi_H k_B T (1 + X_O)}$$

(29)

If $\chi_O > 0$, then the hydrodynamic wind drags oxygen. The carbon dioxide fractionation factor can then be similarly calculated:

$$\chi_{CO_2} = \frac{1 - g (m_{CO_2} - m_H) b_{H-CO_2} / (\Phi_H k_B T) + b_{H-CO_2} X_O (1 - \chi_O) + b_{H-CO_2} X_O \chi_O / b_{O-CO_2}}{1 + b_{H-CO_2} X_O / b_{O-CO_2}}$$

(30)

If $\chi_{CO_2} > 0$ then carbon dioxide is dragged along in the hydrodynamic wind and the corresponding escape fluxes (mol/m²/s) of O and CO₂ are given by the following expressions::

$$\Phi_O = \Phi_H X_O \chi_O$$
$$\Phi_{CO_2} = \Phi_H X_{CO_2} \chi_{CO_2}$$

(31)

Alternatively, if $\chi_O > 0$ and $\chi_{CO_2} < 0$, then oxygen is dragged but not carbon dioxide, and the corresponding escape fluxes are as follows:

$$\Phi_H = \Phi / (m_H + m_O X_O \chi_O)$$
$$\Phi_O = (\Phi - \Phi_H m_H) / m_O$$
$$\Phi_{CO_2} = 0.0$$

(32)

If equation (29) yields $\chi_O < 0$, then then the XUV-driven escape flux is too small to drag oxygen. Carbon dioxide fractionation must be recalculated with $\chi_O = 0$. If $\chi_{CO_2} < 0.0$ then the only escaping gas in hydrogen:

$$\Phi_H = \Phi / m_H$$
$$\Phi_O = \Phi_{CO_2} = 0.0$$

(33)

In rare case where $\chi_O < 0$ and $\chi_{CO_2} > 0$, then loss rates as follows:

$$\Phi_{CO_2} = \Phi_H \chi_k X_{CO_2}$$
$$\Phi_O = 0.0 \tag{34}$$

It is convenient to convert these molecular escape rates to molar escape rates:

$$\varphi_{XUV-H} = \Phi_H m_H / \mu_H$$
$$\varphi_{XUV-O} = \Phi_O m_O / \mu_O \tag{35}$$
$$\varphi_{XUV-CO_2} = \Phi_{CO_2} m_{CO_2} / \mu_{CO_2}$$

If the diffusion limited escape flux exceeds the XUV-driven H loss, then the diffusion limited flux is adjusted downwards:

$$\varphi_{diff}^{'} = \min\left\{ \varphi_{XUV-H} / 2, \varphi_{diff} \right\} \tag{36}$$

Finally, we combine our expressions for diffusion-limit hydrogen escape and XUV-limited escape fluxes to obtain general expressions for the escape fluxes of hydrogen, $E_H$ (mol/s), oxygen $E_O$ (mol/s), and carbon dioxide, $E_{CO_2}$ (mol/s):

$$E_H = \left( \frac{w_1 2\varphi_{diff}^{'} + w_2 \varphi_{XUV-H}}{w_1 + w_2} \right) 4\pi r_p^{\,2}$$
$$E_O = \left( \frac{w_2 \varphi_{XUV-O}}{w_1 + w_2} \right) 4\pi r_p^{\,2} \tag{37}$$
$$E_{CO_2} = \left( \frac{w_2 \varphi_{XUV-CO_2}}{w_1 + w_2} \right) 4\pi r_p^{\,2}$$

Here, the weightings $w_1$ and $w_2$ are a function of the H-abundances in the upper atmosphere:

$$w_1 = \lambda_{tra} \left( 2/3 - X_H \right)^3$$
$$w_2 = X_H^{\,3} \tag{38}$$

These weighting functions ensure diffusion-limited H escape for low stratospheric abundances, and a smoothly transition to XUV-driven escape as the upper atmosphere becomes steam dominated. The precise transition abundance is unknown and will, in general, depend on conductive and radiative cooling of the upper atmosphere as well as downward diffusive transport. Here, it is represented by the free parameter, $\lambda_{tra}$, which ranges from $10^{-2}$ to $10^2$ and is sampled uniformly in log space.

The efficiency of hydrodynamic escape is parameterized by loosely following the approach of Wordsworth et al. (2018). If the XUV-stellar flux is insufficient to drag oxygen, then the efficiency is equal to a constant, $\varepsilon_{lowXUV}$, which is randomly sampled from 1% to 30%. Alternatively, if the XUV-stellar flux exceeds what is required to drag O, then some portion of the excess energy, $\zeta$, goes into driving further escape, whereas the rest, $1-\zeta$, is assumed to be efficiently radiated away. The efficiency factor, $\zeta$, is randomly sampled from 0-100% for complete generality. This leads to the following function for the efficiency of hydrodynamic escape:

$$\varepsilon(F_{XUV}, X_O, \zeta, \varepsilon_{lowXUV}) = \begin{cases} \varepsilon_{lowXUV}, \text{ where } \dfrac{\varepsilon_{lowXUV} F_{XUV} r_p}{4GM_P} < m_H \dfrac{g(m_O - m_H)b_{H-O}}{k_B T(1 + X_O)} \\[4mm] \zeta \varepsilon_{lowXUV} + (1-\zeta)\left(\dfrac{m_H \dfrac{g(m_O - m_H)b_{H-O}}{k_B T(1 + X_O)}}{\dfrac{F_{XUV} r_p}{4GM_P}}\right), \text{ where } \dfrac{\varepsilon_{lowXUV} F_{XUV} r_p}{4GM_P} \geq m_H \dfrac{g(m_O - m_H)b_{H-O}}{k_B T(1 + X_O)} \end{cases}$$

$$(39)$$

Fig. S6 illustrates our escape parameterization, with Monte Carlo outputs plotted for appropriate ranges in uncertain escape parameters.

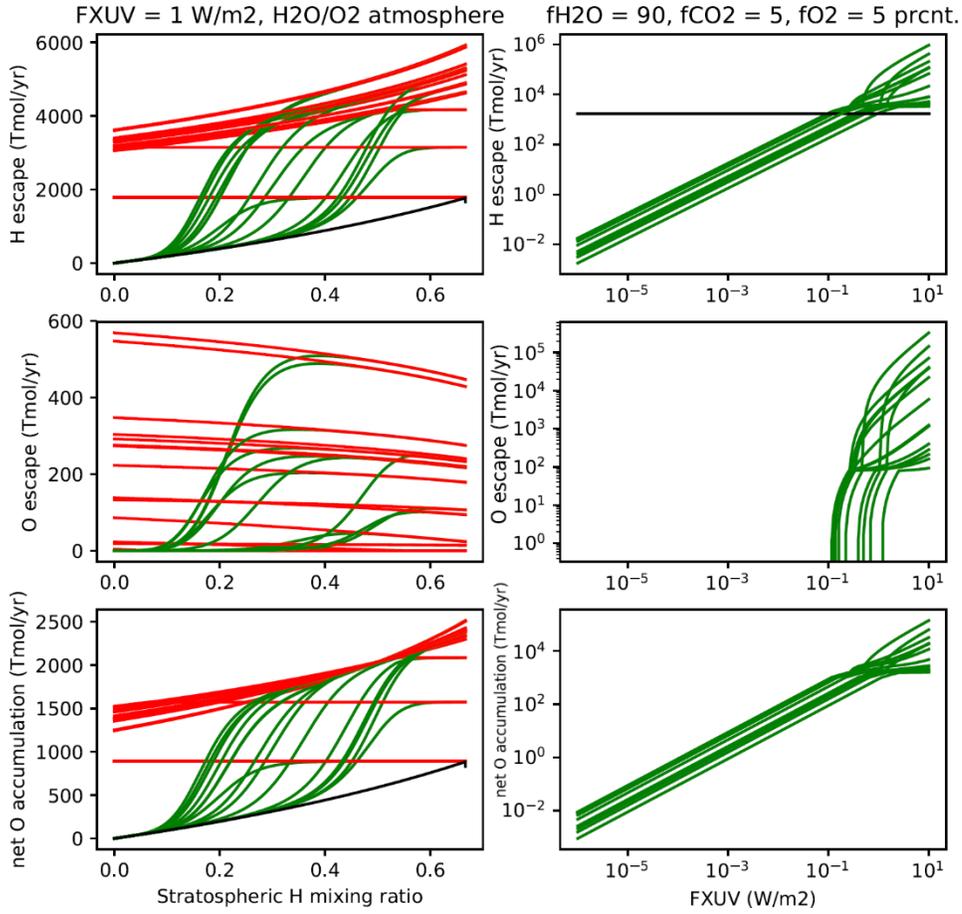

Fig. S6: Illustrative examples of the escape parameterization. Each line denotes a different calculation, where uncertain escape parameters $\varepsilon_{lowXUV}$ and $\lambda_{tra}$ have been randomly sampled. Black lines denote the cold trap diffusion limit (eq. (26)), red lines denote XUV-driven hydrodynamic escape (eq. (35)), and green lines show the weighted combination that is the escape parameterization in our model (eq. (37)). On the left-hand side, the stellar XUV flux is held constant, and escape fluxes are plotted as a function of the stratospheric H mixing ratio (for an atmosphere without $CO_2$). On the right-hand side, the composition of the upper atmosphere is held constant, and escape fluxes are plotted as a function of the stellar XUV flux received by the planet.

A.10) Solid-state evolution: Outgassing and crustal production

The outgassing model follows that described in Wogan et al. (2020) where we calculate redox-dependent speciation of volatiles between melt and gas phase. Given mantle concentrations of $H_2O$ and $CO_2$ (by mass), corresponding melt fractions can be calculated assuming accumulated fractional melting:

$$fr_{melt-H_2O} = \frac{\left(1-(1-\bar{\psi})^{1/k_{H_2O}}\right)}{\bar{\psi}} \frac{M_{solid-H_2O}}{M_{mantle}} \tag{40}$$

$$fr_{melt-CO_2} = \frac{\left(1-(1-\bar{\psi})^{1/k_{CO_2}}\right)}{\bar{\psi}} \frac{M_{solid-CO_2}}{M_{mantle}} \tag{41}$$

Here $\bar{\psi}$ is the average melt fraction over the portion of the mantle where melting occurs (see below). These melt fractions, along with mantle oxygen fugacity (eq. (19)) and magma chamber pressure-temperature conditions are used as inputs to the outgassing calculations. For subaerial outgassing, the outgassing pressure is the pressure overburden of the atmospheric inventory, whereas for submarine outgassing, the pressure overburden is the atmospheric + ocean inventory. All outgassing is assumed to occur at the solidus temperature. Given these inputs, the outgassing thermochemical equilibrium model (Wogan et al. 2020) outputs gaseous mixing ratios, $f_i$, for outgassed $CO_2$, $H_2O$, $H_2$, $CO$, and $CH_4$, as well as the moles of gas per total moles of melt plus gaseous species, $\alpha_G$. Note that the outgassing model does not consider the evolution of melt composition and oxygen fugacity along a degassing path; instead, we assume that the melt oxygen fugacity is buffered to that of the source rock, and that outgassed volatiles are determined by the equilibrium gas phase mixing ratios at surface pressure. Moreover, the molecular oxygen fraction of the degassed mixed is conservatively assumed to be negligible; the only possible source of atmospheric oxygen in the model is H escape.

Gaseous mixing ratios can be convolved with melt production, $MP$, (m$^3$/s, described below) to calculate outgassing fluxes, $V_i$ (mol/s), for each species:

$$V_i = MP \times \rho_m \times \frac{f_i \alpha_G \vartheta_m}{1 - \alpha_G} \tag{42}$$

Here, $\vartheta_m$ = 15.5 mol magma/kg magma is the inverse molar mass of magma, which is assumed to be constant. The overall $O_2$-consumption sink from outgassed volatiles is the summation of reducing species:

$$V_{O_2-sink} = 0.5V_{H_2} + 0.5V_{CO} + 2V_{CH_4} \tag{43}$$

Overall outgassing fluxes are the combination of subaerial and submarine contributions, weighted by the surface land fraction, $LF$ (see below):

$$V_i = V_i^{subaerial} LF + V_i^{submarine}(1 - LF) \tag{44}$$

To obtain mass fluxes, these molar outgassing fluxes must be weighted by their respective molecular masses:

$$\begin{aligned} F_{outgas-CO_2} &= \mu_{CO_2} V_{CO_2} \\ F_{outgas-H_2O} &= \mu_{H_2O} V_{H_2O} \end{aligned} \tag{45}$$

At any given time, the average melt fraction of freshly produced crust is given by integrating the melt fraction from the radius at which mantle temperature equals the solidus to the surface:

$$\bar{\psi} = \frac{\int_{r=r(T=T_{solidus})}^{r=r_p} \psi\left(T(r), T_{solidus}(P_{overburden}, r), T_{liquidus}(P_{overburden}, r)\right) \times 4\pi r^2 dr}{\int_{r=r(T=T_{solidus})}^{r=r_p} 4\pi r^2 dr} \tag{46}$$

Here, the melt fraction at any given radius is given by the following expression:

$$\psi(T(r), T_{solidus}, T_{liquidus}) = \begin{cases} 0, \ T(r) < T_{solidus}(P_{overburden}, r) \\ 1, \ T(r) > T_{liquidus}(P_{overburden}, r) \\ \dfrac{T(r) - T_{solidus}(P_{overburden}, r)}{T_{liquidus}(P_{overburden}, r) - T_{solidus}(P_{overburden}, r)}, \ otherwise \end{cases} \tag{47}$$

Refer above for expressions for the solidus and liquidus.

We assume the crust thickness, $d_{crust}$ (m), is equal to the thickness of the melt layer, which can be calculated from the average melt fraction and total melt volume:

$$d_{crust} = r_p - \left(r_p^3 - \frac{3}{4\pi}\bar{\psi} \int_{r=r(T=T_{solidus})}^{r=r_p} 4\pi r^2 dr\right)^{1/3} \tag{48}$$

Once crustal depth is known, the melt production rate, $MP$ (m³/s) can be calculated from the plate velocity, $v_{plate}$ (m/s), and ridge length, $l_{ridge}$ (m):

$$MP = l_{ridge} \times d_{crust} \times v_{plate} \qquad (49)$$

However, from the expression for half-space cooling of oceanic crust (e.g. Kite et al. 2009; Turcotte & Schubert 2002, Sec 4.15), we can relate interior heatflow to ridge length and spreading rate as follows:

$$\left( \frac{q_m 4\pi r_p^2}{2k\left(T_{surf} - T_p\right)} \right)^2 = \frac{4\pi r_p^2 \times l_{ridge} \times v_{plate}}{\pi \kappa} \qquad (50)$$

Substituting this into eq. (49), melt production can be calculated from only the heatflow and crustal depth:

$$MP = \left( \frac{q_m 4\pi r_p^2}{2k\left(T_{surf} - T_p\right)} \right)^2 \frac{\pi \kappa}{4\pi r_p^2} d_{crust} \qquad (51)$$

Note that we are assuming the entire planetary area is involved in plate tectonics, which might cause us to underestimate melt production slightly. Plate velocity can be estimated by assuming a plausible ridge length, $3\pi r_p$:

$$v_{plate} = \frac{MP}{d_{crust} l_{ridge}} = \frac{MP}{d_{crust} 3\pi r_p} \qquad (52)$$

Plate velocity is only used for calculating seafloor weathering rates. Outgassing fluxes and crustal sinks of oxygen all depend on melt production. Note that for Venus and stagnant lid exoplanets the assumption of plate tectonics may overestimate melt production. Future versions of the model will explore a stagnant lid regime, but plate tectonics ought to maximize melt production and therefore maximize geological sinks of oxygen, which are the focus of this study.

A.11) Solid-state evolution: Weathering

Carbon is transferred from surface reservoirs to the interior via silicate weathering. Silicate weathering is the combination of continental weathering and seafloor weathering. To estimate this partitioning, we need to calculate the average depth of the ocean and corresponding land fraction. Following Cowan and Abbot (2014), we assume that there is a maximum ocean depth, $d_{ocean-\max}$ (m), above which there are no submerged continents.

$$d_{ocean} = (1 - fr_{atmo-H_2O}) M_{fluid-H_2O}$$
$$d_{ocean-\max} = 11400(9.8 / g) \qquad (53)$$

Given these two quantities, we approximate the planetary hypsometric curve (proportion of land as a function of elevation) with a power law, and use this to calculate average land fraction, $LF$, and land fraction relative to the modern Earth, $RLF$:

$$LF = \max\left\{0, 1 - \left(d_{ocean} \,/\, d_{ocean-\max}\right)^{0.25}\right\}$$

$$RLF = \frac{LF}{1 - \left(2.5 \,/\, 11.4\right)^{0.25}}$$

<div align="right">(54)</div>

Clearly these are crude approximations, but they capture the fact the planets with a few Earth oceans ought to have some subaerial land, but that for large water inventories all crust is submerged. In any case, total land fraction does not have a large impact on weathering feedbacks (Abbot et al. 2012).

Continental weathering fluxes, $W_{cont}$ (kg/s), and seafloor weathering fluxes, $W_{SF}$ (kg/s), are given by expressions similar to those described in Krissansen-Totton et al. (2018):

$$W_{SF} = \frac{W_{coef}}{4} \times 10^{\left(-0.3(pH-7.7)\right)} \left(\frac{v_{plate}}{3\,cm\,/\,yr}\right) \exp\left(-\frac{E_{SF}}{8.314}\left(\frac{1}{T_{deep}} - \frac{1}{285}\right)\right)$$

<div align="right">(55)</div>

$$W_{cont} = W_{coef} \times RLF \times \left(\frac{pCO_2}{350\,ppm}\right)^{\gamma} \exp\left(-\frac{\left(T_{surf} - 285\right)}{T_{efold}}\right)$$

<div align="right">(56)</div>

Here, the multiplicative factor $W_{coef}$ = 4000 kg/s is chosen to approximately reproduce modern Earth fluxes (or rather is the weathering flux required to balance mantle-derived $CO_2$ outgassing). Unknown, randomly sampled parameters include the temperature dependence of continental weathering, $T_{efold}$ = 5-30 K and the $CO_2$ dependence of continental weathering, $\gamma$ = 0.1-0.5. The temperature dependence of seafloor weathering, $E_{SF}$ = 90 kJ/mol, and pH-dependence of seafloor weathering are assumed to be known constants. More sophisticated weathering parameterizations that account for kinetic dependencies, thermodynamic solute concentration limits, and a precipitation-limited runoff dependence have been proposed (Graham & Pierrehumbert 2020), but given the uncertainties in geological parameters that feed into such models, our simple $CO_2$ and temperature dependent formulation is adequate for providing a crude thermostat.

Total weathering will be the summation of continental and seafloor weathering, weighted by the fraction of liquid water at the surface, $(1 - fr_{atmo-H_2O})$. This ensures that weathering tends to zero as oceans evaporate. We also include a possible supply limit to weathering as an unknown variable that could potentially limit the rate at which dense $pCO_2$ atmospheres are sequestered if the supply of erodible rock is low. This supply limit, $W_{sup-lim}$ = $10^5$ to $10^7$ (kg/s), is not coupled to crustal production since not all newly produced crust will necessarily be delivered to the surface to be weathered. The overall expression for $CO_2$-removal via weathering is therefore:

$$F_{Weather-CO_2} = \min\left\{W_{sup-lim}, f_{liquid-H_2O-fraction}\left(W_{cont} + W_{SF}\right)\right\}$$

<div align="right">(57)</div>

Because we are modeling a plate tectonics regime, we assume that all carbon dioxide removed by weathering is returned to the mantle. This can be seen in equation (16), whereby the weathering flux, $F_{Weather-CO_2}$, transfers carbon from the surface fluid reservoir to the interior mantle reservoir.

### A.12) Solid-state evolution: Deep water cycle

The transfer of water from the mantle to the surface is already specified by $F_{outgas-H_2O}$. In-gassing of water from the surface to the interior is controlled by water-rock reactions and tectonic processes. There is considerable uncertainty in the function dependencies of ingassing reactions (e.g. Cowan & Abbot 2014; Komacek & Abbot 2016; Schaefer & Sasselov 2015). Here, we loosely follow the approach in these papers, but introduce several unknown parameters to account for different deep hydrological cycle feedbacks.

The maximum depth in the crust within which hydration reactions may occur is approximated by:

$$d_{hydr} = k \frac{T_p - T_{surf}}{q_m} \left( \frac{973 - T_{surf}}{T_p - T_{surf}} \right) = k \frac{\left( 973 - T_{surf} \right)}{q_m} \tag{58}$$

Here, 973 K is the maximum surface temperature for serpentine stability (Schaefer & Sasselov 2015), and $k = 4.2$ W/m/K is the thermal conductivity of silicates (c.f. equation (4)). Since we assume no hydration occurs below the crust, it is also helpful to define the fractional depth of hydration as the ratio of the hydration depth to the crustal depth, or 1.0 (whichever is smaller):

$$f_{hydr-depth} = \min \left\{ d_{hydr} / d_{crust}, 1.0 \right\} \tag{59}$$

Water loss from surface reservoirs can be conceptually partitioned into hydration reactions that add water to the solid interior but do not alter atmospheric redox state, e.g.

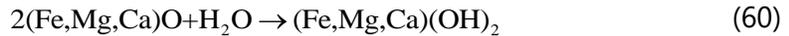
$$2(Fe,Mg,Ca)O + H_2O \rightarrow (Fe,Mg,Ca)(OH)_2 \tag{60}$$

and hydration reactions that oxidize the solid interior and result in outgassed hydrogen (i.e. hydrogen that is lost to space under anoxic conditions, or recombines with atmospheric oxygen under oxidizing conditions):

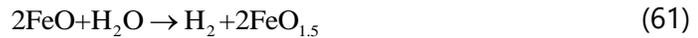
$$2FeO + H_2O \rightarrow H_2 + 2FeO_{1.5} \tag{61}$$

Hydration reactions that add water to the interior, equation (60), are parameterized as follows:

$$F_{H_2O-hydr} = \min \left\{ \frac{d_{ocean}}{d_{max-ocean}}, 1.0 \right\} f_{hydr-depth} fr_{hydr-frac} MP\rho_m \max \left\{ 0, 1 - \frac{M_{solid-H_2O}}{M_{solid-H_2O-max}} \right\} \tag{62}$$

Here, $fr_{hydr-frac} = 10^{-3}$ to 0.03 (sampled uniformly in log space) is the unknown efficiency of hydration reactions. We are assuming that, at most, hydrated crust is 3% water by mass (Schaefer & Sasselov 2015), but this number could be much less depending on the tectonic regime and cracking of the crust. Some unknown portion of this crustal water may be returned to the surface via arc volcanism, for instance. After considering this efficiency factor, we assumed all hydrated crust is returned to the mantle. We assume a linear dependence on ocean depth for as long as there is emerged land, then no ocean depth dependence beyond this. Additionally, we assume that the return of water to the interior tapers off as the water content of the mantle approaches its maximum value, $M_{solid-H_2O-max}$ (kg). This unknown variable is sampled randomly from 0.5-15 Earth oceans (Komacek & Abbot 2016).

Hydration reactions that oxidize the crust and remove water from surface, but do not add water to the interior (i.e. serpentizing reactions that produce hydrogen, equation (61)) are parameterized as follows:

$$F_{H_2O-serp} = f_{wet-oxid} \min\left\{\frac{d_{ocean}}{d_{max-ocean}}, 1.0\right\} f_{hydr-depth} MP\rho_m x_{Fe}\left(\frac{M_{solid-FeO}}{M_{solid-FeO} + M_{solid-FeO_{1.5}}}\right)$$

(63)

Here, $f_{wet-oxid}$ is another unknown efficiency parameter ($10^{-3}$ to $10^{-1}$) representing the fraction of crustal iron that is oxidized via hydration reactions (Lécuyer & Ricard 1999). This efficiency parameter represents the degree of serpentinization in water-rock reactions. Total ingassing contributions are calculated as follows:

$$F_{ingas-H_2O-loss} = F_{H_2O-hydr} + F_{H_2O-serp}\frac{\mu_{H_2O}}{3\mu_{FeO}}$$

$$F_{ingas-H_2O-gain} = F_{H_2O-hydr}$$

(64)

The former, $F_{ingas-H_2O-loss}$, represents the flux of water lost from the surface reservoir, whereas the latter, $F_{ingas-H_2O-gain}$, represents the flux of water into the interior via subduction of hydrated crust. This transfer of water can also be seen in equation (16). Note that these loss and gain quantities are not necessarily equal in because, as discussed above, $H_2$ produced by serpentinization may be lost to space, thereby permanently removing water from the planet.

A.13) Solid-state evolution: Planetary redox budget

The interior may become oxidized via outgassed of reduced species (discussed above), dry oxidation, and wet oxidation. First, we consider dry (direct) crustal oxidation, which can be represented by the following reaction:

$$2\text{FeO} + 0.5\text{O}_2 \rightarrow 2\text{FeO}_{1.5}$$

(65)

The flux of this dry crustal sink is parameterized by the following equation:

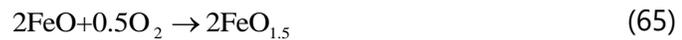

$$F_{dry-oxid} = (1 - RLF) f_{dry-oxid} x_{Fe}\left(\frac{M_{solid-FeO}}{M_{solid-FeO} + M_{solid-FeO_{1.5}}}\right)\rho_m MP_{surface}\frac{\mu_O}{2\mu_{FeO}}$$

(66)

Here, there is a land fraction dependence to ensure that no dry oxidation of the crust occurs when the surface is covered in water. The unknown efficiency parameter $f_{dry-oxid}$ ($10^{-4}$ to 10%) is discussed in the main text, and is the fraction of reduced iron in newly produced crust that is oxidized.

We also modify the melt production term to represent the fact that there is some maximum amount of surface melt accessible to oxidation via diffusion through extrusive lava flows. The diffusivity of oxygen in silicate melts is $D_{ox} \approx 10^{-7}$ cm$^2$/s (Canil & Muehlenbachs 1990), which implies a downward diffusion depth of $\sim \sqrt{D_{ox}t}$ per unit time, $t$. The maximum downward oxygen flux averaged over one year is therefore given by:

$$F_{\max-O_2} = f_{lava} 4\pi r_p^2 \frac{\sqrt{D_{ox} \times 3.15 \times 10^7 \, s}}{3.15 \times 10^7 \, s \times 100 \, cm \, / \, m}$$

$$= f_{lava} \times 2.87 \times 10^5 \, m^3 \, / \, s = f_{lava} \times 9.05 \times 10^3 \, km^3 \, / \, yr$$

(67)

Here, the efficiency factor, $f_{lava}$, is the average fraction of the planetary surface that is continuously molten due to extrusive volcanism after the magma ocean has solidified. Because thermal diffusivity exceeds chemical diffusivity—and because mean surface temperature is below the solidus by definition—extrusive magmas will form a low permeability solid crust as they cool, precluding continuous diffusion of oxygen. Thus, even for extreme rates of resurfacing due to high heatflow, $f_{lava}$ is likely low. For example, on Io, where average internal heatflow is 1-3 W/m$^2$ (Veeder et al. 2012), only a few km$^2$ of the surface is estimated to be molten at any given moment (Mura et al. 2020). We uniformly sample $f_{lava}$ = $10^{-4}$ to 1 in log space for full generality. Given this upper limit on melt oxidation, the amount of oxidizable melt is given by:

$$MP_{surface} = \min\left\{MP, \frac{f_{lava} \times 2.87 \times 10^5}{f_{dry-oxid}}\right\}$$

(68)

This avoids unrealistic oxygen sinks during the transition from magma mush to solid mantle where melt volumes are extremely high, but the melt accessible to atmospheric oxygen via diffusion through extrusive magmas limited. The value for $f_{lava}$ does not affect any of our model scenarios except for Scenario 3, where values exceeding 0.01 are required for oxygen accumulation (Fig. S1d).

The wet oxidation of the interior from hydration reactions is already the appropriately weighted serpentinization flux:

$$F_{wet-oxid} = F_{H_2O-serp} \frac{\mu_O}{3\mu_{FeO}}$$

(69)

Recall the term $F_{oxid-fluid}$ represents the total flux of free oxygen lost from the atmosphere-ocean reservoir to the interior, whereas the term $F_{oxid-fluid}$ represents the total flux of free oxygen gained by the interior. Oxidized crust is assumed to be mixed back into the mantle on long timescales (equation (16)) via subduction, or slab delamination. We also assume that the outgassing of reduced species must ultimately be balanced by oxidation of the crust, and so the so net oxidation of the interior equals the reduction of the fluid reservoir:

$$F_{oxid-fluid} = F_{dry-oxid} + F_{wet-oxid} + \mu_{O_2} V_{O_2-sink} = F_{oxid-solid} \tag{70}$$

The only exception to this equality is when atmospheric oxygen levels are very low, and fluxes are modified for numerical reasons (see below).

### Text B. Numerical approach

All code is written in python and available open source ([DOI_TBD]). The system of differential equations was solved explicitly with either RK45 or RK23 using the solve_ivp module in scipy. The maximum timestep was shorter for the magma ocean phase (10000 years) compared to the temperate evolution phase ($10^6$ years).

To avoid sawtoothing and excessive computation time at low reservoir abundances, various adjustments were made to the differential equations to ensure adequate performance. First, if atmospheric $CO_2$ dropped below 50 Pa and if weathering exceeds outgassing, then the time-derivative of surface carbon dioxide was set equal to zero. This may mean climate evolution is slightly inaccurate at low $pCO_2$, but the effects are minor. Second, atmospheric oxygen is similarly prevented from dropping below 0.1 Pa. If atmospheric oxygen is below 0.1 Pa and if oxygen production via escape is less than oxygen consumption, then the following adjustments are made. The atmosphere is assumed to be in an anoxic steady state and so oxygen atmospheric sinks are set equal to the escape source. However, since oxygen production via escape is less than oxygen sinks (outgassing and $H_2$, wet and dry oxidation reactions), then the interior is assumed to be oxidized by the difference as the excess reductants ($H_2$) escape to space. This might slightly overestimate mantle oxidation because some reductants (e.g. CO) will get photochemically oxidized rather than escape, but the redox budget of the atmosphere is not directly affected, and the effects on mantle redox evolution are negligible.

### Text C. Venus Model validation:

To further validate the model, we demonstrate that it can broadly reproduce the known atmospheric evolution of Venus. To model Venus, all parameters are kept the same as for Earth except for planet radius, mass, planet-star separation, and albedo parameterization (see Table S2). Fig. S7 shows all model outputs that reproduce modern Venus, which is defined to be atmospheric oxygen < $10^{15}$ kg (~0.2 mbar), atmospheric $CO_2$ exceeding $2\times10^{20}$ kg (40 bar), no surface water ocean, and atmospheric $H_2O$ < $2\times10^{16}$ kg (~3 mbar).

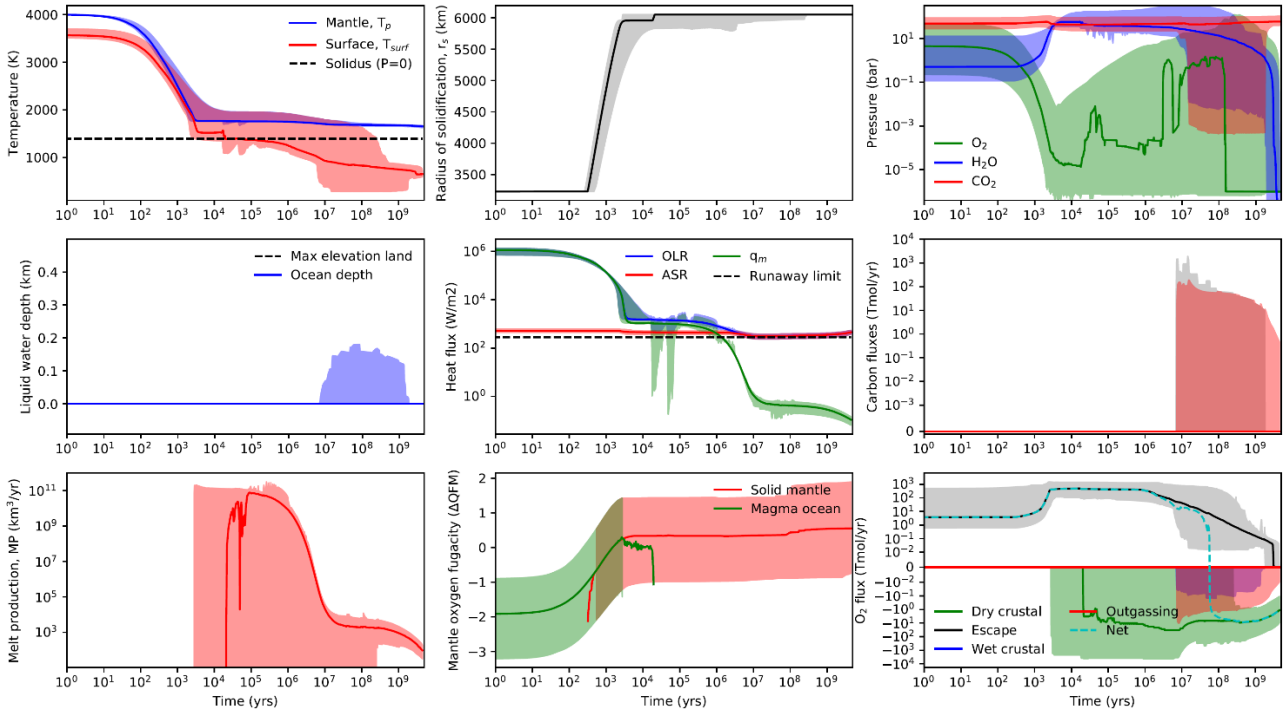

Fig. S7: Model runs that reproduce modern Venus conditions. Note that there are two qualitatively different histories that can reproduce modern Venus (top left). Either Venus was always in a runway greenhouse phase and never condensed a surface ocean, or Venus maintained a temperate surface for several Gyr before transitioning back to runaway greenhouse as solar insolation increased.

Note that our Venus model somewhat overpredicts modern day Venusian heat flow and melt production because we assume a plate tectonics model (Nimmo & McKenzie 1998). We save a more detailed comparative study of solar system planets with stagnant lid tectonics and resurfacing events for future study.

In Fig. S8 we plot key parameter values for model runs that successfully reproduce modern Venus. Initial volatile inventories are likely small (Fig. S8a) and dry crustal oxidation must be relatively efficient (Fig. S8c).

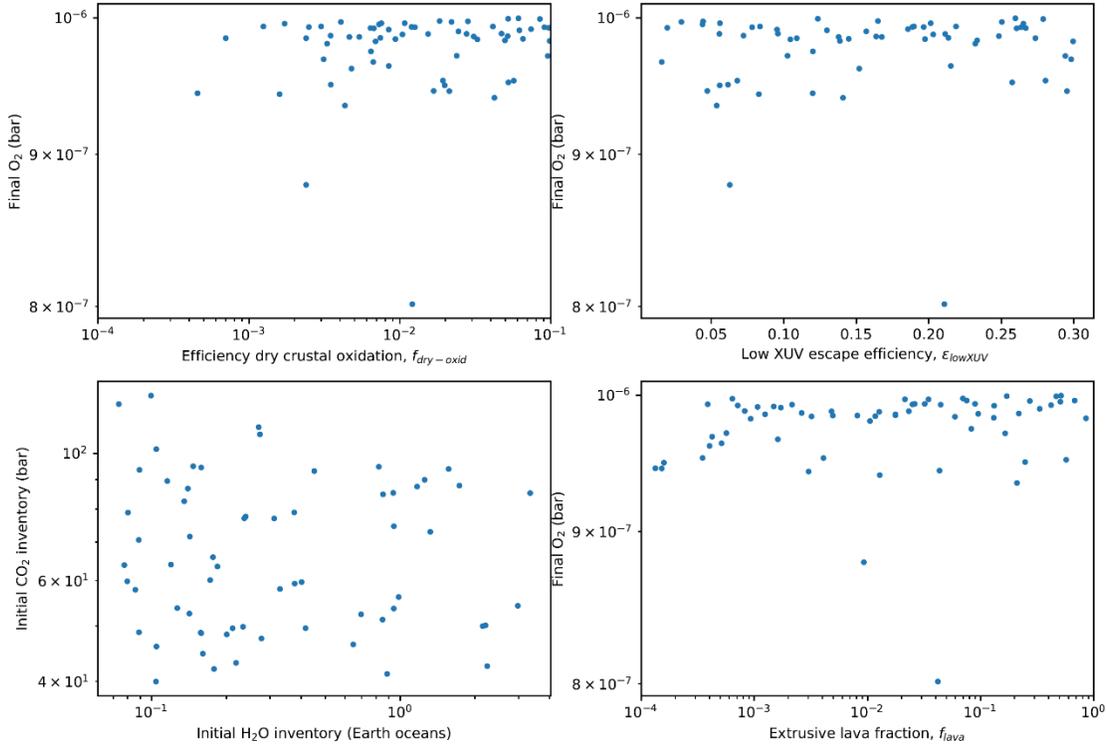

Fig. S8: Parameter values for individual model runs in Fig. S7. These shows the require initial volatile inventories (bottom left) and dry crustal oxidation efficiency (top left) required to reproduce modern Venus. A broad range of XUV escape efficiencies (top right) and fractional molten areas (bottom right) are permissible.

## Text D. Hydrous Mantle Sensitivity Test

Here, we test whether modifying the solidus for hydrated mantles affects oxygen accumulation in the waterworlds scenario. Following Katz et al. (2003) we modify our expressions for the solidus and liquidus as follows:

$$T_{solidus}(r, P_{Overburden}) = \frac{T_1(r)\exp\left(10^{-5}(-r_p + r + 10^5)\right) + T_2(r)\exp\left(10^{-5}(r_p - r - 10^5)\right)}{\exp\left(10^{-5}(-r_p + r + 10^5)\right) + \exp\left(10^{-5}(r_p - r - 10^5)\right)}$$

$$T_1 = \max\left\{1170, 1420 + 104.42\times10^{-9}\,g\,\rho_m\left(r_p - r\right) + 104.42\times10^{-9}P_{overburden} - 4.74\times10^4\left(M_{solid-H_2O}/M_{mantle}\right)^{0.75}\right\}$$

$$T_2 = \max\left\{1170, 1825 + 26.53\times10^{-9}\,g\,\rho_m\left(r_p - r\right) + 26.53\times10^{-9}P_{overburden} - 4.74\times10^4\left(M_{solid-H_2O}/M_{mantle}\right)^{0.75}\right\}$$

$$T_{liquidus}(r, P_{Overburden}) = T_{solidus}(r, P_{Overburden}) + 600$$

$$(71)$$

This modification is a crude polynomial fit to Fig. 3 in Katz et al. (2003). It accounts for the depression of the solidus as the water content of the mantle increases, but the solidus never drops below about 900°C, which is the approximate water saturation limit. Waterworld calculations repeated with this hydrated solidus are shown in Fig. S9. Although crustal production is elevated compared to the anhydrous solidus case in the main text, outcomes are qualitatively similar to Fig. 5 in the main text. Significant abiotic oxygen accumulation becomes increasingly likely for initial water inventories exceeding 100 Earth oceans (Fig. S10).

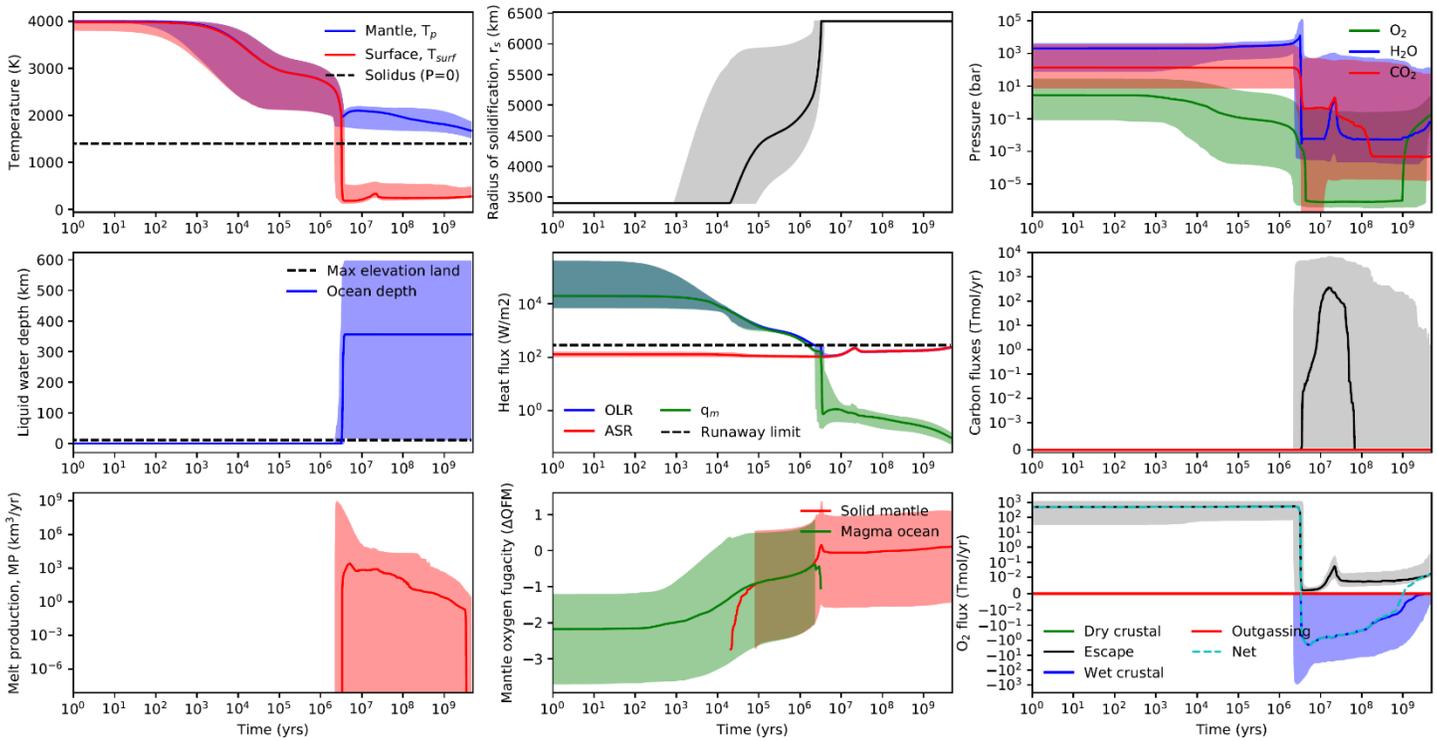

Fig. S9: Same as Fig. 5 in the main text except the solidus decreases with mantle hydration. Early crustal production is elevated (bottom left), but outcomes are qualitatively similar. Oxygen sinks are shut down by the pressure overburden after a few billion years and oxygen accumulates.

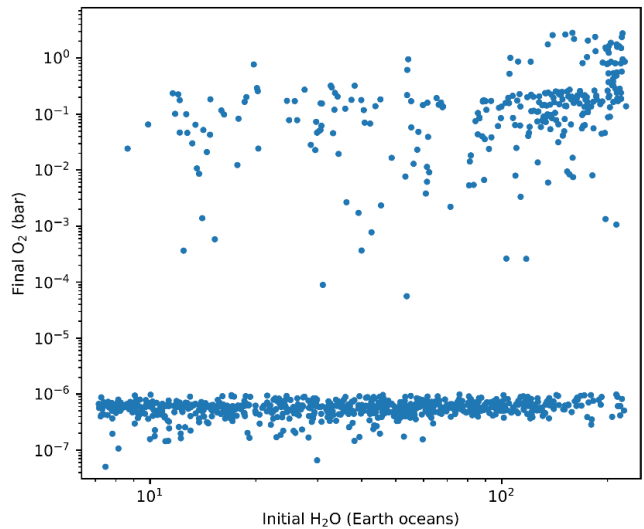

Fig. S10: Same as Fig. 6 in the main text except the solidus decreases with mantle hydration. Abiotic oxygen accumulation is somewhat less frequent, and occurs at higher initial water inventories, but results are qualitatively the same.

**Text E. Impact Ejecta O₂-sinks**

Here, we test whether the delivery of reducing materials from impactors could potentially draw down oxygen produced in desertworld scenarios. We introduce an impactor flux, $F_{imp}$ (kg/yr), that diminishes exponentially with time:

$$F_{imp} = F_{imp}^0 \exp\left(t/t_{decay}\right) \tag{72}$$

Here, the coefficient $F_{imp}^0$ is randomly sampled (in log space) from $10^{11}$ to $10^{14.5}$ kg/yr and the decay time, $t_{decay}$ (Gyr), is randomly sampled from 0.06 to 0.14 Gyr. These ranges are adopted because they approximately reproduce plausible estimates for impactor fluxes in the Hadean and early Archean, both with and without a late heavy bombardment (Kadoya et al. 2020). Additionally, we assume that impactors are 30% metallic iron by mass, and that 100% of this iron is completely oxidized to ferric iron instantaneously upon impact, depleting atmospheric oxygen. Fig. S11 shows our desertworld calculations repeated with this impactor flux. We find oxygen accumulation and retention for several Gyr is still possible, although only when the total impactor flux is low. Fig. S12 shows oxygen accumulation after 4.5 Gyr as a function of total impactor flux. Abiotic oxygen may accumulation for impactor mass fluxes < $10^{20}$ kg.

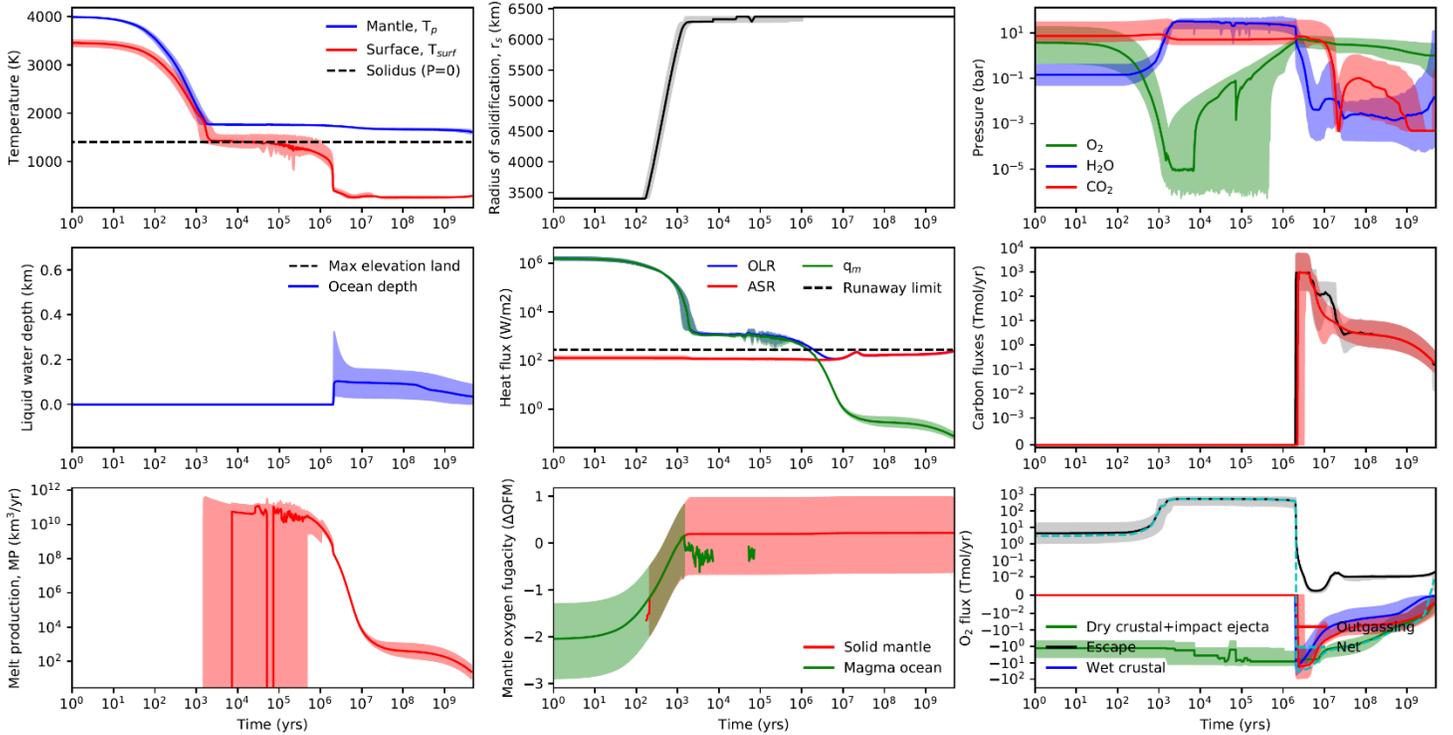

Fig. S11: Same as Fig. 7 in the main text except impact ejecta sinks for oxygen have been added. Retention of abiotic oxygen is still possible if impactor fluxes are low.

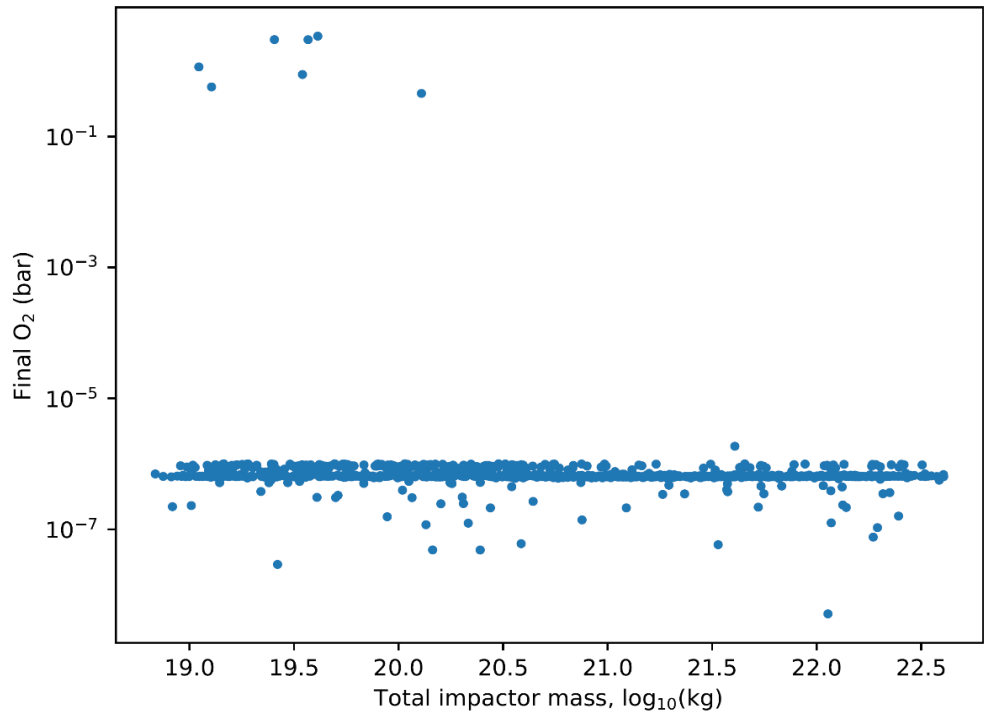

Fig. S12: Scenario 3 abiotic oxygen accumulation as a function of total impactor flux. Large impactor fluxes preclude the retention of abiotic oxidation if all impactor material is efficiently oxidized.

**Text F. Stratospheric Temperature Sensitivity Test**

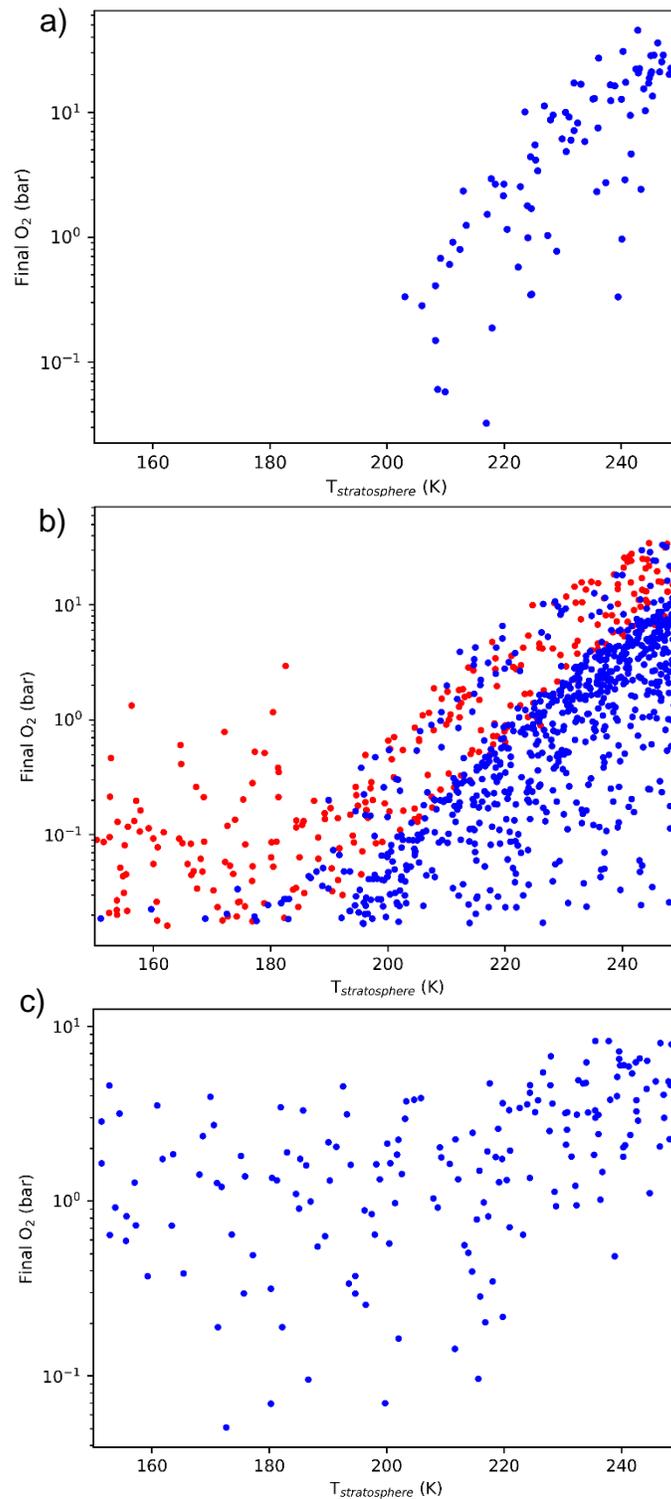

**Fig. S13: Sensitivity of false positive results to stratospheric temperature**.
Calculations in main text were repeated, but stratospheric temperature is a free variable
that is randomly sampled from 150 K to 250 K. Final oxygen accumulation after 4.5 Gyrs
is plotted as a function of stratospheric temperature. Subplot (a) shows all Scenario 1

false positives model runs, whereas (b) shows all Scenario 2 false positives, and (c) shows all Scenario 3 false positives. In (b) red dots denote leftover oxygen after the magma ocean, whereas blue dots show model runs where all oxygen produced during the magma ocean phase is sequestered in the mantle, and oxygen builds up subsequently, as described in the main text.

Fig. S13 shows the sensitivity of each false positive scenario to stratospheric temperature. Each dot is a model run representing an oxygen false positive. For the first scenario (Fig. S13a), abiotic oxygen only occurs when stratospheric temperature exceeds ~200 K. This is because, at lower temperatures, the cold trap becomes more effective and H escape (and therefore O accumulation) is throttled.

For the waterworld scenario (Fig. S13b) oxygen accumulation may occur at any stratospheric temperature. However, this is more probable—and abiotic oxygen abundances are greater—at higher stratospheric temperatures. On waterworlds, cold stratospheres are not necessarily expected because an $N_2$-dominated atmosphere with low $CO_2$ is a probable outcome (Fig. 5), especially if continuous $CO_2$-drawdown via weathering occurs (Nakayama et al. 2019). Moreover, modest oxygen accumulation would result in ozone formation, that would further warm the stratosphere, potentially resulting in a positive $O_2$-accumulation feedback that is not considered here. Note that there are two subclasses of oxygen false positives in Fig. S13b, denoted by red and blue dots. The blue dots show model runs where oxygen accumulated during the initial magma ocean is completely sequestered in the mantle upon magma ocean solidification, whereas red dots denote scenarios whereby appreciable oxygen is left over after magma ocean solidification due to the high pressure-temperature conditions of the overburden suppressed solidus.

The third, desertworld scenario (Fig. S13c) is largely insensitive to stratospheric temperature. This is because water loss and oxygen accumulation occur immediately after magma ocean solidification while the steam-dominated atmosphere persists. There is no effective cold trap in the steam-dominated atmosphere and so escape fluxes are insensitive to stratospheric temperature.

**Text G. Reducing Mantle Sensitivity Test**

The nominal model assumes Earth-sized planets undergo rapid core formation with mantles that quickly approach ~FMQ±2. While this is a common assumption when modeling magma ocean evolution of the early Earth (e.g. Zahnle et al. 2007; Zahnle et al. 2010) and of terrestrial exoplanets (e.g. Schaefer et al. 2016), this may not be true for all terrestrial planets. To investigate the effects of a more reduced initial mantle, sensitivity tests were performed whereby the initial magma ocean was endowed with a smaller amount of free O ($0.5 \times 10^{21}$ to $1.5 \times 10^{21}$ kg), such that after 4.5 Gyrs of evolution, mantle oxygen fugacity is closer to the iron-wüstite buffer than FMQ. Additionally, following Ortenzi et al. (2020), we modified the outgassing model such that the melt-solid partitioning of carbon is controlled by graphite saturation. The concentration of carbon dissolved in a graphite-saturated melt in redox state dependent:

$$X_{CO_3,grahite-sat} = \left(K_{1,graphite}K_{2,graphite}fO_2\right) / \left(1 + K_{1,graphite}K_{2,graphite}fO_2\right)$$

$$X_{CO_2,grahite-sat} = \left(44/36.594\right)X_{CO_3,grahite-sat} \ / \ \left(1 - \left(1 - 44/36.594\right)X_{CO_3,grahite-sat}\right)$$

(73)

Here, $fO_2$ is mantle oxygen fugacity, and we are converting between dissolved carbonate and carbon dioxide concentrations. The temperature and pressure-dependent equilibrium constants are defined as follows:

$$log_{10}\left(K_{1,graphite}\right) = 40.07639 - 2.53932\times10^{-2}\mathrm{T} + 5.27096\times10^{-6}\mathrm{T}^2 + 0.0267\left(\mathrm{P}-1\right)/\mathrm{T}$$

$$log_{10}\left(K_{2,graphite}\right) = -6.24763 - 282.56/\mathrm{T} - 0.119242\left(\mathrm{P}-1000\right)/\mathrm{T}$$

(74)

To calculate melt concentrations for outgassing calculations, we take the minimum of the concentrations in equations (41) and (73):

$$fr_{melt-CO_2} = \min\left\{X_{CO_2,grahite-sat}, \frac{\left(1-(1-\bar{\psi})^{1/k_{CO_2}}\right)}{\bar{\psi}}\frac{M_{solid-CO_2}}{M_{mantle}}\right\}$$

(75)

Taking the minimum ensures that graphite saturation does not overestimate dissolved carbon concentrations under oxidizing conditions and when the total carbon content in the mantle is low.

Finally, while we do not explicitly account for graphite precipitation during magma ocean solidification, we set $k_{CO_2} = 1.0$ in equation (16) to allow for greater retention of carbon in the mantle. It should be emphasized that these modifications do not constitute a fully consistent model of reduced mantle planetary evolution because the radiative transfer model does not allow for CO and $H_2$ dominated atmospheres. However, for post magma ocean evolution they are adequate approximations.

Fig. S14 is identical to Fig. 2 in the main text except for the reducing mantle initial conditions and other assumptions described above. Once again, an anoxic atmosphere is assured after 4.5 Gyrs of evolution because crustal oxygen sinks overwhelm oxygen sources. Fig. S15 is the reduced mantle equivalent of Fig. 3 in the main text showing high $CO_2$:$H_2O$ oxygen false positives. This scenario is largely unchanged by a reducing mantle; magmatic outgassing does not occur due to the high pressures and low mantle volatile concentrations following magma ocean solidification. Gradual oxygen accumulation may occur after several Gyrs of H loss to space. Fig. S16 is the reduced mantle equivalent of Fig. 5 in the main text showing waterworld oxygen false positives. The pressure overburden of a large surface ocean once again shuts down crustal production after ~1 Gyr, thereby removing all crustal oxygen sinks and permitting atmospheric oxygen to accumulate. Fig. S17 is the reduced mantle equivalent of Fig. 7 in the main text showing desertworld oxygen false positives. Although Scenario 3 is apparently unchanged by having a lower mantle redox, a fully self-consistent model that accounted for the high

CO-H₂ content of the originally degassed atmosphere would likely preclude early O₂ accumulation, in practice.

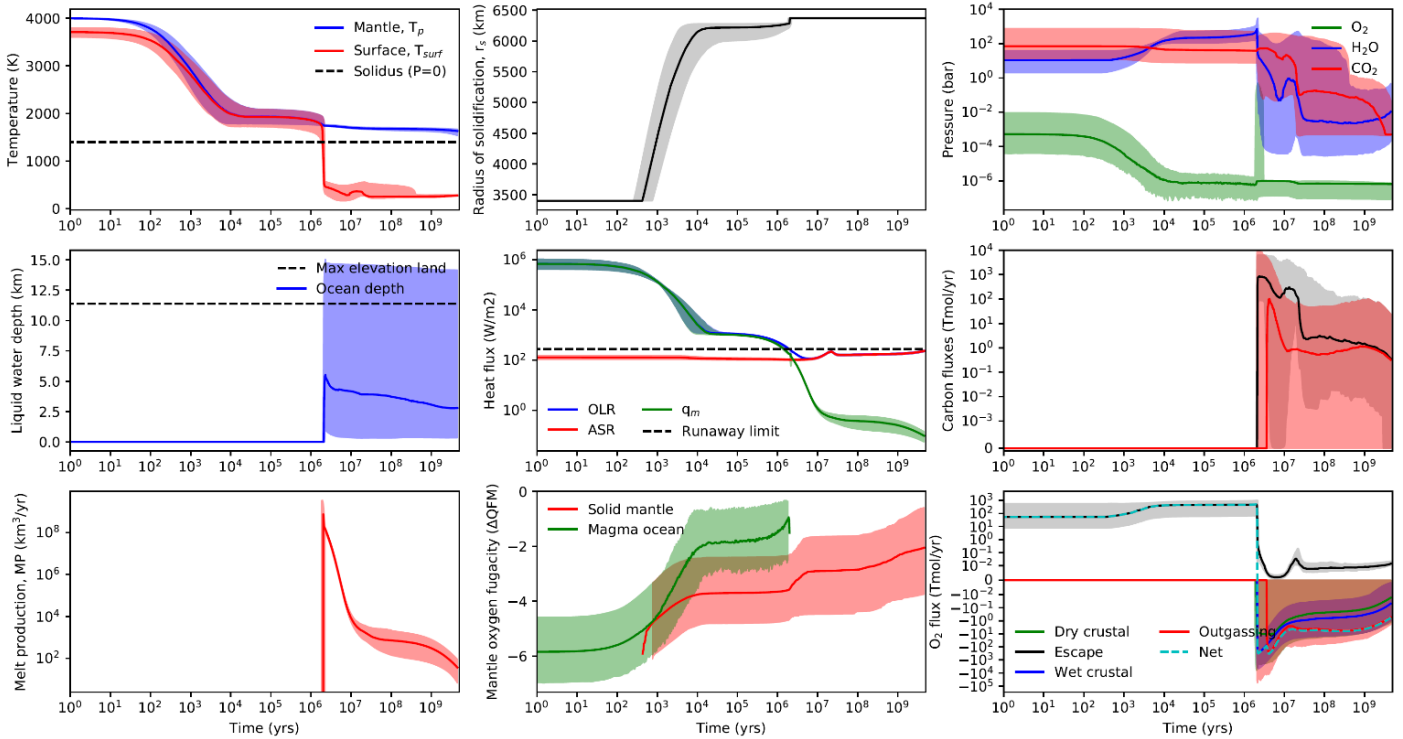

Fig. S14: Nominal Earth evolution with a more reduced initial mantle. This is identical to Fig. 2 in the main text except (i) the initial free oxygen of the mantle is lower (0.5-1.5×10²¹ kg), (ii) graphite saturation in reduced melts is accounted for, (iii) and carbon is partitioned into the solid phase during magma ocean solidification.

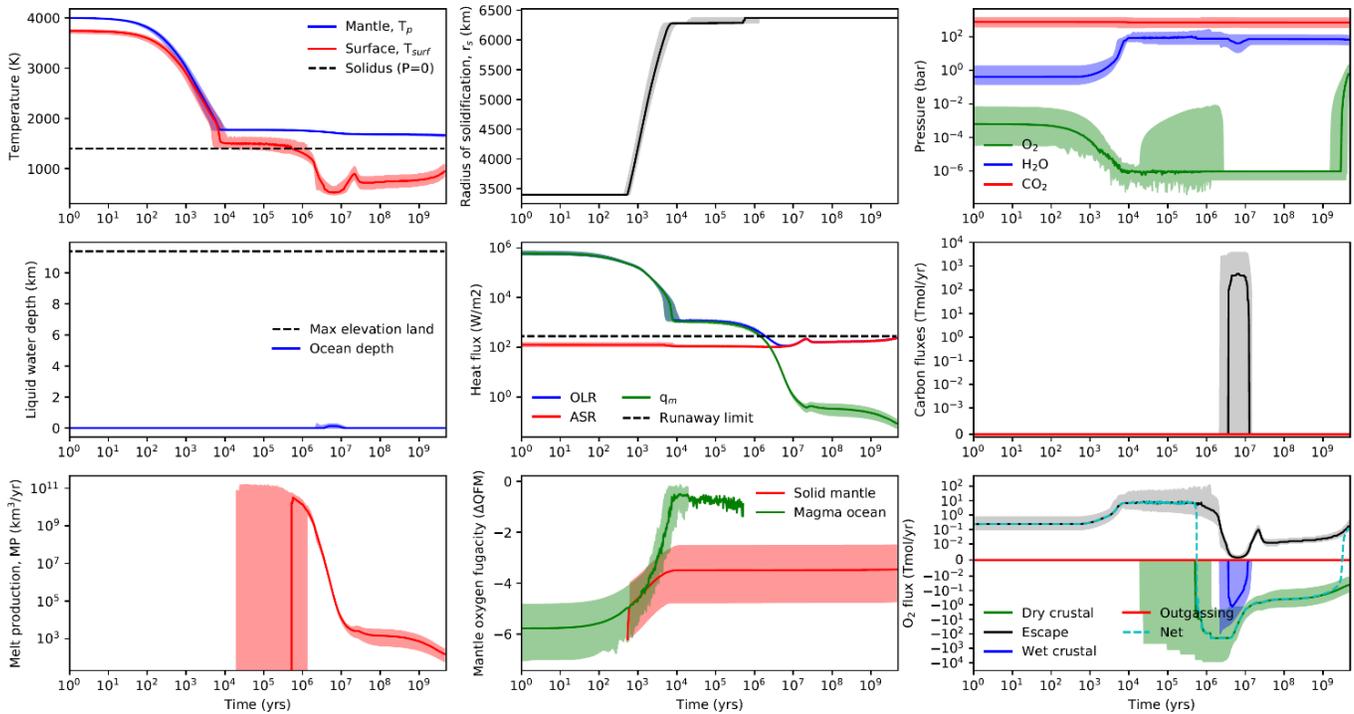

Fig. S15: Scenario 1 oxygen false positives with a more reduced initial mantle. This is identical to Fig. 3 in the main text except (i) the initial free oxygen of the mantle is lower (0.5-1.5×10²¹ kg), (ii) graphite saturation in reduced melts is accounted for, (iii) and carbon is partitioned into the solid phase during magma ocean solidification. The terminal magma ocean becomes more oxidized than the solid mantle as H escape oxidizes the combined melt-volatile reservoir faster than solidification transfers oxidized material to the mantle. Oxygen accumulation may occur after several Gyr because outgassing of C-bearing volatiles is negligible from the graphite-saturated mantle.

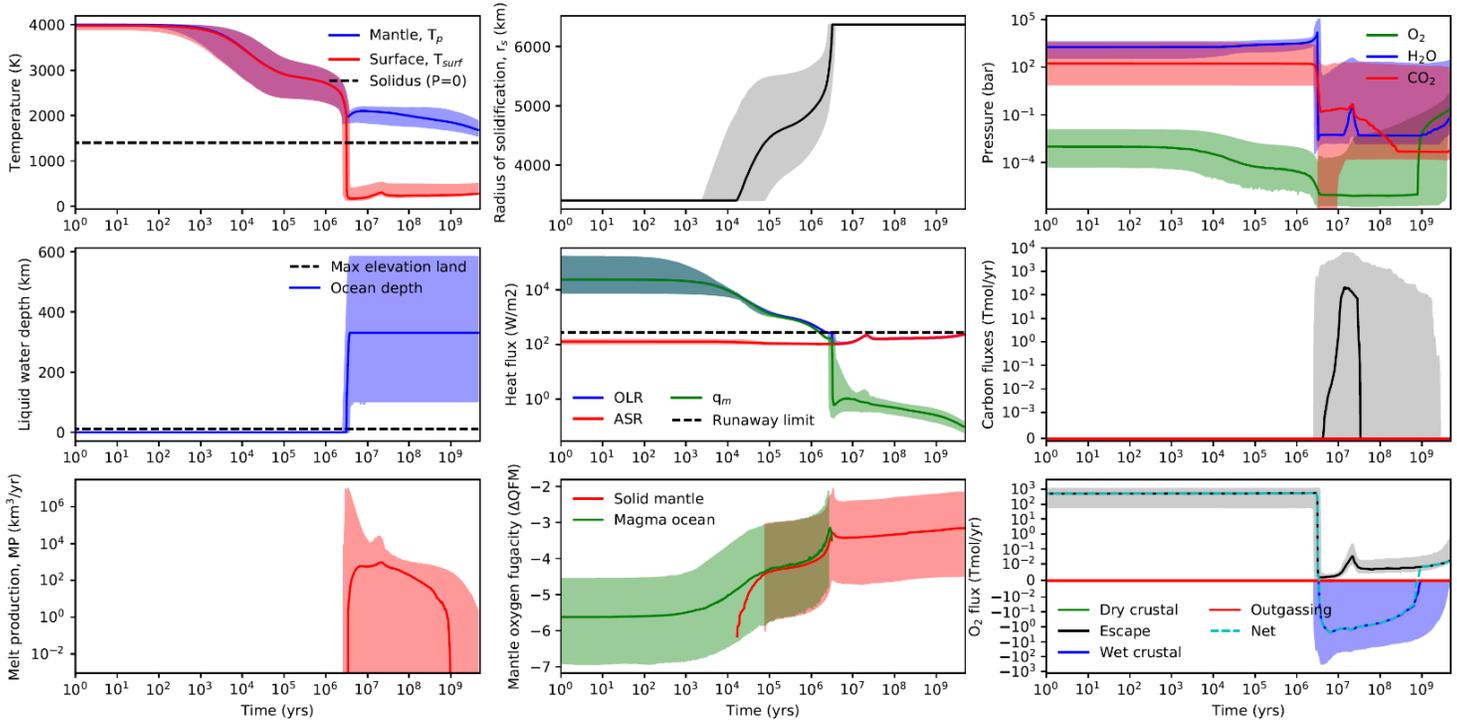

Fig. S16: Scenario 2 oxygen false positives with a more reduced initial mantle. This is identical to Fig. 5 in the main text except (i) the initial free oxygen of the mantle is lower (0.5-1.5×10²¹ kg), (ii) graphite saturation in reduced melts is accounted for, (iii) and carbon is partitioned into the solid phase during magma ocean solidification. Oxygen sinks are suppressed by the large pressure overburden of the surface ocean, the same as in the nominal oxidized-mantle calculations.

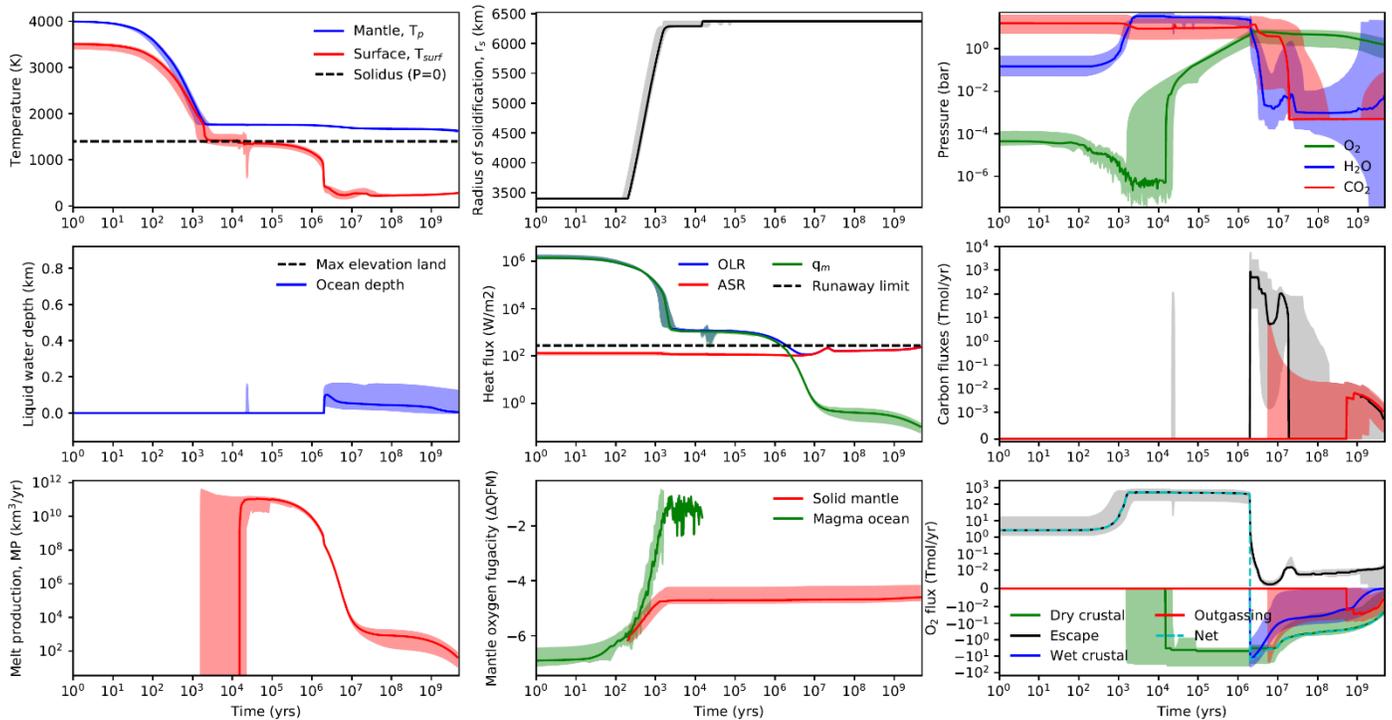

Fig. S17: Scenario 3 oxygen false positives with a more reduced initial mantle. This is identical to Fig. 7 in the main text except (i) the initial free oxygen of the mantle is lower ($0.5$-$1.5 \times 10^{21}$ kg), (ii) graphite saturation in reduced melts is accounted for, (iii) and carbon is partitioned into the solid phase during magma ocean solidification. The terminal magma ocean becomes more oxidized than the solid mantle as H escape oxidizes the combined melt-volatile reservoir faster than solidification transfers oxidized material to the mantle. While the persistence of oxygen is permitted in these calculations, in practice, the early atmosphere is likely too reducing to permit such oxygen accumulation.

**Text H. Stellar Separation Sensitivity Test**

While this study is not an exhaustive exploration of the oxygen false positive parameter space, nominal model calculations were repeated at 1.3 AU to show that oxygen accumulation is not dependent on being close to the inner edge of the habitable zone. Fig. S18 shows all Scenario 2 false positives at 1.3 AU. The increased stellar separation results in lower H escape fluxes, but oxygen accumulation may still occur if crustal sinks are suppressed by pressure overburden. Similarly, Fig. S19 shows all Scenario 3 false positives at 1.3 AU. Oxygen accumulation on desertworlds can similarly occur at larger stellar separations because oxygen accumulation occurs early during the steam-dominated atmosphere. Scenario 1 false positives do not occur at large stellar separations because even under a high $CO_2$ atmosphere, the runaway greenhouse state is not maintained after magma ocean solidification.

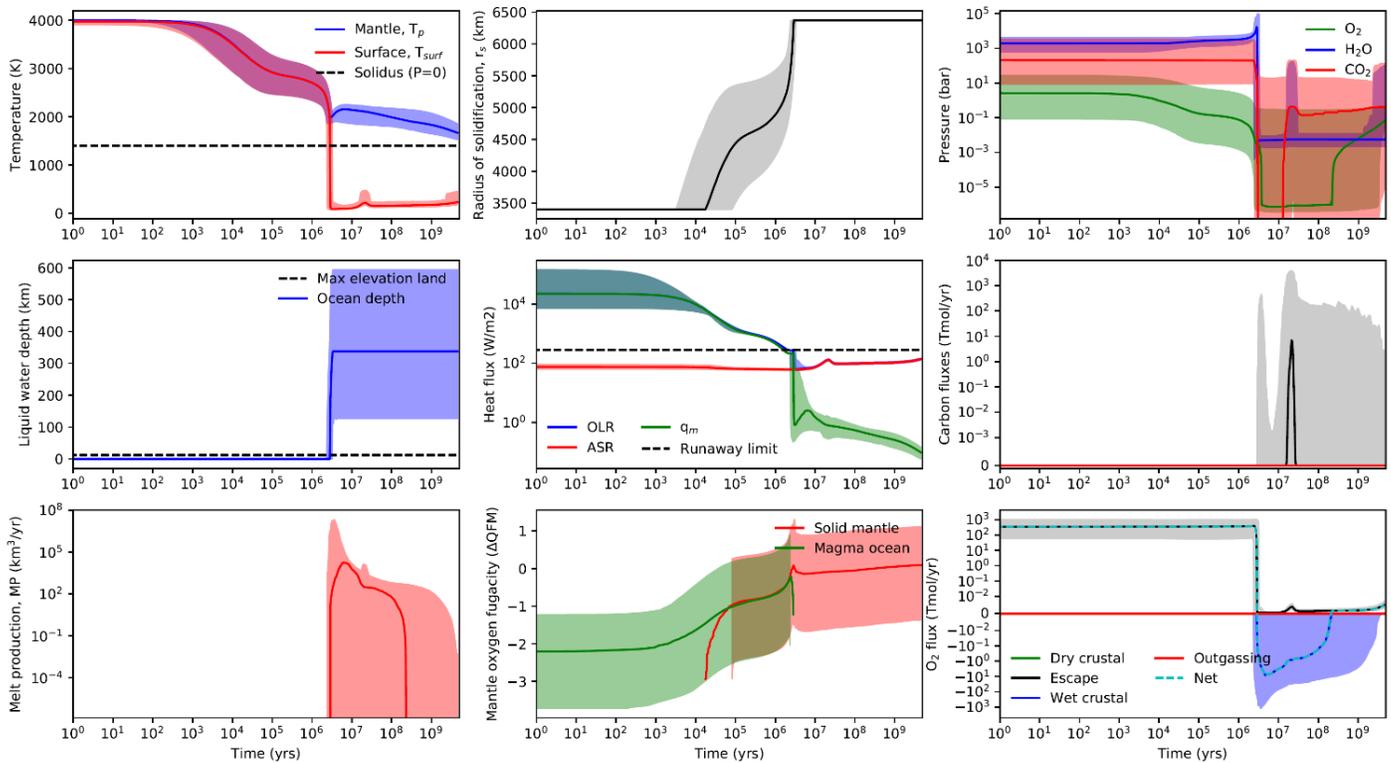

Fig. S18: Identical to Fig. 5 in the main text except the assumed planet-star separation is 1.3 AU. Abiotic oxygen accumulation is still permitted due to overburden pressure suppressing oxygen sinks.

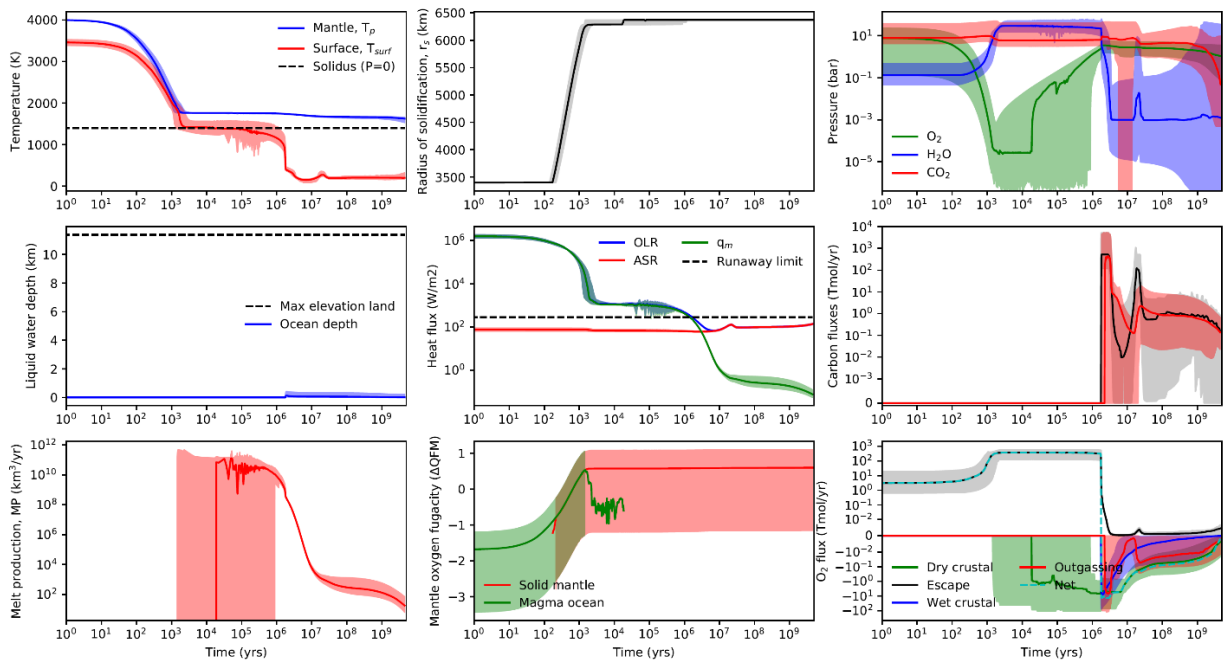

Fig. S19: Identical to Fig. 7 in the main text except the assumed planet-star separation is 1.3 AU. Abiotic oxygen accumulation is still permitted due to early oxygen accumulation during the steam-atmosphere phase.

**Text I. The Effect of Sulfur Outgassing**

Sulfur outgassing and burial may have played an important role in the oxygenation of Earth's atmosphere (e.g. Gaillard et al. 2011; Olson et al. 2019). While including a complete model of sulfur cycling is beyond the scope of this study, we present calculations showing that Earth-like sulfur mantle abundances are unlikely to qualitatively change our conclusions.

Following Gaillard and Scaillet (2014), sulfur speciation is added to our outgassing model by adding the following system of equations to those already described in Wogan et al. (2020):

$$ln(x_{S_2}) = 0.5 \times ln(p_{S_2}) - 0.5 \times \ln(f_{O_2}) - 15.274$$
$$x_{S_2-total}P = (p_{S_2} + 2p_{H_2S} + 2p_{SO_2})\alpha_{gas} + (1 - \alpha_G)x_{S_2}P$$
$$ln(p_{SO_2}) = \ln\left(K_6 f_{O_2}\right) + 0.5 \times ln(p_{S_2})$$
$$0.5 \times ln(p_{S_2}) = -ln(p_{H_2O}) + \ln(K_5 f_{O_2}^{0.5}) + ln(p_{H_2S})$$

$$(76)$$

Here, $p_n$ refers to the gas phase partial pressure of species $n$, $x_{S_2-total}$ is the total sulfur mass fraction of the original melt, $\alpha_G$ is the gas phase mole fraction, $f_{O_2}$ is the oxygen fugacity of the melt. The equilibrium constants $K_5$ and $K_6$ are defined as follows:

$$K_5 = exp(1.90560415 \times 10^4 / T - 0.860366131)$$
$$K_6 = exp(4.35250424 \times 10^4 / T - 8.80403494)$$

$$(77)$$

The equations in (76) are solved simultaneously with the system of equations describing carbon, oxygen, and hydrogen gas-melt speciation (Wogan et al. 2020) to determine total outgassing fluxes, and the overall outgassing oxygen sink:

$$V_{O_2-sink} = 0.5V_{H_2} + 0.5V_{CO} + 2V_{CH_4} + 1.5V_{H_2S} + 2V_{S_2}$$

$$(78)$$

Rather than attempt to model complete sulfur cycling, model outputs from Fig. 2 were post-processed to recalculate outgassing fluxes and oxygen sinks using the above model, assuming constant 300 ppm mantle sulfur abundances (von Gehlen 1992). Fig. S20 shows the results from this calculation comparing oxygen sinks with and without sulfur outgassing. The impact on total oxygen sink fluxes is relatively minor. Fig. S21 shows the same calculations performed on the Scenario 1 false positive outputs from Fig. 3. In this case, the high pressure for the dense $CO_2$ atmosphere prevents the exsolution of sulfur-bearing gases from the melt, and so total oxygen sink fluxes with and without sulfur are barely distinguishable. Moreover, for Scenario 2, the shutdown of crustal production inhibits all crustal sinks, regardless of volatile content.

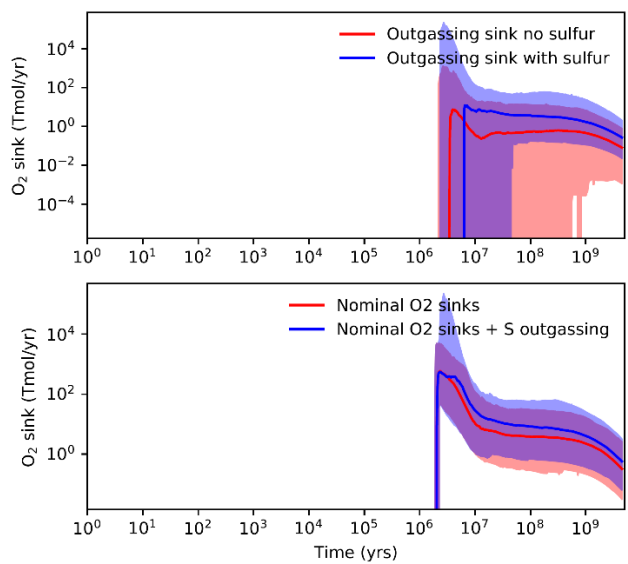

Fig. S20: Comparison of oxygen sinks with and without sulfur-bearing outgassed volatiles. Subplot (a) shows outgassing sinks in the nominal model from Fig. 2 (red) compared to outgassing sinks if sulfur-bearing species are included (blue) and a constant 300 ppm mantle sulfur concentration is assumed (von Gehlen 1992). Subplot (b) shows total oxygen sinks with (blue) and without (red) sulfur-bearing volatiles. While the inclusion of sulfur-bearing species may result in a slightly larger oxygen sink, for Earth-like mantle concentrations the effect of sulfur outgassing is minimal.

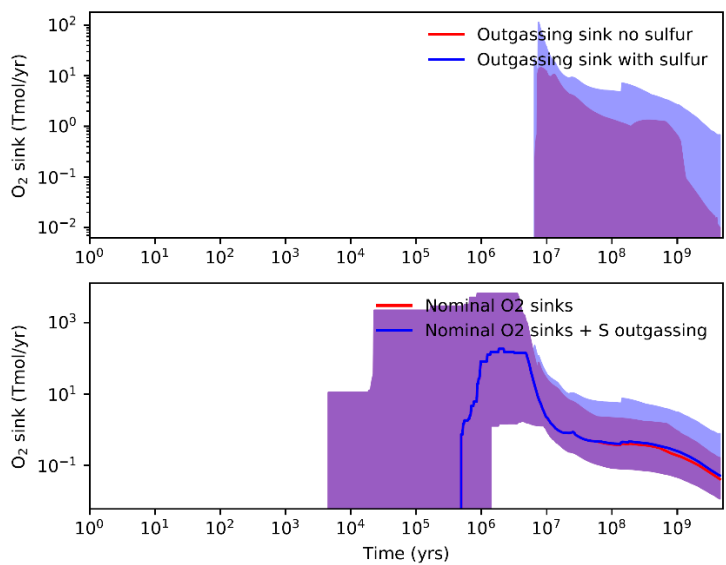

Fig. S21: Comparison of oxygen sinks with and without sulfur-bearing outgassed volatiles. Subplot (a) shows outgassing sinks for Scenario 1 false positives from Fig. 3 (red) compared to outgassing sinks if sulfur-bearing species are included (blue) and a constant 300 ppm mantle sulfur concentration is assumed (von Gehlen 1992). Subplot (b) shows total oxygen sinks with (blue) and without (red) sulfur-bearing volatiles. Including sulfur-bearing species has a negligible effect on outgassing sinks because the pressure of the dense $CO_2$ atmosphere inhibits exsolution of sulfur gases.

**Table S1.** Monte Carlo analysis and uncertain parameter ranges (for nominal Earth model).

| | | Nominal range | References/Notes |
|---|---|---|---|
| Initial conditions | Water | $10^{21}$-$10^{22}$ kg* | Approximately 1 – 10 Earth oceans |
| | Carbon dioxide | $10^{20}$-$10^{22}$ kg* | Approximately 20-2000 bar (if no other atmospheric constituents). |
| | Radiogenic inventory (relative Earth) | 0.33-3.0 | Scalar multiplication of inventories in Lebrun et al. (2013) |
| | Mantle free oxygen | $2\times10^{21}$-$6\times10^{21}$ (kg) | This ensures a post-solidification mantle redox around Quartz-Fayalite-Magnetite buffer. |
| Solar evolution and escape parameters | Early sun rotation rate (relative modern) | 1.8-45 | (Tu et al. 2015) Implies XUV saturation time of 6 – 226 Myrs. |
| | Escape efficiency at low XUV flux, $\varepsilon_{lowXUV}$ | 0.01-0.3 | See escape section. |
| | Transition parameter for cold-trap diffusion limited to XUV-limited escape, $\lambda_{tra}$ | $10^{-2}$ – $10^{2}$ | See escape section. |
| | XUV energy that contributes to XUV escape above hydrodynamic threshold, $\zeta$ | 0-100% | See escape section. |
| Carbon cycle parameters | Temperature-dependence of continental weathering, $T_{efold}$ | 5-30 K | (Krissansen-Totton et al. 2018) |
| | $CO_2$-dependence of continental weathering, $\gamma$ | 0.1-0.5 | (Krissansen-Totton et al. 2018) |
| | Weathering supply limit, $W_{sup-lim}$ | $10^{5}$ – $10^{7}$ kg/s | (Foley 2015) |
| | Ocean calcium concentration, $\left[ Ca^{2+} \right]$ | $10^{-4}$ – $3\times10^{-1}$ mol/kg | See text for explanation (Halevy & Bachan 2017; Kite & Ford 2018) |
| | Ocean carbonate saturation, $\Omega$ | 1-10 | (Zeebe & Westbroek 2003) |
| Interior evolution parameter | Mantle viscosity coefficient, $V_{coef}$ | $10^{1}$ – $10^{3}$ Pa s | Fit to modern heatflow and melt production (see Fig. S3) |
| Crustal sinks oxygen and | Crustal hydration efficiency, $fr_{hydr-frac}$ | $10^{-3}$ to 0.03 | Upper limit wt % $H_2O$ in oceanic crust. Lower limit hydration limited by cracking. |

| hydrological cycle parameters | Dry oxidation efficiency, $f_{dry-oxid}$ | $10^{-4}$ to 10% | Plausible range of processes for Venus (Gillmann et al. 2009) |
|---|---|---|---|
| | Wet oxidation efficiency, $f_{wet-oxid}$ | $10^{-3}$ to $10^{-1}$ | Based on oxidation of Earth's oceanic crust (Lécuyer & Ricard 1999). |
| | Maximum fractional molten area, $f_{lava}$ | $10^{-4}$ to 1.0 | Refer Supplementary Text A.13 |
| | Max mantle water content, $M_{solid-H_2O-max}$ | 0.5-15 Earth oceans | (Cowan & Abbot 2014) |
| Albedo parameters | Hot state albedo, $A_H$ | 0-0.3 | See albedo parameterization. |
| | Cold state albedo, $A_C$ | 0.25-0.35 | See albedo parameterization. |

\* We apply the additional constraint that the initial water inventory is greater than the initial carbon dioxide inventory, in accordance with typical carbonaceous chondrite abundances.

**Table S2**: Changes in Monte Carlo parameters for different abiotic oxygen scenarios and Venus validation:

| | | Nominal range | High $CO_2$:$H_2O$ | Waterworlds | Desertworlds | Venus |
|---|---|---|---|---|---|---|
| Initial cond. | $H_2O$ | $10^{21}$-$10^{22}$ kg* | $10^{20}$-$10^{22}$ kg** | $10^{22}$-$10^{23.5}$ kg | $10^{19.9}$-$10^{20.7}$ kg | $10^{20}$-$10^{22}$ kg |
| | $CO_2$ | $10^{20}$-$10^{22}$ kg* | $10^{20}$-$10^{22}$ kg** | $10^{20}$-$10^{22.5}$ kg | $10^{19.5}$-$10^{20.5}$ kg | $10^{20}$-$10^{22}$ kg |
| Albedo param. | Hot, $A_H$ | 0.0-0.3 | 0.0-0.3 | 0.0-0.3 | 0.0-0.3 | 0-0.3 |
| | Cold, $A_C$ | 0.25-0.35 | 0.25-0.35 | 0.25-0.35 | 0.25-0.35 | 0.2-0.7 |

\* We apply the additional constraint that the initial water inventory is greater than the initial carbon dioxide inventory, in accordance with typical carbonaceous chondrite abundances.
\*\* High $CO_2$:$H_2O$ runs are a subset of this range with initial $CO_2$/$H_2O$>1 (by mass).

**Table S3**: All fixed parameters used in the model.

| Parameter | Value |
|---|---|
| Planetary iron content (silicate mole fraction), $x_{Fe}$ | 0.06 |
| Average silicate density, $\rho_m$ | 4000 kg/m$^3$ |
| Planetary radius, $r_p$ | 6371 km (Earth) or 6052 km (Venus) |
| Core radius, $r_c$ | 3460 km (Earth) or 3230 km (Venus) |
| Planet mass, $M_P$ | 5.972×10$^{24}$ kg (Earth) or 4.867×10$^{24}$ kg (Venus) |
| Latent heat of fusion of silicates, $H_{fusion}$ | 4×10$^5$ J/kg |
| Specific heat of silicates, $c_p$ | 1200 J/kg/K |
| Thermal expansion coefficient for silicates, $\alpha$ | $2\times10^{-5}$ K$^{-1}$ |
| Critical Rayleigh number, $Ra_{crit}$ | 1100 |
| Thermal conductivity of silicates, $k$ | 4.2 W/m/K |
| Thermal diffusivity of silicates, $\kappa$ | $10^{-6}$ m$^2$/s |
| Convective heatflow exponent, $\beta$ | 1/3 |
| Molecular mass of i-th species, $\mu_i$ | Various (kg/mol) |
| Crystal-melt partition coefficient for water, $k_{H_2O}$ | 0.01 |
| Crystal-melt partition coefficient for carbon dioxide, $k_{CO_2}$ | $2\times10^{-3}$ |
| Binary diffusion coefficient of the i-th species through the j-th species, $b_{i-j}$ | Various (mol/m/s) |
| Stratospheric temperature, $T_{strat}$ | 200 K |
| Inverse molar mass of magma, $\vartheta_m$ | 15.5 mol/kg |
| Weathering multiplicative coefficient, $W_{coef}$ | 4000 kg/s |
| Activation energy seafloor weathering, $E_{SF}$ | 90 kJ/mol |
| Mass of i-th species, $m_i$ | Various (kg) |
| Planet-star separation, $D_{planet-star}$ | $1.496\times10^{11}$ m (Earth), $1.047\times10^{11}$ m (Venus) |